\documentclass{aa}

\usepackage{microtype}
\usepackage{amsmath}
\usepackage{color}
\usepackage{graphicx}
\usepackage{hyperref}
\usepackage[varg]{txfonts}
\usepackage{caption}
\usepackage{subcaption}
\usepackage{pifont}

\usepackage{natbib}
\bibpunct{(}{)}{;}{a}{}{,} 
\usepackage{xargs}                      
\usepackage{enumitem}

\graphicspath{ {images/} }

\begin{document}

   \title{Multiwavelength analysis of the X-ray spur and southeast of the Large Magellanic Cloud}

   \subtitle{}

   \author{J. R. Knies\inst{\ref{inst1}} \and M. Sasaki\inst{\ref{inst1}} \and Y. Fukui\inst{\ref{inst2}} \and K. Tsuge\inst{\ref{inst2}}  \and F. Haberl\inst{\ref{inst3}} \and S. Points\inst{\ref{inst4}} \and P. J. Kavanagh\inst{\ref{inst5}} \and M.~D. Filipovi\'c\inst{\ref{inst6}}}

   \institute{ Dr. Karl Remeis Observatory and ECAP, Universit\"at Erlangen-N\"urnberg, Sternwartstra{\ss}e 7, D-96049 Bamberg, Germany\label{inst1} \and
   Department of Physics, Nagoya University, Furo-cho, Chikusa-ku, Nagoya 464-8601, Japan\label{inst2} \and
   Max-Planck-Institut f\"ur extraterrestrische Physik, Gie{\ss}enbachstra{\ss}e 1, D-85741 Garching, Germany\label{inst3} \and
   Cerro Tololo Inter-American Observatory, Casilla 603, La Serena, Chile\label{inst4} \and
   School of Cosmic Physics, Dublin Institute for Advanced Studies, 31 Fitzwilliam Place, Dublin 2, Ireland\label{inst5} \and
   Western Sydney University, Locked Bag 1797, Penrith South DC, NSW 2751, Australia\label{inst6}
	}

   \date{\today}
   \date{Received 25 May 2020 /
Accepted 26 January 2021}

 \abstract
   {}
   {
The giant \ion{H}{ii} region 30 Doradus (30 Dor) located in the eastern part
of the Large Magellanic Cloud is one of the most active star-forming
regions in the Local Group. Studies of \ion{H}{i} data have revealed two large
gas structures which must have collided with each other in the region
around 30 Dor. In X-rays there is extended emission ($\sim 1$~kpc)
south of 30 Dor called the X-ray spur, which appears to be
anticorrelated with the \ion{H}{i} gas.  We study the properties of the hot interstellar medium (ISM) in the X-ray spur and investigate its origin including related interactions in the ISM.}
   {We analyzed new and archival XMM-Newton data of the X-ray spur
and its surroundings to determine the properties of the hot diffuse
plasma. We created detailed plasma property maps by utilizing the Voronoi tessellation algorithm. We also studied \ion{H}{i} and CO data, as well as optical line
emission data of H$\alpha$ and [\ion{S}{ii}], and compared them to the results of the X-ray spectral analysis.}
   {We find evidence of two hot plasma
components with temperatures of $kT_1 \sim 0.2$~keV and $kT_2 \sim
0.5-0.9$~keV, with the hotter component being much more pronounced
near 30 Dor and the X-ray spur. In 30 Dor, the plasma has most likely
been heated by massive stellar winds and supernova remnants. In the
X-ray spur, we find no evidence of heating by stars. Instead, the
X-ray spur must have been compressed and heated by the collision of
the \ion{H}{i} gas. }
   {}


	\titlerunning{Multiwavelength analysis of the X-ray spur}
	\authorrunning{Knies et al.}

  \maketitle

\section{Introduction}
The Large Magellanic Cloud (LMC) is a nearby satellite galaxy of the Milky Way at a distance of $\sim 50$~kpc. Because of the unobstructed view toward it, the LMC is ideal for studying the interstellar medium (ISM) at different wavelengths. Advances in observational techniques in the last decades allow for detailed studies of interactions between different phases of the ISM.
\citet{HI:LMC_rohlfs}  found a system with two distinct \ion{H}{i} components, the L- and D-components, based on a study of the \ion{H}{i} gas in the LMC at 15 arcmin resolution. The D-component covers most of the LMC and is located in the plane of the galaxy disk. The L-component, on the other hand, is much more localized and has radial velocities that are $\sim 30\mathrm{-} 60$~km\,s$^{-1}$ lower than the D-component. The active star-forming region 30 Doradus (30 Dor) is located in the L-component, while the L-component could be located in front of the D-component near 30 Dor. Based on the measured velocities, \citet{HI:LMC_rohlfs} concluded that there was a past encounter between the two components $\sim 10^7$~yr ago.

Recently, \citet{HI:fukui_R136} have analyzed the ATCA+Parkes \ion{H}{i} 21cm survey at 1 arcmin resolution \citep{HI:atca_parkes}, 15 times higher than that by \citet{HI:LMC_rohlfs}, and they confirm the L- and D-components.
\citet{HI:fukui_R136} present that the D-component shows complementary distribution with the L-component \citep[see Fig. 3 of][]{HI:fukui_R136}. Such a complementary distribution is a typical observational signature of triggered high-mass star formation by cloud-cloud collision and has been used in identifying the collision in Galactic super star clusters and \ion{H}{ii} regions \citep[e.g., Westerlund2, NGC3603, M42, and M43]{furukawa_2009,fukui_2014,fukui_2016, fukui_2018_m42}. Additionally, they found a third component, the I-component, with radial velocities between the L- and D-component, indicating interactions between them. They also observed a large elongated complex of molecular clouds seen in CO just west of 30 Doradus, called the CO ridge. The CO ridge is also located toward the \ion{H}{i} L- and I-components \citep{fukui_1999, mizuno_2001, co_lmc}. It is likely that the \ion{H}{i} gas was driven by the tidal force in a close encounter 0.15-0.2 Gyr ago between the LMC and the Small Magellanic Cloud (SMC) as proposed by \citet{HI:smc_lmc_tidal_scenario}. According to the scenario, gas was stripped from the SMC during the encounter and remained bound in the LMC, resulting in ``the highly asymmetric'' \ion{H}{i} and CO gas distribution. Detailed numerical simulations of the galactic interaction by \citet{tidal_simulation2} reproduced the asymmetry, lending support for the scenario.
Additional evidence was presented by \citet{n44_tsuge}, who investigated N44, which is one of the most active star-forming regions in the LMC. They conclude that the collision between LMC gas and the L-component at the location of N44 was similar to R136, but it occurred earlier due to the geometry of the systems.
Based on the Planck/IRAS dust emission, \citet{HI:fukui_R136} and \citet{n44_tsuge} have shown that the dust-to-HI ratio is lower in the regions around R136 and N44 than in the stellar bar of the LMC, indicating that a lower-metallicity gas from the SMC has been mixed into the LMC in a collision. \citet{av_lmc} found low dust-to-\ion{H}{i} ratio near the \ion{H}{i} ridge, which is also consistent with the mixing of the SMC and the LMC gas.

The position of the CO ridge also coincides with a structure observed in X-rays, called the X-ray spur. The X-ray spur was observed by the ROSAT All-Sky survey (RASS) and analyzed by \citet{LMC_xray_shadow} and \citet{spur:2001}. Located just south of the supergiant shell LMC 2 (LMC-SGS~2) and 30 Dor, the spur is a large diffuse X-ray emitting structure with an extent of $\sim 1$~kpc. A uniform foreground absorption was measured for the spur, with a strongly absorbed region just west of it, the ``X-ray shadow''. The X-ray spur also appears to be located where the L- and D-components collide. Therefore, the X-ray spur is a perfect candidate to study the various interactions between the different phases of the ISM. Using new and archival XMM-Newton X-ray data we performed a multiwavelength analysis with additional radio, submillimeter, and optical data to investigate the ISM in the spur and its surroundings.

\section{Data}
\subsection{X-ray}
We use new and archival data obtained with the European Photon Imaging Camera (EPIC) on-board XMM-Newton \citep{xmm_general}.
Most of the observations were made as part of the LMC survey (PI: F. Haberl) which covers most of the X-ray bright regions of the LMC \citep{xmm:lmc_survey1,xmm:lmc_survey2}. The exposure times were on average $\sim 30$~ks which allows spectral analysis of the diffuse emission. Several other observations were pointed at SN 1987A, also with diffuse emission of 30 Dor in the field of view (FOV). For background studies we used three XMM-Newton observations of the south ecliptic pole (SEP). Additionally, we observed the southern tip of the X-ray spur in October 2018 (ObsID: 0820920101, PI: M. Sasaki) with an average exposure time of 32\,ks over all EPIC detectors after background flare filtering. All observations used in this study are listed in Table~\ref{tab:list_xmm_observations}.
\begin{table}
	\centering
	\caption{\label{tab:list_xmm_observations} List of XMM-Newton observations used for spectral analysis. The exposure time was calculated as average over all available data and detectors after filtering of background flares.}
	\begin{tabular}{r|rr|rl}
		ObsID & RA [$\degr$] & Dec [$\degr$] & Exp. [ks] & PI\\
		\hline
		0086770101 & 81.79 & -70.01 & 44.1 & M. Orio\\
		0094410101 & 85.64 & -69.06 & 10.5 & Y. Chu\\
		0094410201 & 85.74 & -69.47 & 10.6 &  Y. Chu\\
		0094411501 & 85.71 & -69.81 & 4.6 & Y. Chu\\
		0125120101 & 85.00 & -69.36 & 29.6 & F. Jansen\\
		0127720201 & 81.17 & -70.23 & 19.0 & F. Jansen\\
		0137551401 & 81.67 & -69.58 & 35.4 & F. Jansen\\
		0148870501 & 84.36 & -70.55 & 21.7 & M. Orio\\
		0201030101 & 86.73 & -69.58 & 9.8 & Y. Chu\\
		0201030201 & 86.73 & -69.19 & 10.1 & Y. Chu\\
		0201030301 & 85.83 & -70.25 & 9.8 & Y. Chu\\
		0304720201 & 84.49 & -70.58 & 9.9 & M. Orio\\
		0402000701 & 83.08 & -70.64 & 19.2 & F. Haberl \\
		0406840301 & 83.95 & -69.27 & 71.5 & F. Haberl\\
		0679380101 & 85.29 & -69.01 & 7.1 & N. Schartel\\
		0690744401 & 84.62 & -68.98 & 21.9 & F. Haberl\\
		0690744601 & 82.69 & -69.15 & 30.6 & F. Haberl\\
		0690744701 & 81.85 & -69.38 & 30.1 & F. Haberl\\
		0690744801 & 82.38 & -69.53 & 28.5 & F. Haberl\\
		0690744901 & 83.14 & -69.44 & 21.7 & F. Haberl\\
		0690745001 & 84.20 & -69.55 & 24.5 & F. Haberl\\
		0690750101 & 83.40 & -69.78 & 27.6 & F. Haberl\\
		0690750201 & 82.48 & -69.81 & 15.0 & F. Haberl\\
		0690750301 & 85.14 & -69.97 & 25.3 & F. Haberl\\
		0690750401 & 84.18 & -69.95 & 26.8 & F. Haberl\\
		0690750501 & 84.68 & -70.23 & 27.1 & F. Haberl\\
		0690750601 & 83.68 & -70.17 & 29.7 & F. Haberl\\
		0690750701 & 82.90 & -70.15 & 25.9 & F. Haberl\\
		0690750801 & 82.30 & -70.39 & 23.2 & F. Haberl\\
		0690751201 & 81.42 & -70.52 & 30.1 & F. Haberl\\
		0690751301 & 87.25 & -69.75 & 25.4 & F. Haberl\\
		0690751401 & 86.58 & -70.02 & 25.7 & F. Haberl\\
		0690751501 & 86.83 & -70.29 & 28.9 & F. Haberl\\
		0690751601 & 85.50 & -70.52 & 26.0 & F. Haberl\\
		0820920101 & 86.19 & -70.52 & 32.2 & M. Sasaki\\
		\hline\hline
		\multicolumn{4}{c}{SEP observations\tablefootmark{$\dagger$}}\\
		\textsc{0162160101} & 90.03 & -66.54 & 11.2 & B. Altieri\\
		0162160301 & 90.05 & -66.54 & 8.4  & B. Altieri\\
		0162160501 & 90.05 & -66.54 & 8.9  & B. Altieri\\
	\end{tabular} 
	\tablefoot{\tablefoottext{$\dagger$}{Used for estimating the local X-ray background.}}
\end{table}
\subsection{\ion{H}{i}}
In this study we used the combined data from the ATCA and Parkes 21\,cm surveys \citep{HI:atca_parkes}. The data have an angular resolution of $\sim 1$~arcmin with a brightness temperature sensitivity of $2.4$~K and velocity resolution of 1.649\,km\,s$^{-1}$. The Galactic rotation was subtracted from the data and processed as discussed in the previous works by \citet{HI:fukui_R136} and \citet{n44_tsuge}. The \ion{H}{i} data were divided into individual component maps by integrating the intensity over the velocity range of the data cube.
\subsection{CO}
For our multiwavelength analysis we also used the \element[][12]{CO} data of the second NANTEN survey of the LMC \citep{co_lmc}. The data were recorded at the ($J = 1 - 0$) transition frequency with the 4m NANTEN telescope, located at the Las Campanas Observatory in Chile, and operated by Nagoya University. The resolution of the data at $115$~GHz is $2.6'$ (half-power beamwidth).
\subsection{Submillimeter}
\label{sec:dust_data}
We used archival data for the dust optical depth $\tau_{\text{353}}$ at 353\,GHz. The optical depth was obtained by a modified black-body fit to the combined Planck and IRAS data \citep{planck_dust} in the range of 353 to 3000\,GHz. The dust maps have a resolution of $\sim 5'$. We used the maps to correlate the X-ray and \ion{H}{i} emission with the optical depth. 
\subsection{Optical}
We used optical narrow band images from the Magellanic Clouds Emission Line Survey (MCELS) in H$\alpha$ at 6563~$\mathring{A}$ and [\ion{S}{ii}] at 6725~$\mathring{A}$. The images were obtained at the Curtis Schmidt Telescope by the Cerro Tololo Inter-American Observatory (CTIO). \citep{mcels}
\section{Analysis}
\subsection{\ion{H}{i}}
\label{sec:HI_images}
\begin{figure*}
	\centering
		\begin{subfigure}[t]{0.49\textwidth}
			\includegraphics[width=\textwidth]{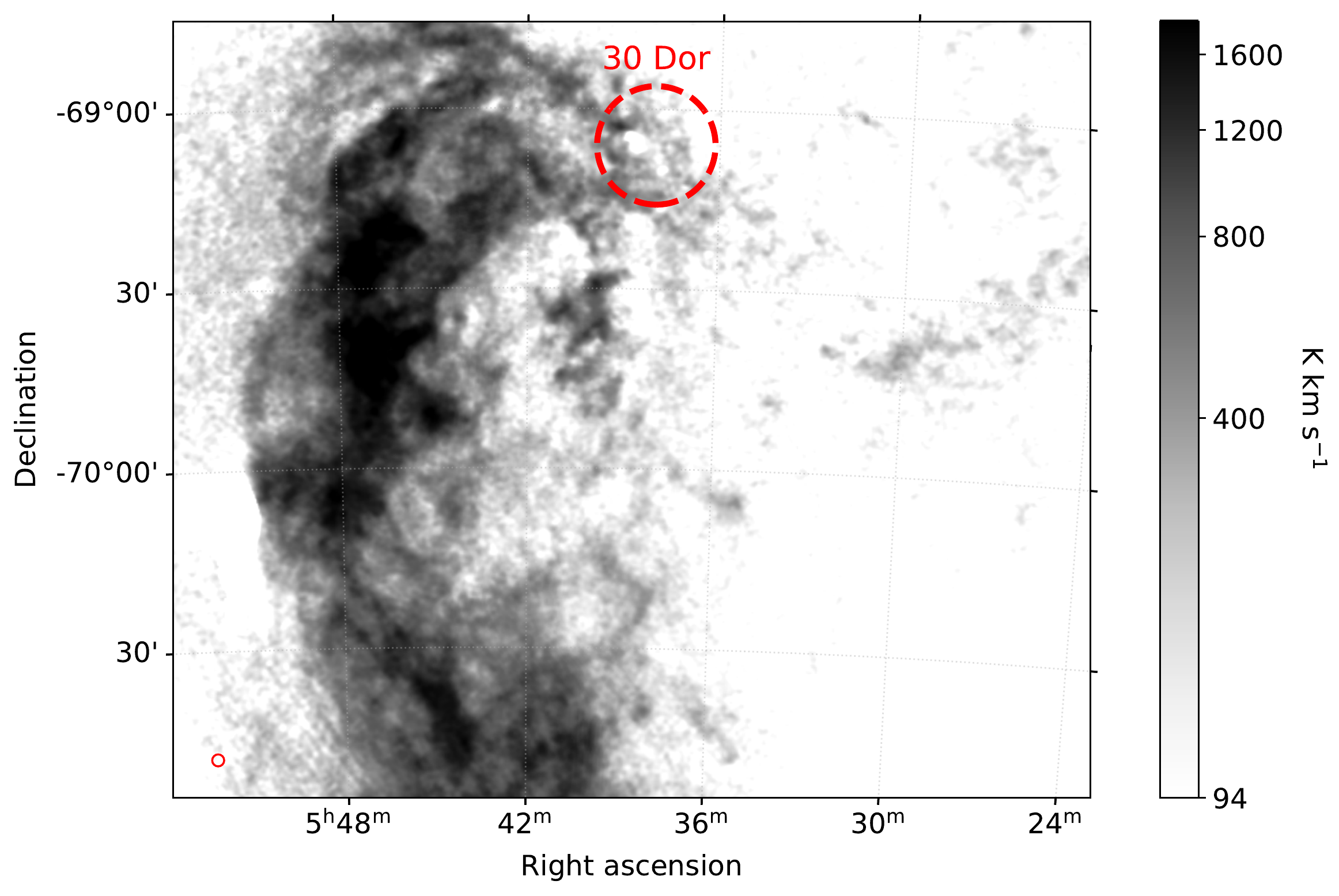}
	\caption{\label{fig:L_comp_inverse}}
		\end{subfigure}
		\hfill
		\begin{subfigure}[t]{0.49\textwidth}
			\includegraphics[width=\textwidth]{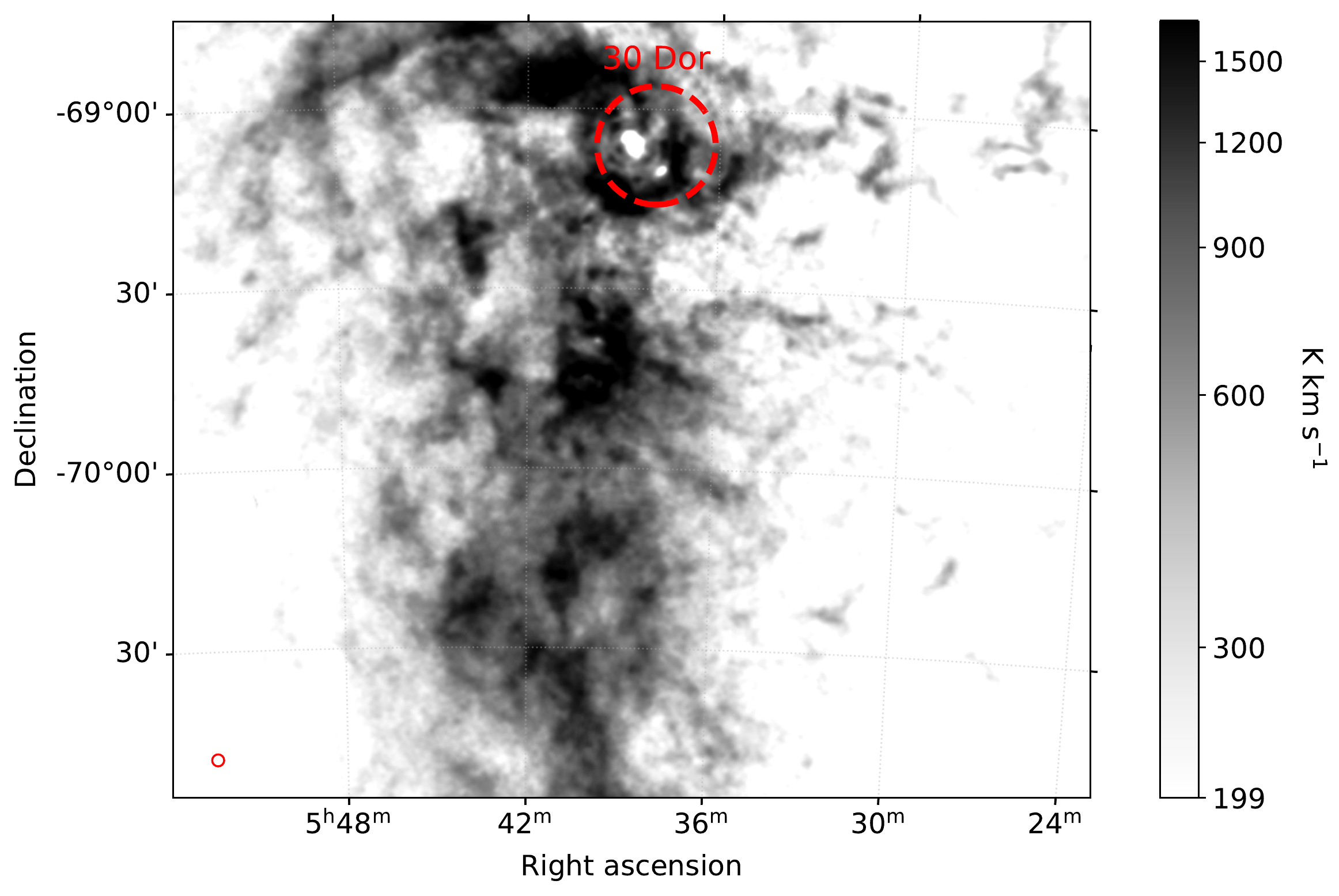}
	\caption{\label{fig:I_comp_inverse}}
		\end{subfigure}
		\\
		\begin{subfigure}[t]{0.49\textwidth}
			\includegraphics[width=\textwidth]{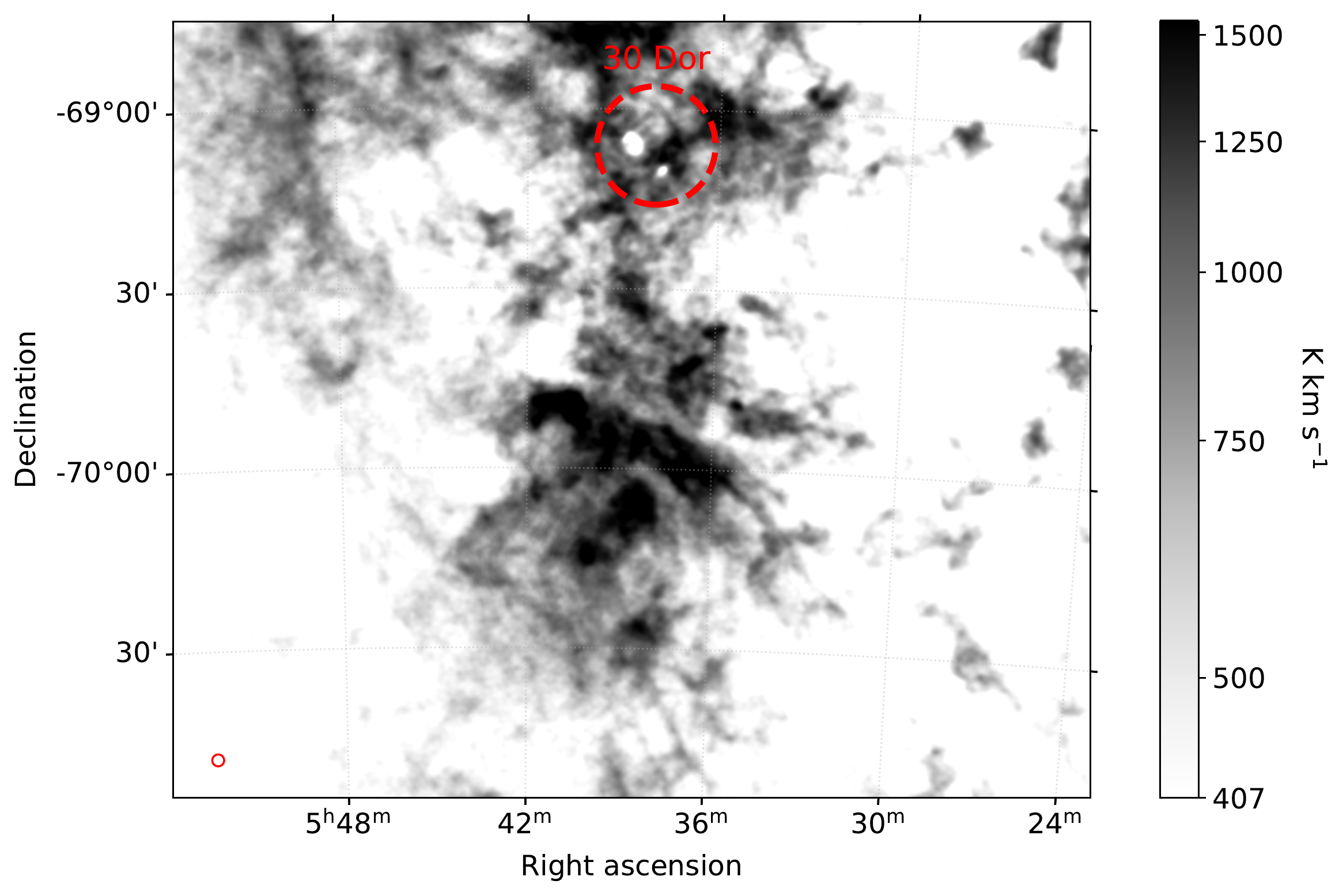}
	\caption{\label{fig:D_comp_inverse}}
		\end{subfigure}
			\caption{\label{fig:HI_images}\ion{H}{i} intensity maps of the different \ion{H}{i} components. The image scales were calculated from the $50\mathrm{-} 99.5$ percentiles. The beam size is indicated with a red circle in the bottom left corner of each panel. (a) L-component, integrated over the velocity range $v_{\mathrm{offset}} = -100$ to $-30$~km\,s$^{-1}$. (b) I-component, integrated over the velocity range $v_{\mathrm{offset}} = -30$ to $-10$~km\,s$^{-1}$. (c) D-component, integrated over the velocity range $v_{\mathrm{offset}} = -10$ to $10$~km\,s$^{-1}$ }
\end{figure*}

\begin{figure*}
	\centering
		\begin{subfigure}[t]{0.49\textwidth}
			\includegraphics[width=\textwidth]{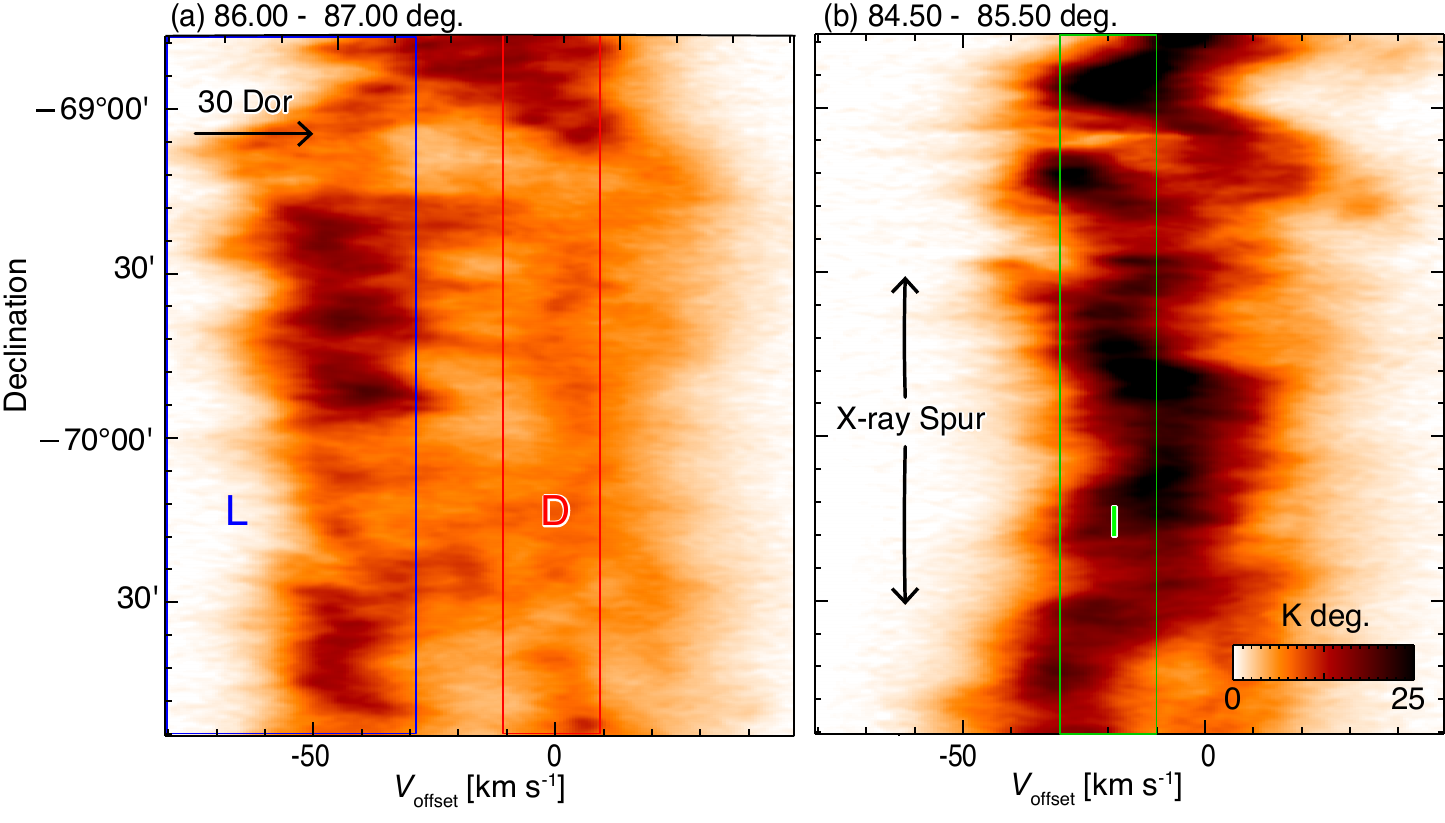}
			\caption{}
		\end{subfigure}
		\hfill
		\begin{subfigure}[t]{0.49\textwidth}
			\includegraphics[width=\textwidth]{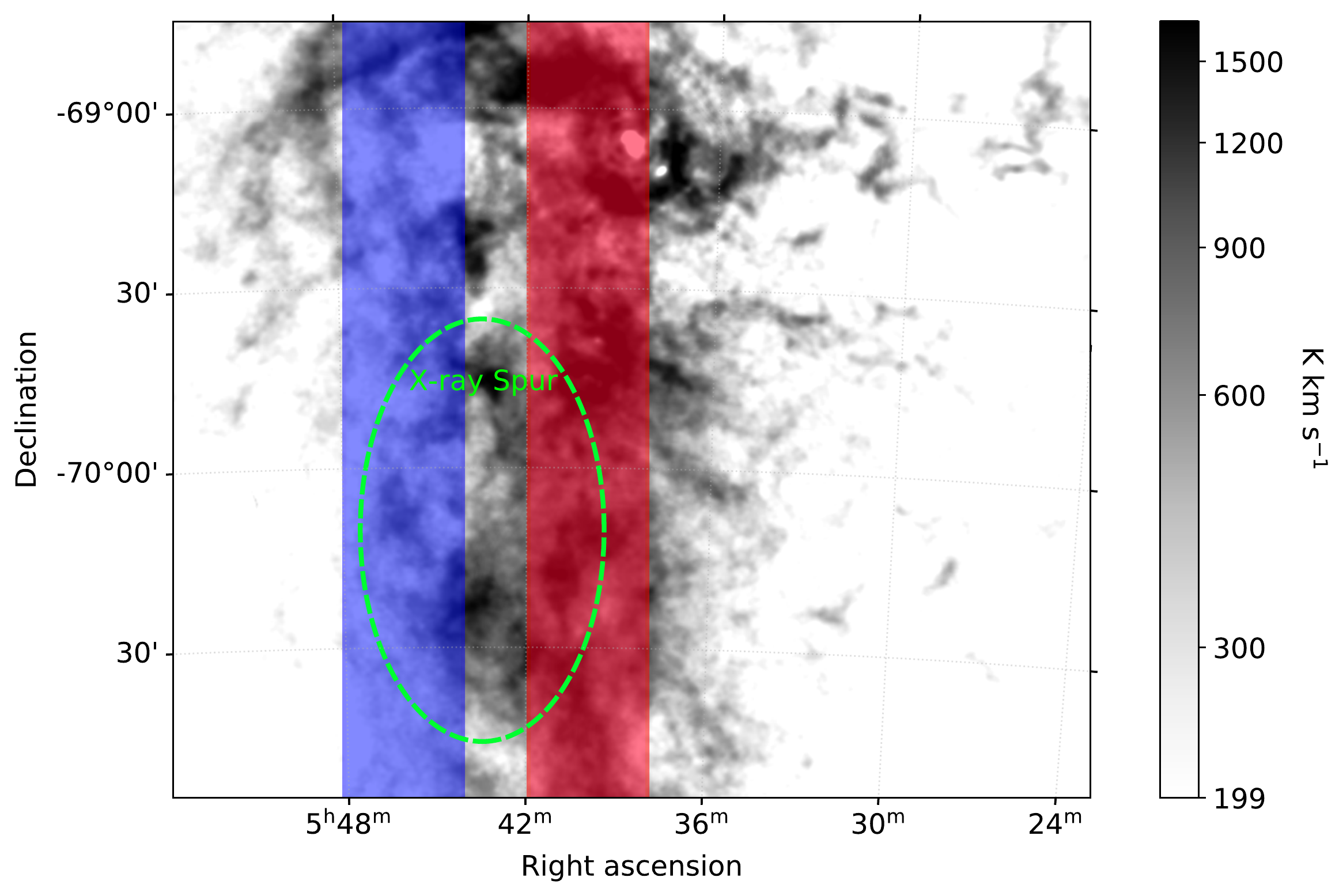}
			\caption{}
		\end{subfigure}
		\caption{\label{fig:pv_diagram} (a) Declination-Velocity diagram of two slices, integrated over the denoted Right ascension range. The approximate position of 30 Dor and the X-ray spur are shown with arrows. The colored boxes indicate the velocity ranges of  the \ion{H}{i} components. The integration slices are indicated in blue (left panel) and red (right panel) in (b), which also shows the \ion{H}{i} I-component integrated in the $-30$ to $-10$~km\,s$^{-1}$ range. The position of the X-ray spur is indicated with a green dashed ellipse.}
\end{figure*}

From the data cubes we created intensity maps for the different \ion{H}{i} components similar to \cite{n44_tsuge}. The intensity maps for the \ion{H}{i} ridge are shown in Fig. \ref{fig:HI_images}. 
The L-component is the low velocity component in the range of $v_{\mathrm{offset}} = -100$ to $-30$~km\,s$^{-1}$. The origin is most likely SMC gas stripped from its galaxy via tidal interaction of a past encounter with the LMC \citep{tidal_simulation1, tidal_simulation2, tidal_simulation3, HI:fukui_R136}.  The D-component or disk component consists of LMC \ion{H}{i} gas and roughly follows the stellar population \citep{HI:LMC_rohlfs} with velocities $v_{\mathrm{offset}}$ between $-10$ to $ 10$~km\,s$^{-1}$. The last component is the recently discovered intermediate component or I-component in between the L- and D-components with $v_{\mathrm{offset}} = -30$ to $-10$~km\,s$^{-1}$ \citep{HI:fukui_R136}. This component is thought to be caused by interactions of the L- and D-components and is especially strong in regions where bridging features in velocity space can be observed.
Indications of interactions between the L- and D-components are clearly visible in position-velocity (PV) space diagrams. An example of two slices near the X-ray spur and \ion{H}{i} ridge is shown in Fig. \ref{fig:pv_diagram}.
\subsection{X-ray}
\label{sec:data_reduction}
\subsubsection{Image production}
\begin{figure}
	\centering
		\includegraphics[width=0.49\textwidth]{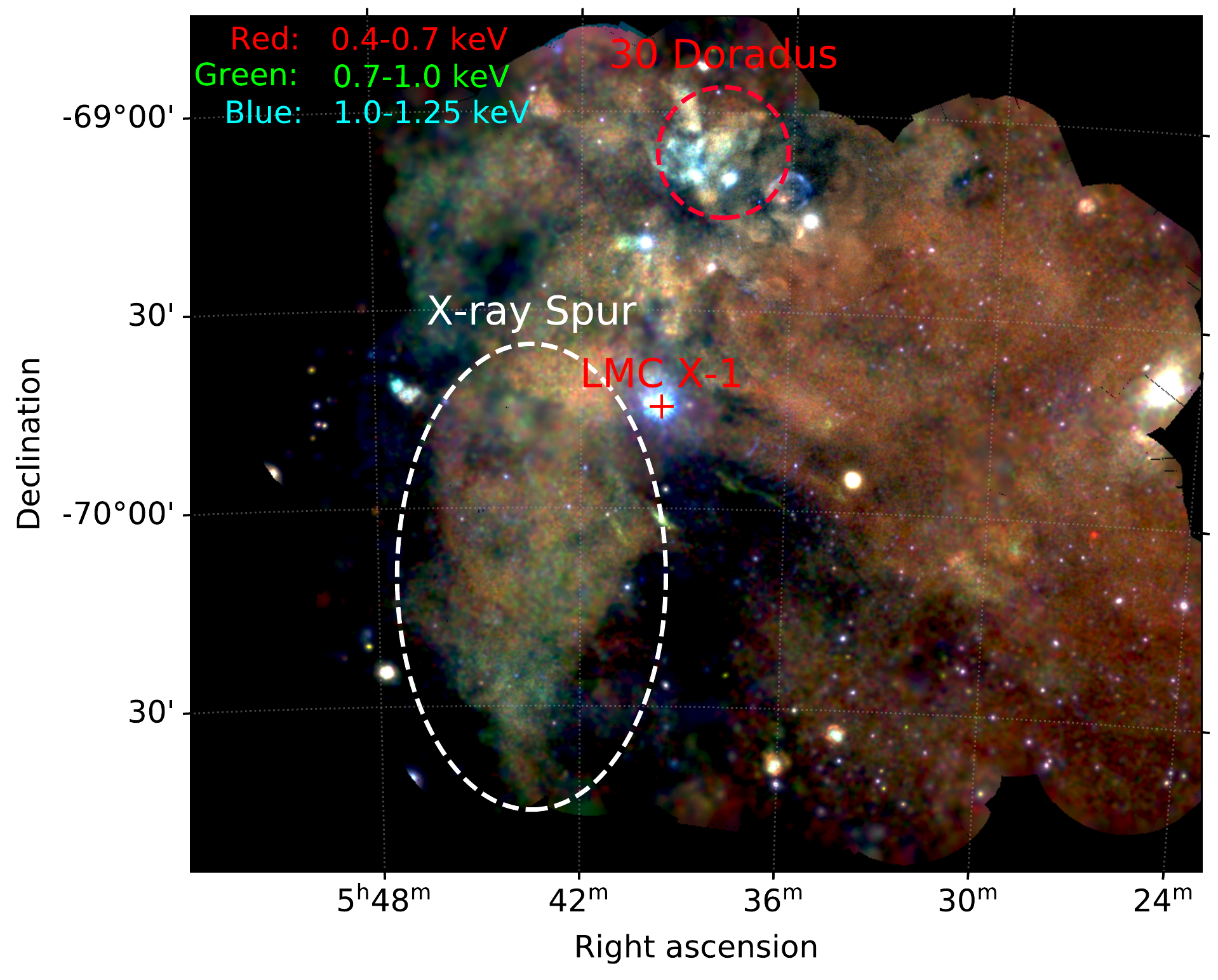}
		\caption{\label{fig:three-color_soft} Soft XMM-Newton three-color mosaic of the spur and its surroundings in the LMC. The X-ray colors correspond to 0.4-0.7\,keV (red), 0.7-1.0\,keV (green), and 1.0-1.25\,keV (blue). Relevant objects in the vicinity are marked for orientation.}  
\end{figure}

\begin{figure*}
	\centering
		\includegraphics[width=\textwidth]{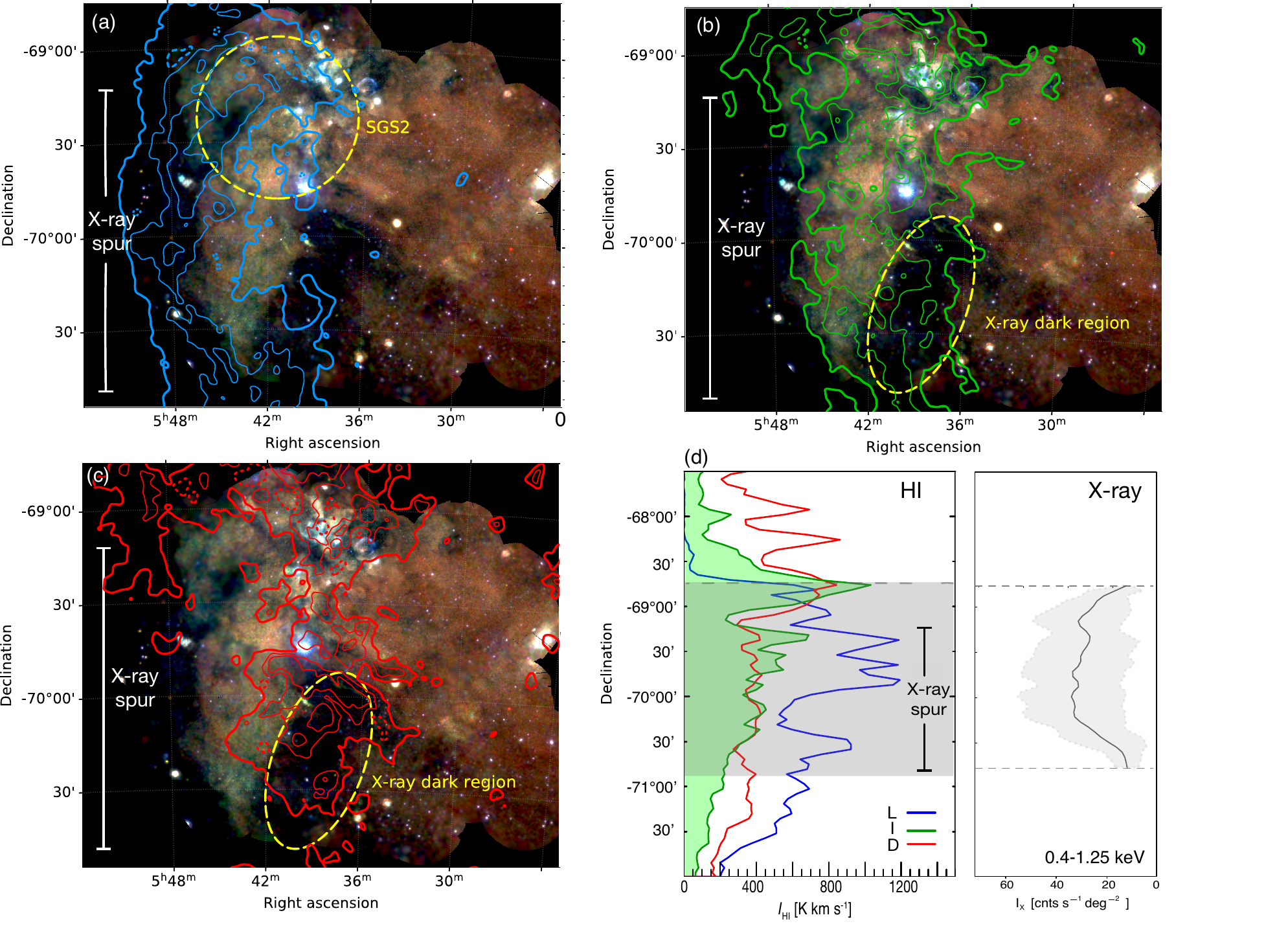}
		\caption{\label{fig:X_HI_cont} Composite of the XMM-Newton three-color mosaic and \ion{H}{i} contours. The X-ray colors correspond to 0.4-0.7\,keV (red), 0.7-1.0\,keV (green) and 1.0-1.25\,keV (blue). The L-component contours are shown in (a) in cyan, the I-component (b) in green and the D-component in red (c). The contours were generated from the $80\mathrm{-} 99$ percentile range of the corresponding emission strength with linear scaling (L-component: $293\mathrm{-}1441$~K\,km\,s$^{-1}$; I-component: $458\mathrm{-}1406$~K\,km\,s$^{-1}$; D-component: $669\mathrm{-}1359$~K\,km\,s$^{-1}$). The lowest level is indicated with thick lines. The LMC-SGS~2 and X-ray dark region are marked with dashed regions. Intensity profiles are shown in (d) which were integrated in Right ascension from RA\,$=5^{\text{\text{h}}}49^{\text{\text{m}}}$ to $5^{\text{\text{h}}}40^{\text{\text{m}}}$. The left panel shows the \ion{H}{i} component intensities and the right panel the X-ray intensity in the 0.4-1.25\,keV range. A detailed description how the brightness profiles were derived is given in Sect. \ref{sec:brightness_profiles}. }  
\end{figure*}

The EPIC data were reduced with the Extended Source Analysis Software \citep[ESAS]{esas_2} which is part of SAS. We followed the procedure described in the XMM ESAS Cookbook\footnote{\url{https://heasarc.gsfc.nasa.gov/docs/xmm/esas/cookbook/xmm-esas.html}} with some modifications.
First, we filtered the data for time intervals where background flares occurred, with thresholds determined from a Gaussian fit to the count rate.
 All time periods with count rates above the threshold were excluded from the analysis using the tasks \textsc{mos-filter} and \textsc{pn-filter}. We excluded anomalous MOS detector CCDs as indicated by the task \textsc{mos-filter} from the analysis.
To remove point source contamination we deviated from the ESAS cookbook and used the \textsc{wavdetect}\footnote{\url{http://cxc.harvard.edu/ciao/ahelp/wavdetect.html}} task which is part of the CIAO software package \citep{ciao}. We calculated a point spread function map with a constant size of 9\arcsec\ for the XMM-Newton data. The detection threshold was set to produce one false detection on average with wavelet scale factors of 1 to 32 to cover various sizes of background and compact sources in the images. 
For some larger bright diffuse sources, such as supernova remnants (SNRs), the point source detection failed and the sources had to be excluded manually.
From the filtered data the spectra and response files were created for each detector with \textsc{mos-spectra} and \textsc{pn-spectra}. Events inside the point source candidate regions were excluded. Additionally, model quiescent particle backgrounds (QPBs) were created from data taken with closed filters with the \textsc{mos-back} and \textsc{pn-back} tasks.
We combined the count-rate images, exposure-, and particle background maps for each detector with the \textsc{comb} task.
 We then created particle background subtracted count-rate images with the \textsc{bin\_image} task. We used a binning factor of 4 for both the image and count-rate uncertainty map. Additionally, smoothed, exposure corrected, and background subtracted images were created for each observation with the \textsc{adapt} task. We used a binning factor of two and a minimum of 50 counts for smoothing. 
 To create X-ray mosaics of all observations we used the \textsc{merge\_comp\_xmm} and \textsc{adapt\_merge} tasks. Again we used a binning factor of 2 and a minimum of 50 counts for smoothing. We also created binned mosaics using the \textsc{bin\_image\_merge} task with a binning of 2.
  The resulting soft X-ray mosaic is shown in Fig. \ref{fig:three-color_soft}.
We also created composite images of the X-ray emission and the different \ion{H}{i} components (Fig. \ref{fig:X_HI_cont}). The contour levels were calculated from the $80-99$ percentile range of the corresponding \ion{H}{i} emission strength.
\subsubsection{Voronoi binning}
\begin{figure}
	\centering
		\includegraphics[width=0.49\textwidth]{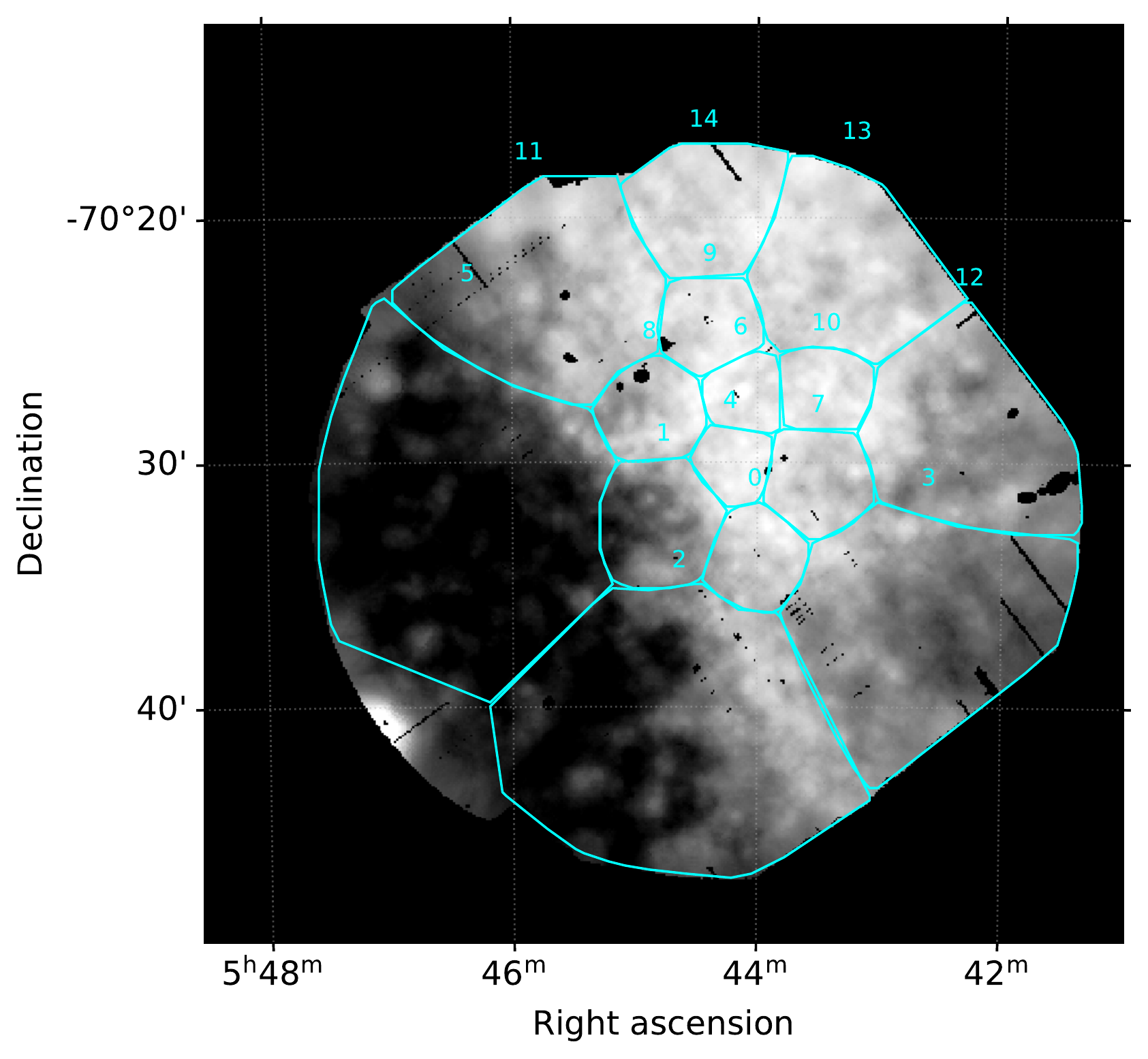}
		\caption{\label{fig:voronoi_regions} Example spatial binning of XMM-Newton data with the Voronoi tessellation algorithm for the new observation (ObsID: 0820920101). The energy range of the image is 0.4-1.25\,keV. }
\end{figure}

\begin{figure}
	\centering
		\includegraphics[width=0.49\textwidth]{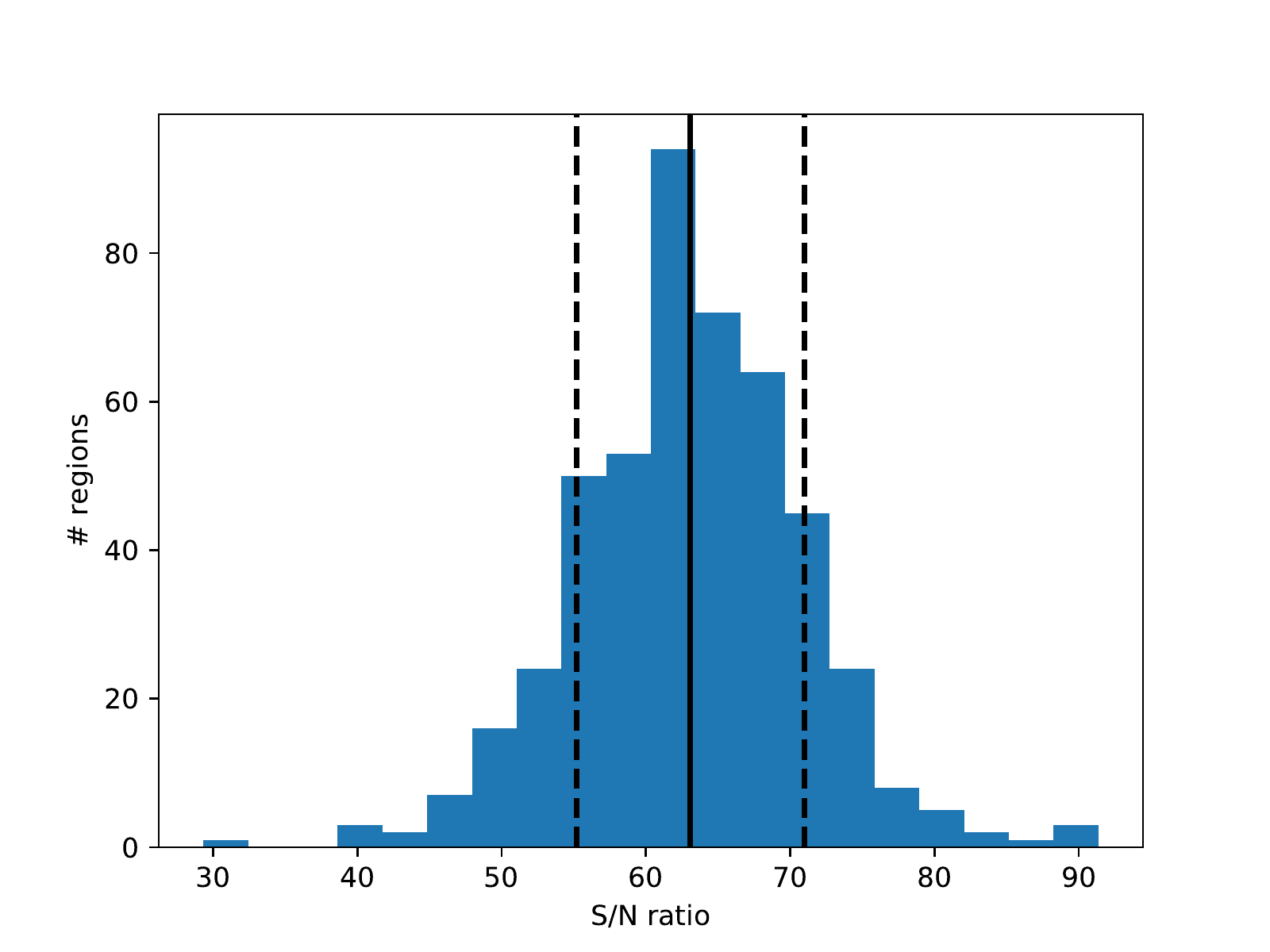}
		\caption{\label{fig:sn_ratios_imgWav} Histogram of the S/N for all tessellates used in the spectral analysis (see also Table~\ref{tab:list_xmm_observations}). The solid line shows the mean S/N and the dashed lines the $1\sigma$ interval.}
\end{figure}

We performed spatial binning of the data using the Voronoi tessellation algorithm \citep{voronoi1}. This algorithm provides an unbiased binning, where each bin has about the same signal-to-noise ratio (S/N).
We implemented the Voronoi tesselation for XMM-Newton EPIC data based on the voronoi binning python scripts by \citet{voronoi1}. The program takes an image created during the data reduction as input. Additionally a mask file is read in to define the limits of the exposed area of the image. The image is then compared with the mask and set to zero where the mask is also zero. As the resulting regions should have about the same S/N, the input images were not exposure corrected, corresponding to the true statistics of the observed spectra. Therefore, we use the combined (added) image of all available EPIC detector images created during the data reduction. The images were binned with a factor of 4, as mentioned above. Additionally, we subtracted the combined simulated particle background from the image.
We used masks which also excluded point source candidates. The image was then binned with the Voronoi tesselation algorithm by \citet{voronoi1}. The noise was estimated from the count rate of the binned image and the exposure per pixel.
 The algorithm returns the assigned bin for each pixel of the image. With this information, we constructed the analysis regions by using a convex hull algorithm. The convex hulls were then converted into a convenient format for further analysis, namely, the \textsc{saoimage ds9} region file with world coordinate system (WCS) coordinates \citep{ds9}. An example of the resulting regions is shown in Fig. \ref{fig:voronoi_regions}. We performed the binning individually for each XMM-Newton EPIC observation and kept overlapping observations separated. 
For the S/N we aimed for values in the range of $\sim 60$ to yield sufficient photon statistics in the spectra with a good spatial resolution. We also tested higher S/N, however, this resulted in larger regions for which the spectra could be averaged despite possible spatial variations. The distribution of S/N for all observations is shown in Fig. \ref{fig:sn_ratios_imgWav}.
\subsubsection{Extraction of spectra}
\label{sec:spectra_extraction}
We extracted spectra for each Voronoi region obtained from the procedure described above. We also extracted spectra in manually defined regions from several observations in the vicinity of the X-ray spur. Those regions covered most of the FOV and were used to verify and cross-check the spectral analysis of the Voronoi tessellates. We call these regions ``large regions'' in the following. We also extracted spectra at the SEP for background estimates. We repeated the spectra extraction tasks \textsc{mos-spectra} and \textsc{pn-spectra} for each region and detector individually to obtain the correct spectra and response files. We also simulated the QPB for each region, using \textsc{mos-back} and \textsc{pn-back}. We filtered the events using the pattern $0\mathrm{-} 4$ for pn and $0\mathrm{-} 12$ for MOS. The spectra were energy binned with a minimum of 30 counts per bin. We also calculated the area of each region in arcsec$^{2}$ with the \textsc{proton\_scale} task. The missing area of the detector gaps and excluded point-sources were taken into account in the calculations.
%
\subsection{Spectral analysis}
We used \textsc{xspec} \citep{xspec_general} version 12.10.1 to fit the spectra obtained after data reduction. We used the latest AtomDB\footnote{\url{http://www.atomdb.org/index.php}} version 3.0.9 with the nonequilibrium ionization (NEI) version 3.0.7. We had to cut the spectra below $0.4$~keV and above $7.0$~keV due to strong noise below and above these energies, respectively.
\label{sec:analysis}
\paragraph{Background components}
\label{sec:spectra_bg}
Since we study diffuse emission with low photon statistic, and take spectra from large parts of the EPIC detectors, we need to model all background components very accurately.
The background components are: local diffuse X-ray background, instrumental background,  and particle background. We modeled all background components similar to the work of \citet{esas_2}.
The X-ray background consists of thermal and nonthermal components.
  We first modeled the Local Bubble emission with an unabsorbed thermal component with a temperature of $kT = 0.1$~keV. 
 To account for emission from the cold- and hot Galactic halo we used two absorbed thermal components with temperatures $kT = 0.1$~keV and $kT = 0.3 \mathrm{-} 0.7$~keV, respectively. The Galactic absorbing hydrogen column density was fixed to the values from the map by \citet{dl_map} at the respective observation position. The LMC absorption is excluded in this map, yielding only the foreground Galactic absorption. Lastly, we used an absorbed powerlaw with a fixed index of $\Gamma = 1.46$ and an initial normalization corresponding to 10.5 photons keV cm$^{-2}$ s$^{-1}$ sr$^{-1}$ to account for the unresolved extragalactic background \citep{powerlaw_index_esas, esas_2}. The absorption $N_{\text{H, extragal}}$ was set free to fit.
 
Instrumental EPIC lines were modeled with Gaussian lines \citep{xmm_instrumental_lines}. For MOS we used two lines at $\sim 1.49$~keV and $\sim 1.75$~keV to account for Al-K and Si-K fluorescence lines. For pn only the line at $\sim 1.49$~keV (Al-K) are present and needs to be modeled. The line width was fixed to zero. Lastly, we introduced up to 6 Gaussian lines to account for solar wind charge exchange (SWCX): \ion{C}{vi} (0.46\,keV), \ion{O}{vii} (0.57\,keV), \ion{O}{viii} (0.65\,keV), \ion{O}{viii} (0.81\,keV), \ion{Ne}{ix} (0.92\,keV), \ion{Ne}{ix} (1.02\,keV) and \ion{Mg}{xi}  (1.35\,keV) \citep{swcx_xmm}. However, in the following spectral analysis we noticed that the emission line models for \ion{C}{vi}, \ion{O}{vii}, and \ion{O}{viii} had the largest effect for reducing residuals. Therefore, we limited the SWCX line emission models (Gaussians) to those elements, also to keep the number of free parameters as low as possible.
The model we used is described by 
\begin{equation}
\begin{split}
\text{skyarea}\times(\text{ Local Bubble} + \text{ instrumental lines}   + \text{ SWCX lines} \\ + \text{ Gal. abs. } N_{\text{H, Gal}}\times (\text{ cold halo} + \text{ hot halo} \\ + \text{ extragal. abs. } N_{\text{H, extragal}}\times\text{ extragal. BKG})) \,\, ,
\end{split}
\label{eq:background_model}
\end{equation}
which translates to the \textsc{xspec} model
\begin{equation}
\begin{split}
\text{CONSTANT}\times(\text{APEC} + \text{GAUSSIANs} + \text{TBABS}\times(\text{APEC} + \\  \text{APEC} + \text{TBABS}\times\text{POWERLAW}))\,\,.
\end{split}
\label{eq:background_model_xspec}
\end{equation}
To account for soft proton contamination, we added another model that consists of a simple power law. This model is not folded through the effective area of the detectors and uses diagonal response matrices. These models were added and linked for each spectrum individually. 
The power law indices were fitted within the range of $\Gamma = 0.2-1.5$. The indices were linked between all spectra of the two MOS detectors. For pn spectra we used an individual power law index, since both detector types have different sensitivities to soft protons. The normalizations were set free to fit independently.
 We carefully tested possible SWCX contamination by fitting one SWCX line at a time. The line width was fixed to zero. If the red. $\chi^2$ did not improve significantly or the normalization became very small, we removed the line from the model. 
\paragraph{Background study}
\label{sec:background_sep}
We used three observations of the SEP to determine the X-ray background near the X-ray spur. The SEP is relatively close to the LMC and has no bright or extended sources in soft X-rays in the FOV of XMM-Newton. We used the model discussed above to simultaneously fit all three observations. We linked all soft X-ray parameters to reduce the number of free parameters. Only the instrumental lines and SWCX lines were free to fit individually for each spectrum, since the intensity varies between the observations. Additionally, we linked the soft proton powerlaw index of the two MOS detectors for each observation. To account for slightly different extraction areas for each spectrum, we normalized each to units of arcmin$^{-2}$. We only used one Gaussian line at $\sim 0.65$~keV to account for possible SWCX. The other SWCX lines were shown to have no significant impact on the fit statistic \citep{diss_warth}.

The best-fit model yielded a good agreement with the data with $\chi^2/\text{d.o.f} = 2659/2375 = 1.12$.
For the Local Bubble thermal component we obtain a normalization of $2.81\cdot 10^{-6}$~cm$^{-5}$. For the cold Galactic halo we obtain $6.59\cdot 10^{-6}$~cm$^{-5}$ and for the hot Galactic halo $8.39\cdot 10^{-7}$~cm$^{-5}$ and a temperature of $kT = 0.3$~keV. For the absorption of the extragalactic background we obtain $N_H = 0.63 \cdot 10^{22}$~cm$^{-2}$. We used the fit results from the SEP spectra to constrain the X-ray background parameters while fitting the LMC spectra. We used the $90\%$ confidence interval (CI) as fit parameter constraints, which were calculated using the \textit{error} and \textit{steppar} commands.
\paragraph{Straylight}
\label{sec:spectra_straylight}
Nearly all observations we analyzed were pointed near the high mass X-ray binary (HMXB) LMC X-1 \citep{lmc_x1_basic}. We observed strong single reflections from straylight of this source in the majority of the observations.  Only 0.2\% of the effective area\footnote{\url{https://xmm-tools.cosmos.esa.int/external/xmm_user_support/documentation/uhb/xmmstray.html}} causes stray light. However, LMC X-1 is a very bright source and the stray light is indeed noticeable, especially for energies $> 1.25$~keV. One strategy would be to exclude the parts of the FOV where straylight is clearly visible. However, we would lose a significant amount of area for each detector and more importantly, sections of the sky for each observation. In addition, even after excluding these areas, there would still be some unknown amount of contamination left in the remaining data. Therefore, we decided to model the straylight emission instead of excluding it. We assume that the straylight roughly corresponds to the spectrum of the source itself. According to \citet{2010_hanke_lmc_x1} the spectrum of LMC X-1 is well described by the DISKBB\footnote{\url{https://heasarc.gsfc.nasa.gov/xanadu/xspec/manual/node160.html}} model which describes emission from an accretion disk. A previous study by \citet{lmc_x1_diskbb_values} showed that the temperature parameter varies for LMC X-1. We constrained the parameter $T_\text{in}$ to previously derived values - with a small additional leeway - to $0.82-1.05$~keV. The temperature was linked for all spectra of one observation since the temperature can be assumed to be constant on these timescales. The normalization was let free for each spectrum since the amount of straylight varied greatly between each region. However, we drastically reduced the amount of free parameters by fixing the normalizations to zero for regions that were not affected by straylight. This was achieved by ``flagging'' regions which, completely or partially, overlapped with the part of the FOV that was affected by straylight. This large area safely included all contaminated regions. In total, $\sim 60\%$ of all regions were flagged as possibly being affected by straylight. We also compared the strong straylight emission in the higher energy bands ($E > 1.25$~keV) with the normalization of the DISKBB component in the spectral fits and obtained a good agreement in the relative strength for most regions.

Another possible strong straylight source in the vicinity is the soft source N132D. We inspected all regions within a 90$'$ arcmin radius of this source for possible straylight contamination. According to \citet{N132D_hitomi}, spectral fits to Hitomi data yield a collisional ionization equilibrium (CIE) temperature of $\sim 0.7$~keV for this source. Since we divided each observation with the tessellates we were able to compare regions in the (roughly) half of the FOV in the direction of SNR N132D and regions in the other, straylight-unaffected part of the FOV.
We observe no systematic effect in the spectral fit results for regions located toward N132D.
\paragraph{Model for the X-ray spur:}
\label{sec:spectra_source}
\begin{table*}
	\centering
	\caption{\label{tab:linked_parameters} Schematic overview of the model components that were linked between the different regions and/or observations for the two-component model. A detailed description of the spectral model is given in the text.
	}
	\begin{tabular}{l|c|ccc}
		&  & \multicolumn{3}{c}{linked with same} \\
		Parameter & fixed & detector\tablefootmark{c} & region & observation\\
		\hline
		 &  &  &  &   \\
		 Area normalization &  \ding{51} &  &  &   \\
		 Detector Lines &  &  &  &   \\	
		 \multicolumn{1}{r|}{Energy\tablefootmark{b}} &  & \ding{51} &  &   \\
		 \multicolumn{1}{r|}{Norm\tablefootmark{e}} &  & \ding{51} &  &   \\
		 \hline
		 SWCX Lines &  &  &  &   \\	
		 \multicolumn{1}{r|}{Energy} &  &  &  &  \ding{51} \\
		 \multicolumn{1}{r|}{Norm\tablefootmark{e}} &  & \ding{51} &  &   \\
		 \hline
		 Soft Proton BKG &  &  &  &\\
		 \multicolumn{1}{r|}{Index\tablefootmark{b}} &  & \ding{51} &  &   \\
		 \multicolumn{1}{r|}{Norm\tablefootmark{e}} &  & \ding{51} &  &   \\
		 \hline
		 Local bubble\tablefootmark{a} & \ding{51} &  &  & \ding{51}  \\
		 \hline
		 Galactic Absorption &  \ding{51} &  &  &  \\
		 \hline		 
		 Cold Halo\tablefootmark{a} & \ding{51} &  &  & \ding{51}  \\		 	 
		 \hline
		 Hot halo\tablefootmark{a} & \ding{51} &  &  & \ding{51}   \\
		 \hline
		 LMC Absorption &  &  & \ding{51}  &   \\
		 \hline
		 Straylight DISKBB &  &  &   & \\
		 \multicolumn{1}{r|}{$T_{\text{in}}$} &  &  &  & \ding{51}  \\
		 \multicolumn{1}{r|}{Norm\tablefootmark{d, e}} &  &  &  &   \\
		 \hline
		 Diffuse Component &   &  &  &   \\
		 \multicolumn{1}{r|}{kT} &  &  & \ding{51} &   \\
		 \multicolumn{1}{r|}{Norm\tablefootmark{e}} &  &  & \ding{51} &   \\
		 \hline
		 2nd Diffuse Component &  &  &  &   \\
		 \multicolumn{1}{r|}{kT} &  &  & \ding{51} &   \\
		 \multicolumn{1}{r|}{Norm\tablefootmark{e}} &  &  & \ding{51} &   \\
		 \hline
		 LMC BKG Absorption &  &  &  & \ding{51}  \\
		 \hline
		 Extragal. BKG\tablefootmark{a} & \ding{51} &  &  & \ding{51}  \\	 
	\end{tabular}
	\tablefoot{
	\tablefoottext{a}{Parameters constrained to the $1\sigma$ uncertainty range from the fit results of the background study.} \\
	\tablefoottext{b}{Parameter also linked between the MOS1 and MOS2 detectors.} \\
	\tablefoottext{c}{Same detector means MOS1 with MOS1, MOS2 with MOS2 and pn with pn if not stated otherwise.}\\
	\tablefoottext{d}{Fixed to zero if not flagged as straylight affected region (see text) and set free to fit otherwise.}\\
	\tablefoottext{e}{Normalization of the model.}
	}
\end{table*}
We tried several different models for the X-ray emitting plasma of the X-ray spur and the surrounding area.
First, we used the simple APEC model for a plasma in CIE. We also tried the nonequilibrium ionization model NEI. Additionally, a combination of two APEC models were tried, with one at lower ($kT < 0.5$~keV) and one at higher temperature ($kT > 0.5$~keV). In contrast to the background study, we also need to account for absorption by the LMC. Since we do not know the exact geometry and therefore location of the plasma within the LMC, we split the absorption column by the LMC into two components, one in front of the diffuse plasma ($N_{\text{H,LMC}}$) and one behind ($N_{\text{H,BG}}$) used for the absorption of the extragalactic background. The element abundances of these components were fixed to 0.5\,solar, that is to say average LMC abundances, using the TBVARABS model. These two components replace the $N_{\text{H,extragal}}$ mentioned in Eq. \ref{eq:background_model}. For the X-ray background model we used the results from the SEP data. The X-ray background parameters were constrained to the previous fit results within the $1\sigma$ uncertainties. 

\paragraph{Model for the tessellates:}
Due to the voronoi binning and the desired spatial resolution of the bins, we obtain a large number of regions for each observation. Moreover, we have up to three spectra for each region because of the different EPIC detectors, which then results in up to several hundred free fit parameters.
In order to reduce the vast amount of free parameters, we linked them whenever possible and reasonable. A schematic overview of all linked parameters is given in Table~\ref{tab:linked_parameters}.

We linked all parameters of the X-ray spur emission models separately for each region. This way, we reduced the number of free parameters while keeping the spatial resolution achieved by the voronoi binning. 
We repeated this for the foreground LMC absorption component in order to be sensitive to variations between the different regions. All X-ray background parameters were constrained to the value ranges obtained by the fit to the SEP data. 
The temperature parameter $T_{\mathrm{in}}$ of the straylight model was linked between all region spectra, and the normalization was free to fit individually if the corresponding region was straylight flagged. Otherwise the normalization was fixed to zero.
To further reduce the number of free parameters, we also linked the normalizations of the soft proton powerlaw models of the same detector types, for example, pn to pn, MOS1 to MOS1 etc. We corrected for different region sizes with a multiplicative factor. 

We linked the foreground absorption for all regions of an observation to constrain the model initially. After an initial fit we froze all parameters of the instrumental lines. We then fitted the individual foreground absorption parameters for each region. We also added SWCX lines for all spectra, limited to the lines described previously. The spectra were then fit again and manually verified if the red. $\chi^2$ was worse than $\sim 1.2$. In some cases, the emission model for the hot plasma in the LMC tended to unreasonably low or high $kT$ values; this problem was solved by ``resetting'' the parameter and an additional fit in most cases. When the fit no longer improved, we saved the results and calculated the uncertainties for all important parameters, that are the LMC plasma emission model parameters and foreground/background LMC absorption. The 90\% CIs were calculated using the \textsc{error} command of \textsc{xspec}. In case the \textsc{error} results indicated problems, the uncertainties were calculated with the \textsc{steppar} command.
\subsection{Spectral fit results}
\label{sec:spectra_results}
\subsubsection{Manually selected ``large regions''}
\begin{figure}
	\centering
		\includegraphics[width=0.49\textwidth]{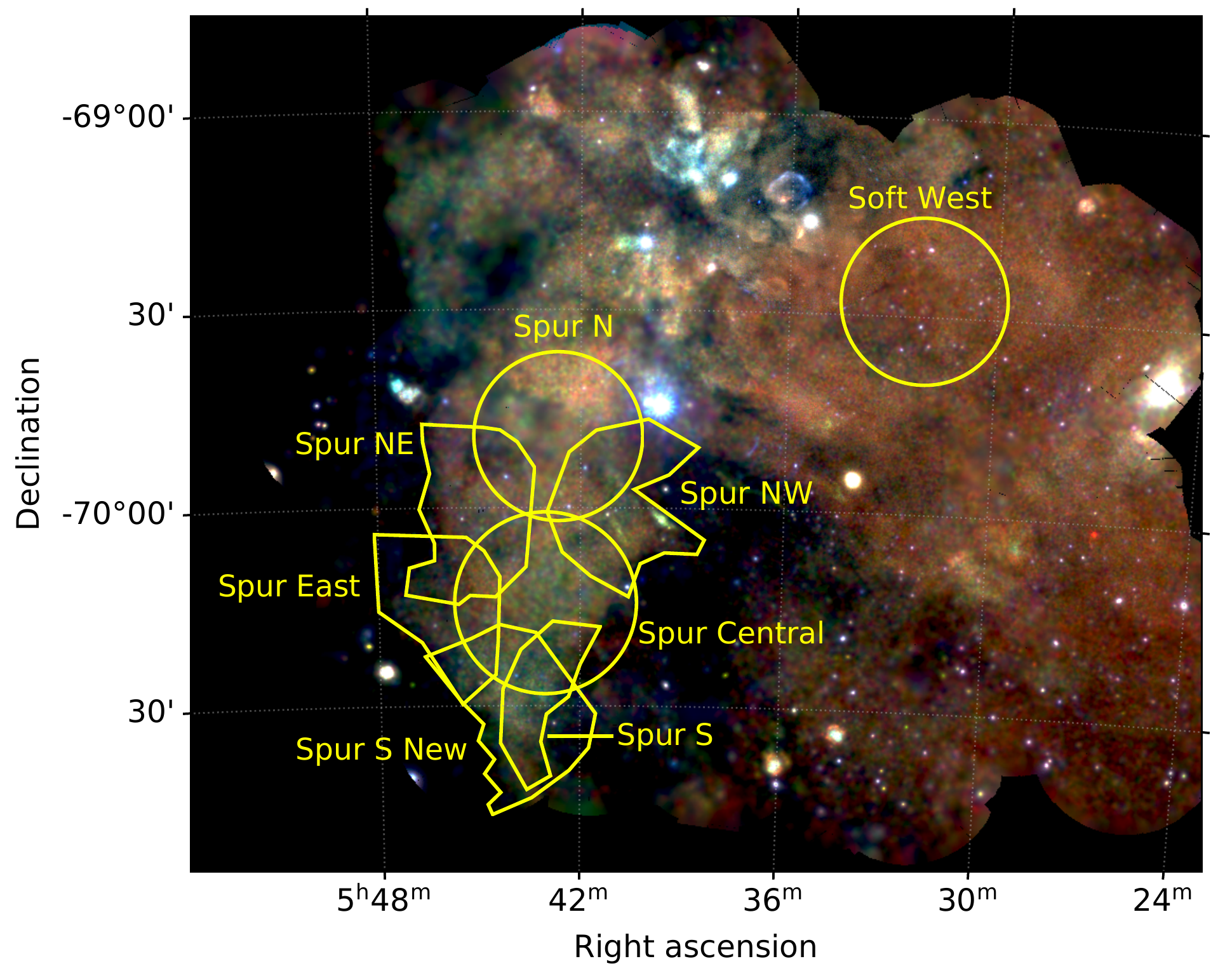}
		\caption{\label{fig:big_spectra_regions} Soft XMM-Newton three-color mosaic with 0.4-0.7\,keV (red), 0.7-1.0\,keV (green) and 1.0-1.25\,keV (blue) and the extraction regions of the ``large region'' spectra.}
\end{figure}

\begin{figure*}
	\centering
		\begin{subfigure}[t]{0.49\textwidth}
			\includegraphics[width=\textwidth]{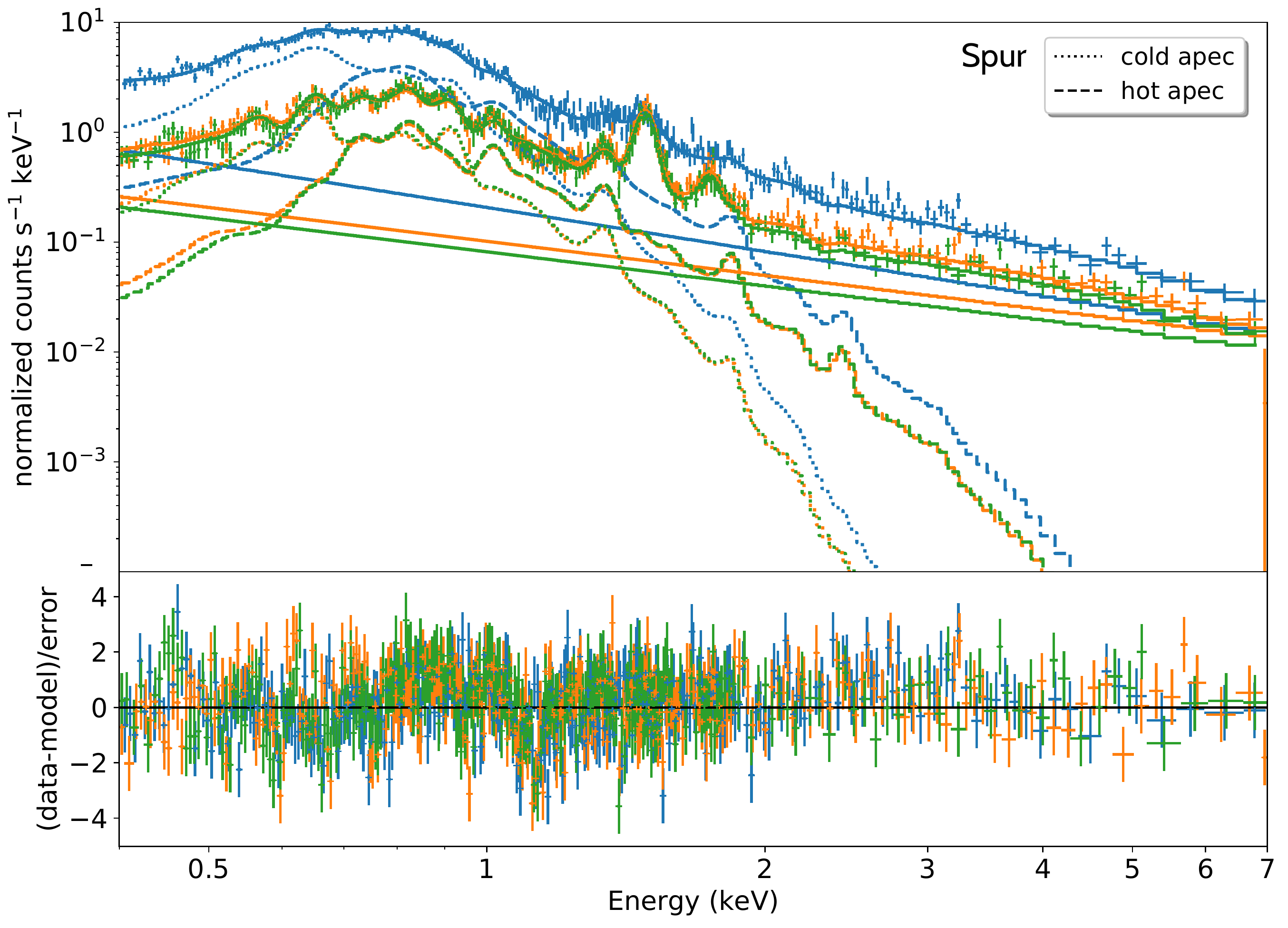}
	\caption{\label{fig:spectrum_spur}}
		\end{subfigure}
		\hfill
		\begin{subfigure}[t]{0.49\textwidth}
			\includegraphics[width=\textwidth]{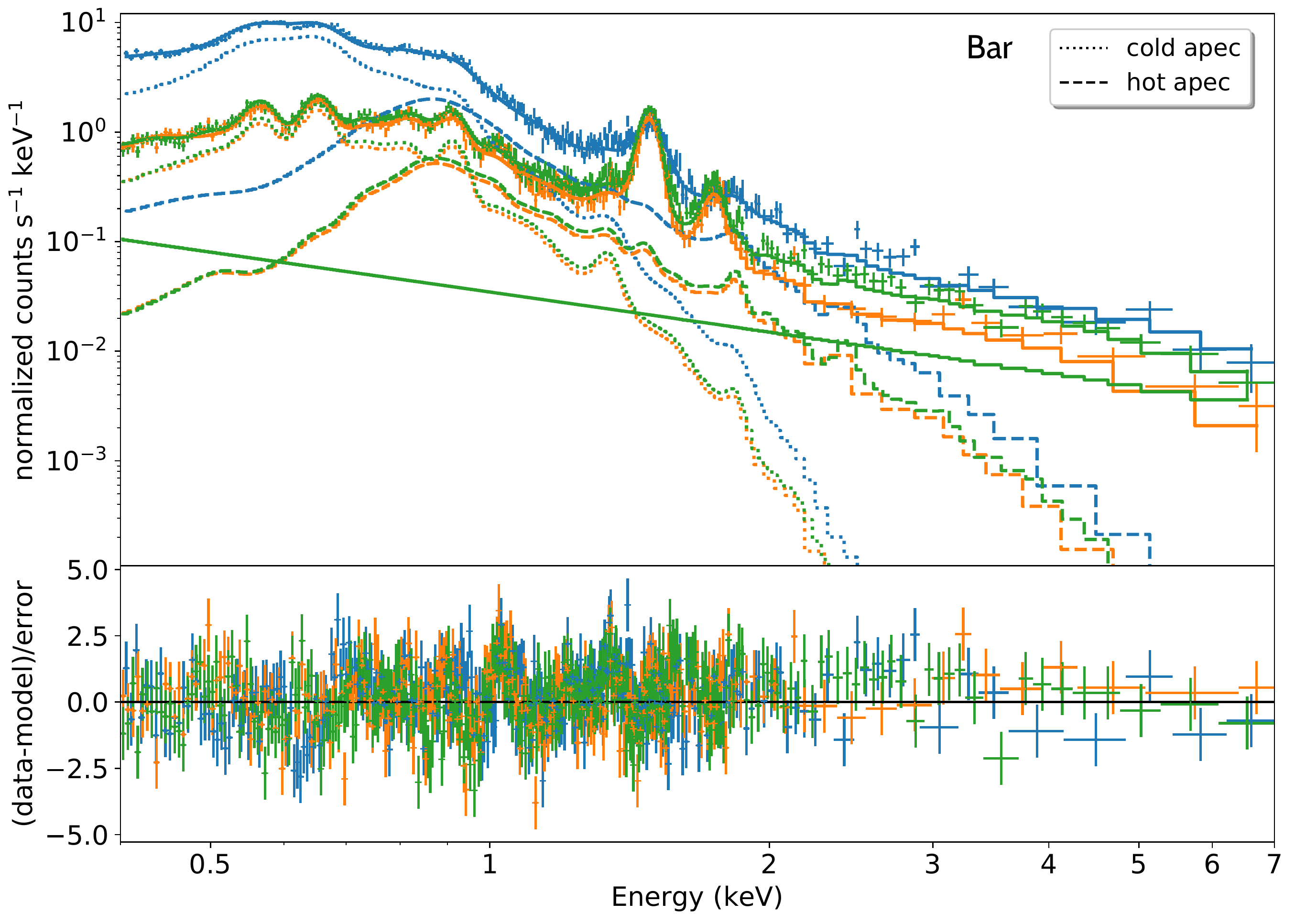}
	\caption{\label{fig:spectrum_west}}
		\end{subfigure}
		\caption{Spectra of the ``Spur Central'' region in the X-ray spur shown in (a) and for the ``Soft West'' region outside of the X-ray spur and 30 Dor in (b).  The pn spectrum and model are shown in blue, while the MOS1 and MOS2 models and spectra are shown in green and yellow, respectively. The diffuse plasma model components are indicated with thick dashed and dotted lines in the corresponding detector color (see legend). The lower panel shows the residuals between model and data. The background was not subtracted but modeled simultaneously (see text). For better presentation, the data were rebinned visually with either 50 counts or $5\sigma$. \label{fig:spectra_on_off_spur}}
\end{figure*}

We obtain the best fit results for the spectra of the manually selected large region spectra (see Sect. \ref{sec:spectra_extraction}) with two CIE model APEC components for the diffuse emission. The extraction regions are shown in Fig. \ref{fig:big_spectra_regions} and the individual fit results are given in Table~\ref{tab:big_spectra_params}. We show two example spectra in Fig.~\ref{fig:spectra_on_off_spur}, one inside and one outside of the X-ray spur. For most observations, adding SWCX lines to the model improved the fit statistics slightly, while leaving the plasma parameters unchanged within uncertainties. 
  When using a single APEC, we obtain a significantly higher temperature for spectra in the X-ray spur compared to the spectra extracted from the ``Soft West'' region.
 Using the best-fit two APEC model, we notice that the lower temperature APEC component tends to values of $kT_1 \sim 0.2$~keV. This is consistent with the single-APEC temperature obtained for the ``Soft West'' region inside the stellar bar of the LMC. The origin of this low-temperature thermal emission is most likely the  undisturbed hot interstellar gas and unresolved stellar sources. The second, higher temperature APEC tends to values of $kT_2\sim 0.5\mathrm{-}0.9$~keV.
 The hot plasma component is weaker by roughly an order of magnitude outside of the X-ray spur. Only in the eastern part of the spur - the ``Spur NE'' and ``Spur E'' regions - the hot component seems to be comparably weak. The low temperature component is also weaker by about the same factor in these regions, however.
  The foreground absorption parameter in the central spur (``Spur NW'', ``Spur Central'', and ``Spur S'') is larger by a factor of two compared to the ``Soft West''.
  In the eastern part of the spur, where the hot component is weaker, we also obtain a very high background absorption. For the new observation in the very south (``Spur S New'') we also obtain significant background absorption.
\begin{table*}
\renewcommand{\arraystretch}{1.25}
	\centering
	\caption{\label{tab:big_spectra_params} Spectral fit parameters of the manually selected ``large region'' spectra. The extraction regions are show in Fig. \ref{fig:big_spectra_regions}. The uncertainties correspond to the $90\%$ CI of the fit parameters. The fit parameters were obtained with assuming two APECs for the soft diffuse emission in the LMC.}
	\begin{tabular}{ll|rrrrrr|rr}
		Region & ObsID & $N_{\text{H,LMC}}\,$ & $kT_{1}\,$ & $\text{norm}_{1}\,$ & $kT_{2}\,$ & $\text{norm}_{2}\,$ & $N_{\text{H,BG}}\,$ & red. $\chi^2$ & d.o.f.\\
		&  & [$10^{22}$~cm$^{-2}$] & [keV] & [$10^{-5}$~cm$^{-5}$] & [keV] & [$10^{-5}$~cm$^{-5}$] & [$10^{22}$~cm$^{-2}$] &  & \\
		\hline
		Spur N & 0094410701 & $0.55^{+0.12}_{-0.21}$ & $0.24^{+0.03}_{-0.02}$ & $9.55^{+4.33}_{-3.26}$ & $0.65^{+0.10}_{-0.05}$ & $1.87^{+0.49}_{-0.63}$ & $0$ & $1.08$ & $471$\\
		Spur NW & 0690750301 & $0.69^{+0.02}_{-0.06}$ & $0.19^{+0.01}_{-0.02}$ & $8.33^{+1.94}_{-1.80}$ & $0.42^{+0.01}_{-0.03}$ & $2.84^{+0.89}_{-0.88}$ & $0$ & $1.16$ & $1048$\\
		Spur NE & 0690751401 & $0.35^{+0.07}_{-0.11}$ & $0.23^{+0.01}_{-0.01}$ & $4.38^{+0.59}_{-1.22}$ & $0.87^{+0.02}_{-0.04}$ & $0.46^{+0.07}_{-0.03}$ & $5.57^{+3.69}_{-3.29}$ & $1.21$ & $1668$\\
		Spur Central & 0201030301 & $0.69^{+0.04}_{-0.04}$ & $0.20^{+0.01}_{-0.01}$ & $15.69^{+2.40}_{-0.30}$ & $0.46^{+0.01}_{-0.02}$ & $2.50^{+0.10}_{-0.29}$ & $0$ & $1.11$ & $1385$\\
		Spur East & 0690751501 & $0.22^{+0.13}_{-0.08}$ & $0.23^{+0.01}_{-0.02}$ & $2.12^{+0.59}_{-0.48}$ & $0.73^{+0.04}_{-0.03}$ & $0.35^{+0.01}_{-0.01}$ & $3.43^{+3.86}_{-1.76}$ & $1.16$ & $1567$\\
		Spur S & 0690751601 & $0.67^{+0.12}_{-0.13}$ & $0.23^{+0.02}_{-0.02}$ & $8.62^{+3.99}_{-2.50}$ & $0.59^{+0.05}_{-0.03}$ & $1.46^{+0.74}_{-0.48}$ & $1.02$\tablefootmark{$\dagger$} & $1.15$ & $826$\\
		Spur S New & 0820920101 & $0.49^{+0.01}_{-0.05}$ & $0.20^{+0.01}_{-0.01}$ & $5.00^{+0.16}_{-0.12}$ & $0.60^{+0.01}_{-0.01}$ & $1.53^{+0.03}_{-0.03}$ & $2.59^{+0.77}_{-0.68}$ & $1.08$ & $2043$ \\
		Soft West & 0690744901 & $0.36^{+0.05}_{-0.10}$ & $0.20^{+0.01}_{-0.01}$ & $9.53^{+1.41}_{-2.63}$ & $0.78^{+0.03}_{-0.02}$ & $0.73^{+0.35}_{-0.56}$ & $0$ & $1.14$ & $1721$\\
	\end{tabular} 
	\tablefoot{\tablefoottext{$\dagger$}{Missing uncertainties indicate that no reliable upper and/or lower limits were found.}
	}
\end{table*}
\subsubsection{Tessellates}
For the tessellate regions, we obtain a good fit for the LMC emission with one CIE model APEC in the region west of the X-ray spur. However, for many observations the data are not fit well with this model, especially in the eastern part of the mosaic where 30 Dor and the X-ray spur are located. Therefore, we used two APEC model components for the diffuse emission, as previously done for the large regions. The fit statistics improved drastically compared to the single APEC model. We were able to further improve the fit statistics by adding SWCX lines to this model.
 
 The difference between using one and two APECs is much smaller in the west, outside 30 Dor and the X-ray spur. Based on the results from the foregoing manually selected region spectral analysis, we constrained the lower temperature APEC to the range $kT_1 = 0.17\mathrm{-} 0.21$~keV.
 With this, the second APEC consistently yields higher temperatures of $kT_2 \sim 0.5 \mathrm{-} 0.9$~keV and appears well constrained. The results agree well with the fit results of the manually selected regions. 
 
\begin{figure}
	\centering
		\includegraphics[width=0.49\textwidth]{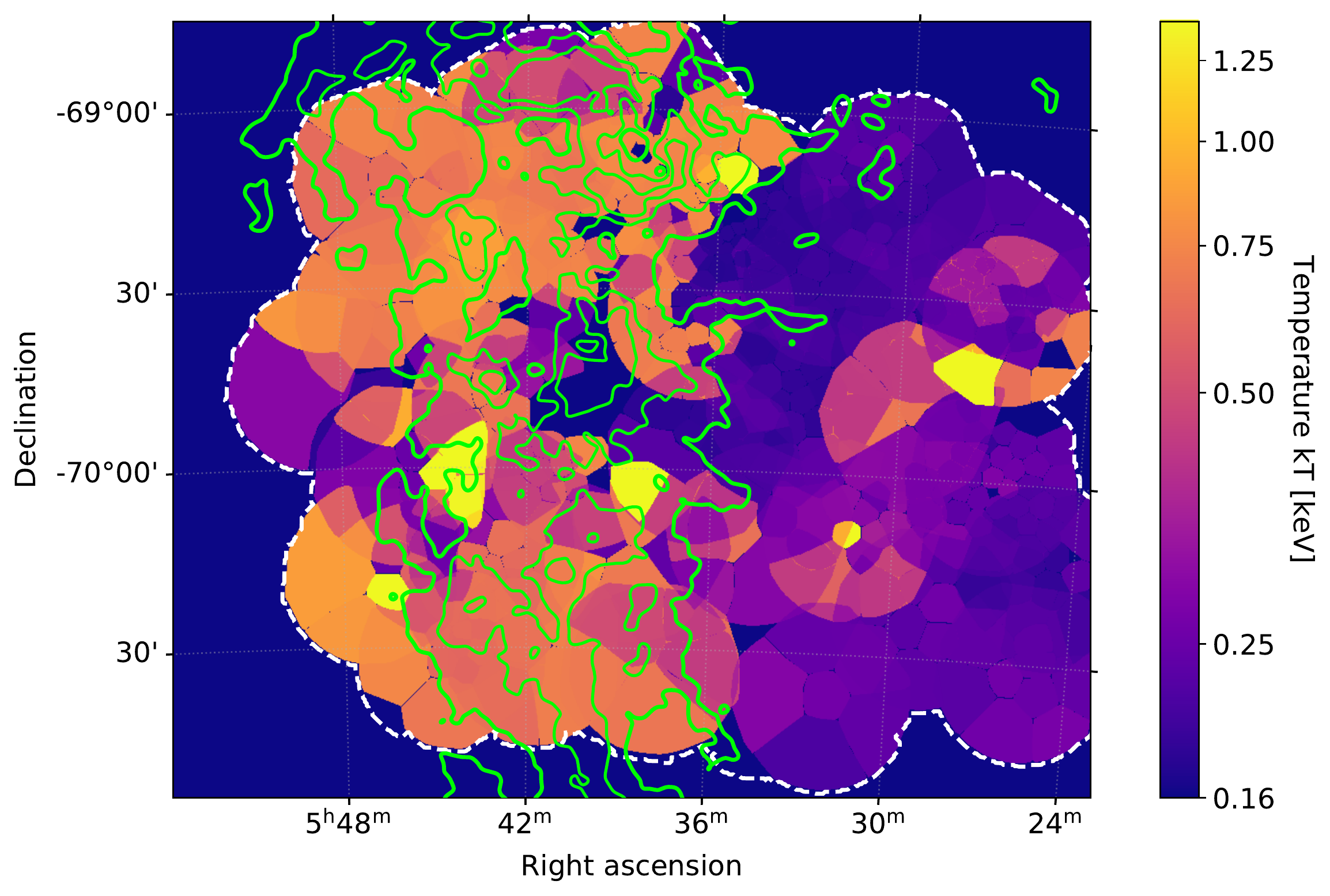}
		\caption{\label{fig:kT_confidence} Map of the plasma temperature parameter $kT$ overlayed with \ion{H}{i} I-component contours (green) in the $80\mathrm{-} 99$ percentile range ($458\mathrm{-}1406$~K\,km\,s$^{-1}$) when using the single APEC model. The area covered by the spectral analysis is indicated with dashed white lines.}
\end{figure}

\begin{figure*}
	\centering
		\begin{subfigure}[t]{0.49\textwidth}
		\centering
		\includegraphics[width=\textwidth]{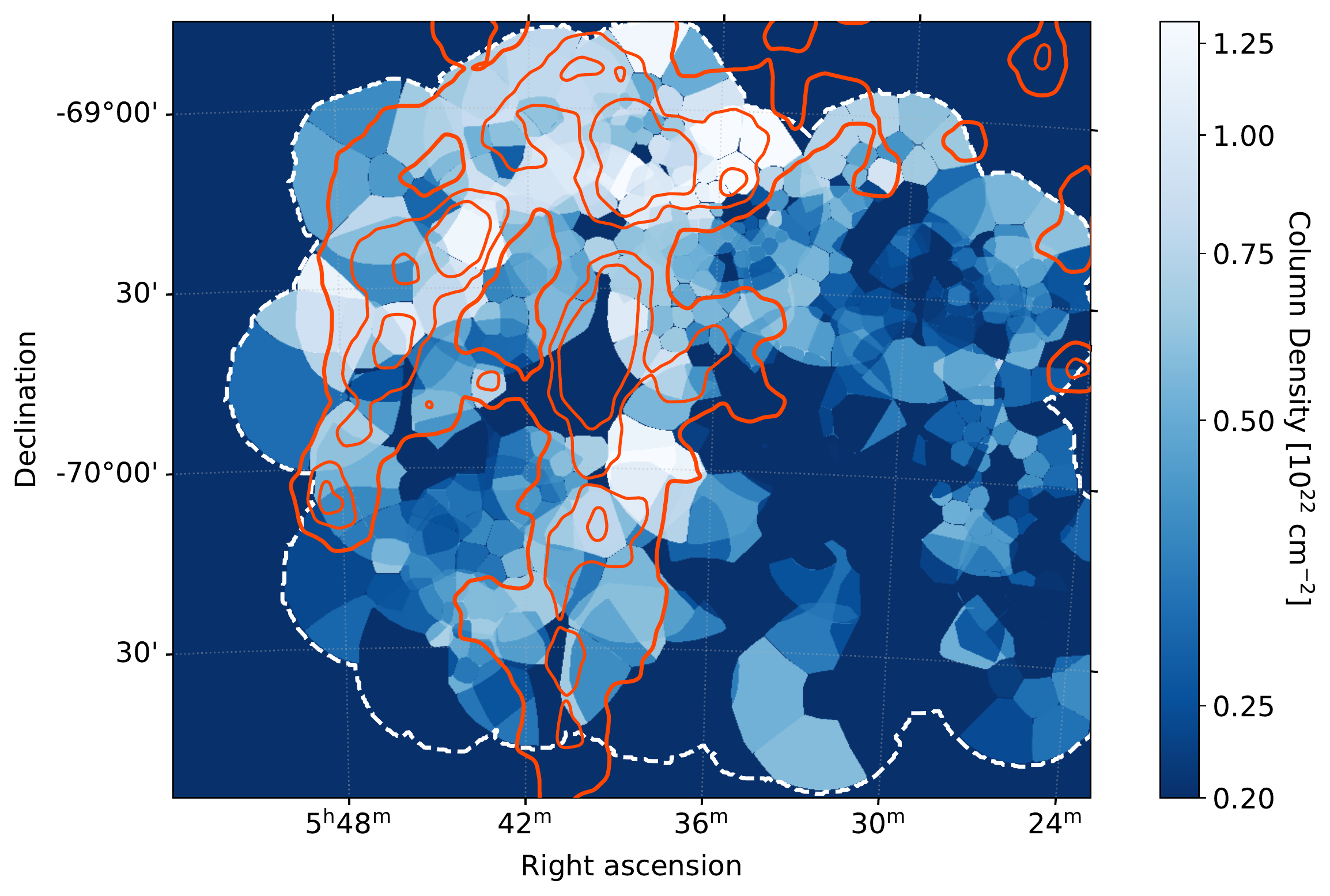}
			\caption{\label{fig:nH}}
	\end{subfigure}
	\hfill
	\begin{subfigure}[t]{0.49\textwidth}
		\centering
		\includegraphics[width=\textwidth]{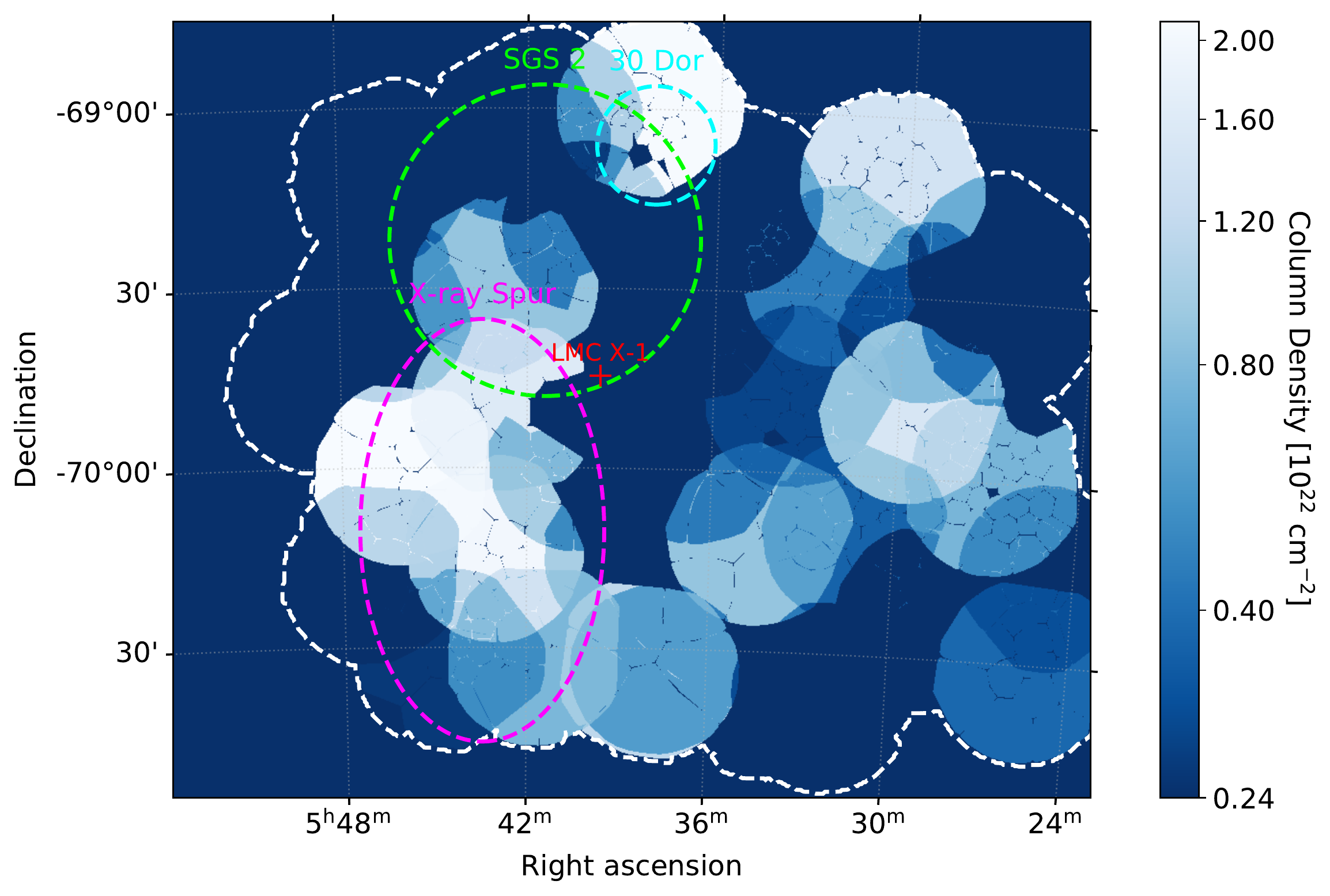}
		\caption{\label{fig:nH_bkg}}
	\end{subfigure}
	\caption{(a) Map of the lower limit ($90\%$ CI) foreground X-ray absorbing hydrogen column density equivalent $N_{\mathrm{H}}$\  in units of $10^{22}$~cm$^{-2}$. The tessellate spectra were fit with the two-APEC model. The contours of the dust 				optical depth are overlayed in orange, where the contours levels correspond to the $80\mathrm{-} 95$ percentile range ($2.0\cdot 10^{-5}\mathrm{-}6.4\cdot 10^{-5}$) of the image shown in Fig. \ref{fig:tau353_map}. 
	(b)  Map of the lower limit ($90\%$ CI) background X-ray absorbing hydrogen column density equivalent $N_{\text{H,BG}}$ in units of $10^{22}$~cm$^{-2}$. The tessellate spectra were fit with the two-APEC model. The positions of the X-ray spur, LMC-SGS~2 and LMC X-1 are marked for orientation. We only obtain one $N_{\text{H,BG}}$ value per observation since the parameter was linked between the tessellates. The fit values were averaged for regions where observations overlap. The area covered by the spectral analysis is indicated with dashed white lines both in (a) and (b).}
\end{figure*}

\begin{figure}
	\centering
		\includegraphics[width=0.49\textwidth]{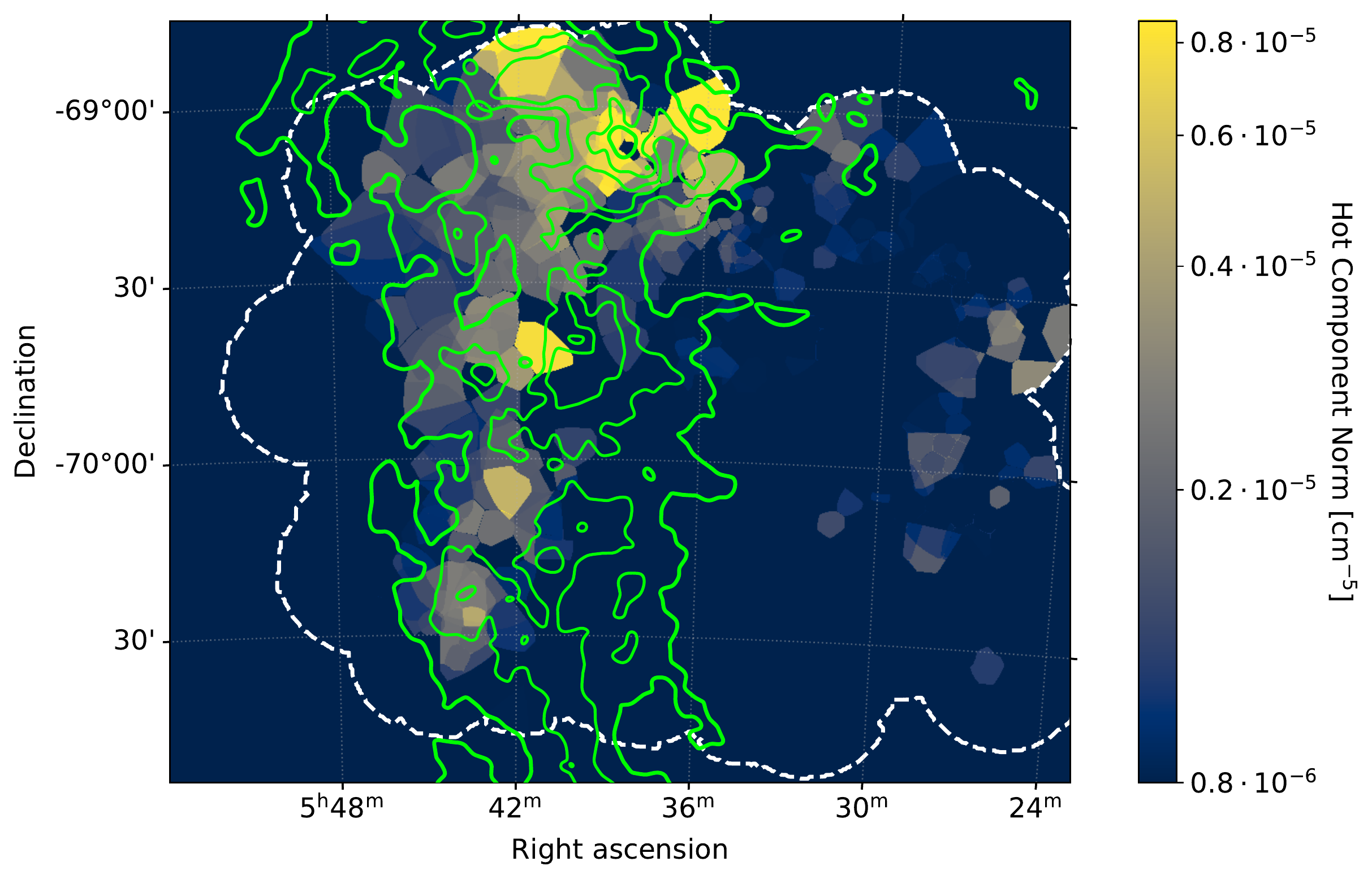}
		\caption{\label{fig:two_apec_norm_confidence} Lower limit norm of the second hot APEC component ($kT\sim 0.6\mathrm{-} 0.9$~keV) when using the two-APEC model. Overlayed in green are \ion{H}{i} I-component contours in the $80\mathrm{-} 99$ percentile range ($458\mathrm{-}1406$~K\,km\,s$^{-1}$). The area covered by the spectral analysis is indicated with dashed white lines.}
\end{figure}

  To visualize the distribution of the spectral properties of the LMC plasma, we created maps for the most important fit parameters as well as upper- and lower limit maps ($90\%$ CI). For simplicity we show the temperature map of the single APEC model, which can be interpreted as an ``average'' temperature, compared to the two APEC model. All other maps show the parameters for the best-fit two APEC model.
The map for the temperature parameter is shown in Fig. \ref{fig:kT_confidence} and for the foreground absorption $N_{\text{H,LMC}}$ inside the LMC in Fig. \ref{fig:nH}.  The background absorption $N_{\text{H,BG}}$ is shown in Fig. \ref{fig:nH_bkg}.
 We only show regions, where we were able to obtain reliable upper and lower limits for the respective parameters. 
 In a few isolated cases, the fits tended toward either very high or very low temperatures when using the single APEC model. This is most likely caused by residual point source contamination or straylight from sources other than LMC X-1. Most isolated hot temperature regions correlate with a massive stellar population, which could also explain the fit results.
 When using the two-APEC model, the normalization of the cool component maps the X-ray brightness seen in the soft X-ray mosaic (Fig. \ref{fig:three-color_soft}) well. 
  The lower limit normalization of the hot component is shown in Fig. \ref{fig:two_apec_norm_confidence}. Since the temperature parameter of this component is relatively homogeneous, the strength of the component is a better measure for the temperature of the plasma. The $90\%$ CI lower limit of the hot component normalization is already area normalized to units of arcmin$^2$. To the west of the spur and R136 the hot component normalization is much weaker by at least one order of magnitude in comparison. On the other hand, at the position of the X-ray spur, LMC-SGS~2, and 30 Dor, the hot component normalization is the highest. 

We calculated the mean fit parameter values from the parameter maps previously shown in defined regions. To obtain uncertainties of the mean, we took the average of the lower- and upper limit parameter maps for the respective fit parameter. For the $kT$ map, we excluded values outside the $3\mathrm{-} 97$ percentile range to reduce the effect of the previously discussed outliers.
We observe low X-ray temperatures west of 30 Dor, with an average of $kT = 0.27^{+0.03}_{-0.01}$~keV. This is in agreement with the X-ray color in the three-color mosaic, where the area appears red in color (Fig. \ref{fig:three-color_soft}). In contrast, the temperatures around 30 Dor and R136 are much higher with an average of $kT= 0.66^{+0.05}_{-0.02}$~keV. For the X-ray spur we obtain significantly higher temperatures compared to the western part, with an average temperature of $kT= 0.64^{+0.13}_{-0.05}$~keV. 

For the foreground LMC absorption shown in Fig. \ref{fig:nH}, we notice a gradient from north to south, with increasing column densities toward 30 Dor. We show the dust optical depth contours overlayed on the parameter map, which seem to agree very well with our fit results. We also get high absorption east and west to the spur, in direct correlation with the shown dust optical depth contours. The foreground absorption column in the central spur is on average $N_{\text{H,LMC}} = (0.51^{+0.24}_{-0.08})\cdot 10^{22}$~cm$^{-2}$, with higher absorption east and west of the spur of $N_{\text{H,LMC}} > 0.7\cdot 10^{22}$~cm$^{-2}$. In the vicinity of 30 Dor we obtain the highest column densities of $N_{\text{H,LMC}} = (0.96^{+0.25}_{-0.12}) \cdot 10^{22}$~cm$^{-2}$. In the northwest the column densities are slightly lower with $N_{\text{H,LMC}} = (0.42^{+0.19}_{-0.09}) \cdot 10^{22}$~cm$^{-2}$ and the lowest in the southwest with $N_{\text{H,LMC}} = (0.29^{+0.13}_{-0.03}) \cdot 10^{22}$~cm$^{-2}$.

The distribution of the background absorbing column density (Fig. \ref{fig:nH_bkg}) appears to be different from the $N_{\mathrm{H,LMC}}$\ in front of the plasma. We obtain the highest background absorption in the X-ray spur and 30 Dor with values of $N_{\mathrm{H,BG}} = (1.31^{+0.79}_{-0.27}) \cdot 10^{22}$~cm$^{-2}$. In the western part we also obtain background absorption, albeit weaker, with $N_{\mathrm{H,BG}} = (0.80^{+0.35}_{-0.08}) \cdot 10^{22}$~cm$^{-2}$.

\begin{figure}
		\includegraphics[width=0.49\textwidth]{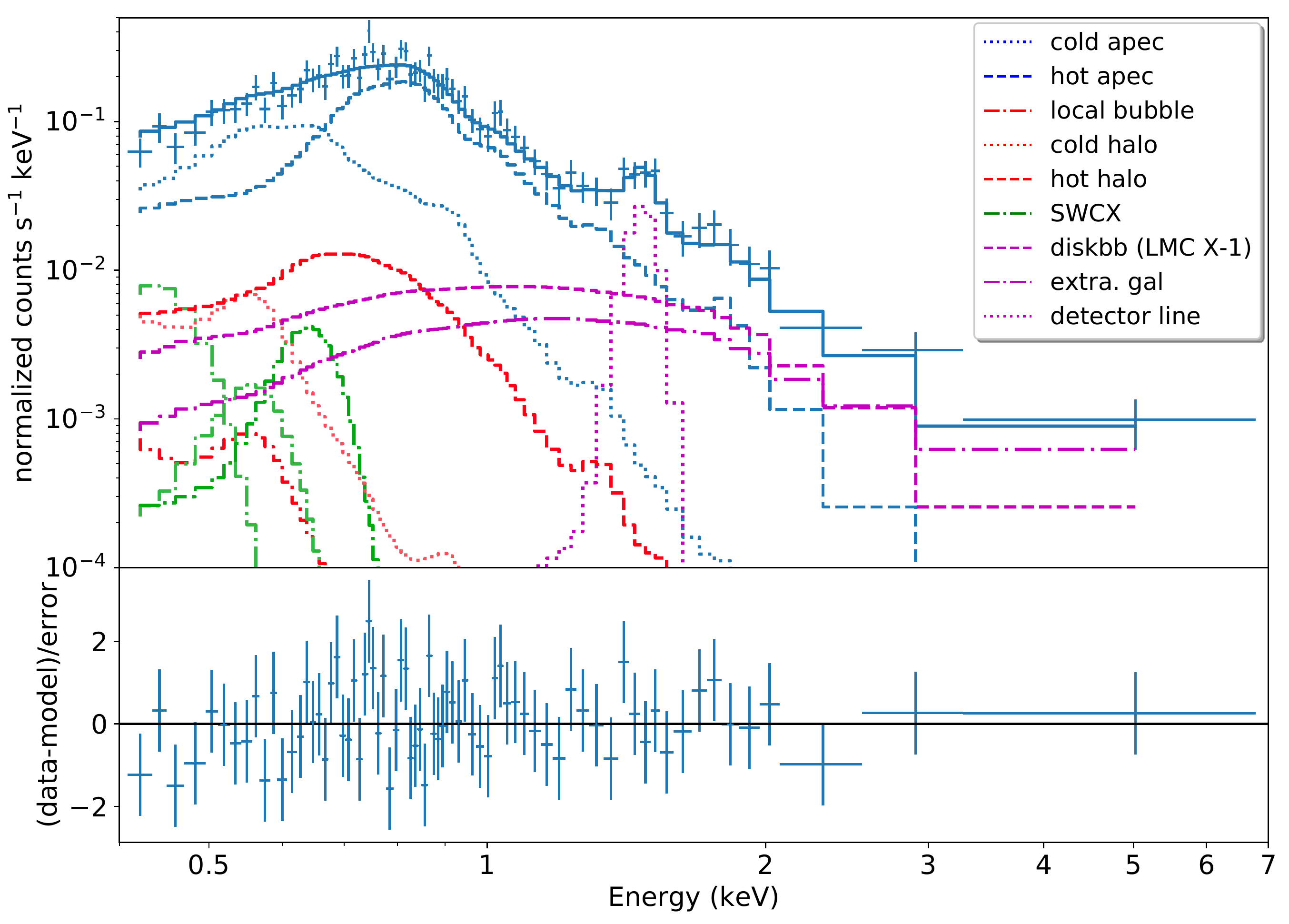}
			\caption{\label{fig:spectrum_new_observation_nei_swcx} XMM-Newton EPIC pn spectrum and model of the central tessellate of the new observation. The residuals are shown in the lower panel. The different model components are shown with dashed, dotted, and dash-dotted lines as indicated in the legend. The fit model uses two APEC components to account for the diffuse X-ray emission (cold and hot apec in legend). For better presentation, the data were rebinned visually with either 30 counts or $3\sigma$.}
\end{figure}

We also tried to fit an NEI model, which yielded similar good fits to the data compared to the two APEC model. However, the fits were much less constrained due to the additional ionization timescale parameter $\tau$. This parameter varied from region to region, even within the FOV of one observation. Furthermore, this leads to strong variations in the fitted diffuse plasma temperature. In short, the available data yield good results assuming CIE, but lack the necessary statistics to obtain reliable fits with the more complex NEI model.
\paragraph{New X-ray spur observation:}
Among the various archival data we analyzed,
we also performed a new observation with XMM-Newton EPIC of the southern tip of the X-ray spur (ObsID: 0820920101). This observation was analyzed in the same way as the others. However since the data have not been published so far, we describe the results in more detail here. The average effective exposure time over all detectors after data reduction was $\sim 32$~ks. The observation was pointed at RA $= 86.19\degr$ and Dec $= -70.52\degr$.
  The spectrum of the central tessellate, using the two APEC model to fit the data, is shown in Fig. \ref{fig:spectrum_new_observation_nei_swcx}. The soft X-ray emission in the FOV is shown in Fig. \ref{fig:newobs_cont_ratio}. The soft emission of the X-ray spur clearly extends further to the south. In the lower half of the FOV the emission appears arc-shaped. The eastern part of the FOV shows a very low X-ray brightness, dominated by background.
  There are two massive stars located in the FOV of the observation, the O-star UCAC2 1449308 and WR star Brey 97.
  We observe a slightly harder X-ray color in the vicinity of the WR star.
%
\subsection{Physical properties of the hot X-ray gas}
\label{sec:phys_prop}
Based on the spectral fit results, we estimated the physical properties of the X-ray emitting gas. We limit our estimates to the second, hot plasma component discussed above, since the lower temperature plasma is also detected in the stellar bar. The lower temperature component corresponds to the diffuse emission consisting of the undisturbed hot ISM in equilibrium and unresolved stellar emission. With these estimates we can gain further insight into the physical conditions and origin of this hot plasma component.

We calculated the unabsorbed flux $F_X$ for each tessellate region for the second - hot - APEC component using the CFLUX\footnote{\url{https://heasarc.gsfc.nasa.gov/xanadu/xspec/manual/node280.html}} model. The 90\% confidence range was calculated using the error command of XSPEC. From this we also calculated the luminosity 
\begin{equation}
L_X = 4\pi D^2 F_X
\label{eq:lum}
\end{equation}
with using the distance $D=50$ kpc for the LMC. We normalized both the flux and luminosity by the area of the respective region to obtain the surface flux $F_X^A$ and brightness $L_X^A$.

Further, the physical conditions of the plasma can be derived from the 
 normalization of the APEC model component, given as
\begin{equation}
K = \frac{10^{-14}}{4\pi D^2_{\mathrm{LMC}} }  \int n_e n_H f dV \,[\mathrm{cm}^{-5}] \mathrm{ ,}
\label{eq:norm}
\end{equation}
with the electron and hydrogen densities $n_e$ and $n_H$, respectively, and the filling factor $f$.
We used the ratio of $n_e \approx 1.21 n_H$ for the LMC abundance of $0.5$ solar  \citep{lmc_n158}.
With this, we can estimate the hydrogen density from the normalization with
\begin{equation}
n_H = \sqrt{\frac{4\pi K D^2 10^{14}}{1.21 V}} f^{-1/2} \,[\mathrm{cm}^{-3}]
\label{eq:n_H}
\end{equation}
Assuming an ideal gas, the pressure can then be estimated from the temperature of the plasma, which we obtained from the spectral fits:
\begin{equation}
P/k_B = 2.31 n_H T_X f^{-1/2} \, [\mathrm{cm}^{-3} K] \mathrm{ ,}
\label{eq:P}
\end{equation}
with the X-ray temperature $T_x$ and the Boltzmann constant $k_B$.
And with the pressure $P$ we can estimate the thermal energy $E$ stored in the plasma with
\begin{equation}
E = \frac{3}{2} P V f^{1/2} \,[\mathrm{erg}]
\label{eq:E}
\end{equation}
Additionally, the mass of the plasma can be calculated using
\begin{equation}
M = 2.31 n_H m_H \mu V f^{-1/2} \,[\mathrm{g}]
\label{eq:M}
\end{equation}
with the $m_H$ and $\mu$ being the hydrogen mass and molecular weight of the ionized gas, respectively.

We used the spectral fit results of the tessellates for the calculations, as those results cover most of the X-ray spur and the surroundings. The derived values are given in Table~\ref{tab:phys_prop}. We derived the results for each region individually and then took the average of the resulting map over regions which correspond to the large scale structures in the vicinity. The upper and lower limits were calculated by taking the average of the upper and lower limit map of the respective derived parameter.

The volume $V$ was estimated by calculating the area of each tessellate polygon region using the Shapely\footnote{\url{https://shapely.readthedocs.io/en/latest/manual.html\#polygons}} python package. Multiplied with the depth of the X-ray emitting plasma, we obtain the volume of each region. We varied the depth for the volumes, depending on the region the tessellates are located in, to account for the complex geometry of the ISM. We took the values given by \citet{spur:2001} and additionally, we estimated the depth of the 30 Dor region with $z = 300$ pc based on the extent of the structure. The assumed depth is by far the largest factor of uncertainty in the calculations, however, the order of magnitude should be approximately correct.

The thermal energy $E$ and mass $M$ of the plasma was also normalized to the volume in $cm^{3}$. Otherwise we would obtain the total numbers for each tessellate region, which would not be comparable between different regions.
We give the results without assuming any specific filling factor, as this parameter is largely unknown to us. Based on the true value of the factor the results could change slightly. If we compare, for example, a filling factor of $f=1$ versus $f=0.3$, the pressure, hydrogen density and mass would increase by a factor $\sim 3$ while the thermal energy would decrease by that factor for $f=0.3$. 

To gain further insight into the geometry of the hot plasma, we compared the absorption column $N_{\mathrm{H}}^X$ from the spectral fits with the column density $N_H$ of the \ion{H}{i} data. This ratio $N_{\mathrm{H}}^X/N_H$ is a measure how deeply the plasma is embedded inside the X-ray absorbing material \citep{snrs_lmc_maggi}. A value smaller or close to unity suggests that the plasma is located more to the front and consequently a value larger then unity that the plasma is located to the back.
We obtain the column density $N_H$ from the \ion{H}{i} velocity maps by using
\begin{equation}
N_H = 1.823 \times 10^{18} \int T_b \, d\nu  \,[\mathrm{cm}^{-2}]
\label{eq:hi}
\end{equation}
with the brightness temperature $T_b$, assuming that the gas is optically thin \citep{dl_map}.

 The values are also given in Table~\ref{tab:phys_prop}.
For the spur we obtain the lowest $N_{\mathrm{H}}^X/N_H$ fraction. In comparison, the region close to the bar of the LMC, West, has a ratio about three times larger than for the spur. For the structures north of the spur, LMC-SGS~2 and 30 Dor, we also obtain larger rations, albeit still significantly smaller than in the region West. Further results are discussed below.

\begin{table*}
\centering
        \caption{\label{tab:phys_prop}Physical properties of the diffuse emission (second, hot apec component). Ideal gas assumed.}
\renewcommand{\arraystretch}{1.3}
                \begin{tabular}{l r r r r r r r}
                        Region & $\log F_X^A$ & $L_X^A\, [10^{33} ] $ & $n_H\, [10^{-2} ] $ & $P \, [10^{5} ] $ & $M \,[10^{-26} ] $ & $E\, [10^{-11} ] $ & $N_{\mathrm{H}}^X/N_H$ \\
                         & erg cm$^{-2}$ s$^{-1}$ arcmin$^{-2}$ & erg s$^{-1}$ arcmin$^{-2}$ & $f^{-0.5}$ cm$^{-3}$ & $f^{-0.5}$ cm$^{-3}$\,K  & $f^{-0.5}$ g cm$^{-3}$ & $f^{0.5}$ erg cm$^{-3}$ & \\
                        \hline
                        Spur & $-13.77 \pm 0.03$ & $5.08 \pm 0.33$ & $0.99^{+0.08}_{-0.15}$ & $1.53^{+0.43}_{-0.34}$ & $2.34^{+0.18}_{-0.35}$ & $3.18^{+0.89}_{-0.70}$ & $1.14$ \\
                        30 Dor & $-13.45 \pm 0.02$ & $10.54 \pm 0.52$ & $1.74^{+0.01}_{-0.08}$ & $2.77^{+0.48}_{-0.38}$ & $4.10^{+0.03}_{-0.19}$ & $5.74^{+0.99}_{-0.80}$ & $2.27$ \\
                        West & $-14.02 \pm 0.03$ & $2.84 \pm 0.22$ & $0.69^{+0.03}_{-0.09}$ & $1.36^{+0.13}_{-0.24}$ & $1.63^{+0.07}_{-0.20}$ & $2.81^{+0.27}_{-0.49}$ & $3.76$ \\
                        Dark Region & $-14.45 \pm 0.03$ & $1.05 \pm 0.09$ & $0.44^{+0.05}_{-0.08}$ & $0.71^{+0.18}_{-0.12}$ & $1.04^{+0.11}_{-0.19}$ & $1.47^{+0.38}_{-0.24}$ & $1.64$ \\
                        LMC-SGS~2 & $-13.68 \pm 0.02$ & $6.24 \pm 0.29$ & $1.24^{+0.09}_{-0.18}$ & $2.04^{+0.43}_{-0.39}$ & $2.92^{+0.22}_{-0.42}$ & $4.23^{+0.90}_{-0.81}$ & $1.76$ \\
                \end{tabular}
\end{table*}

\section{Multiwavelength analysis}
\label{sec:multi_analysis}
\subsection{X-ray vs. \ion{H}{i}}
\label{sec:HI_analysis}
\paragraph{L-Component:}
The \ion{H}{i} emission of the L-component is strongest east of the X-ray spur. Here, the \ion{H}{i} contours almost perfectly trace the edge of the X-ray spur (Fig. \ref{fig:X_HI_cont}a). This is also the case for the X-ray emission of the southern tip of the spur. The arc-shaped emission is strongly anticorrelated with the L-component contours. To the northeast, near LMC-SGS~2, the two extended brighter extended emission regions in the \ion{H}{i} contours perfectly fit into the depressions of X-ray brightness.
  To the very east (RA\,$ > 5^{\mathrm{h}}47^{\text{m}}\,$), the X-ray emission is extremely low. The X-ray emission could either be intrinsically weak there or heavily absorbed. However, since the L-component is the only bright \ion{H}{i} component in this area the column density might not be high enough for absorption.
\paragraph{I-Component:}
For the I-component (Fig. \ref{fig:X_HI_cont}b), the strongest \ion{H}{i} emission covers the regions bright in X-rays in the east of the mosaic. This component is very bright near the massive stellar cluster R136, surrounding the  X-ray bright regions. The dark patch in X-rays, just south of R136, is very well anticorrelated with the highest I-component contours. Another X-ray dark area near RA\,$ = 5^{\mathrm{h}}45^{\text{m}}\,$, Dec\,$= -69\degr 20'$ seems to be complementary to \ion{H}{i}, as well as the dark patches in the vicinity of the HMXB LMC X-1. South of LMC X-1, the I-component overlaps with the X-ray spur with no apparent depression in brightness. However, the western edge of the large X-ray dark region west of the spur is nicely traced by the I-component emission. 
\paragraph{D-Component:}
The \ion{H}{i} D-component (Fig. \ref{fig:X_HI_cont}c) shows no clear correlation or anticorrelation with the northern part of the X-ray mosaic. However, in the south (Dec\,$< -69\degr 50'$) the morphology of the D-component perfectly matches the X-ray dark region west of the X-ray spur. The \ion{H}{i} distribution is clearly anticorrelated with the diffuse X-ray emission at RA\,$ = 5^{\text{h}}40^{\text{m}}$ to $5^{\text{h}}32^{\text{m}} $.
\subsubsection{Correlation calculations}
\begin{figure}
	\centering
		\includegraphics[width=0.49\textwidth]{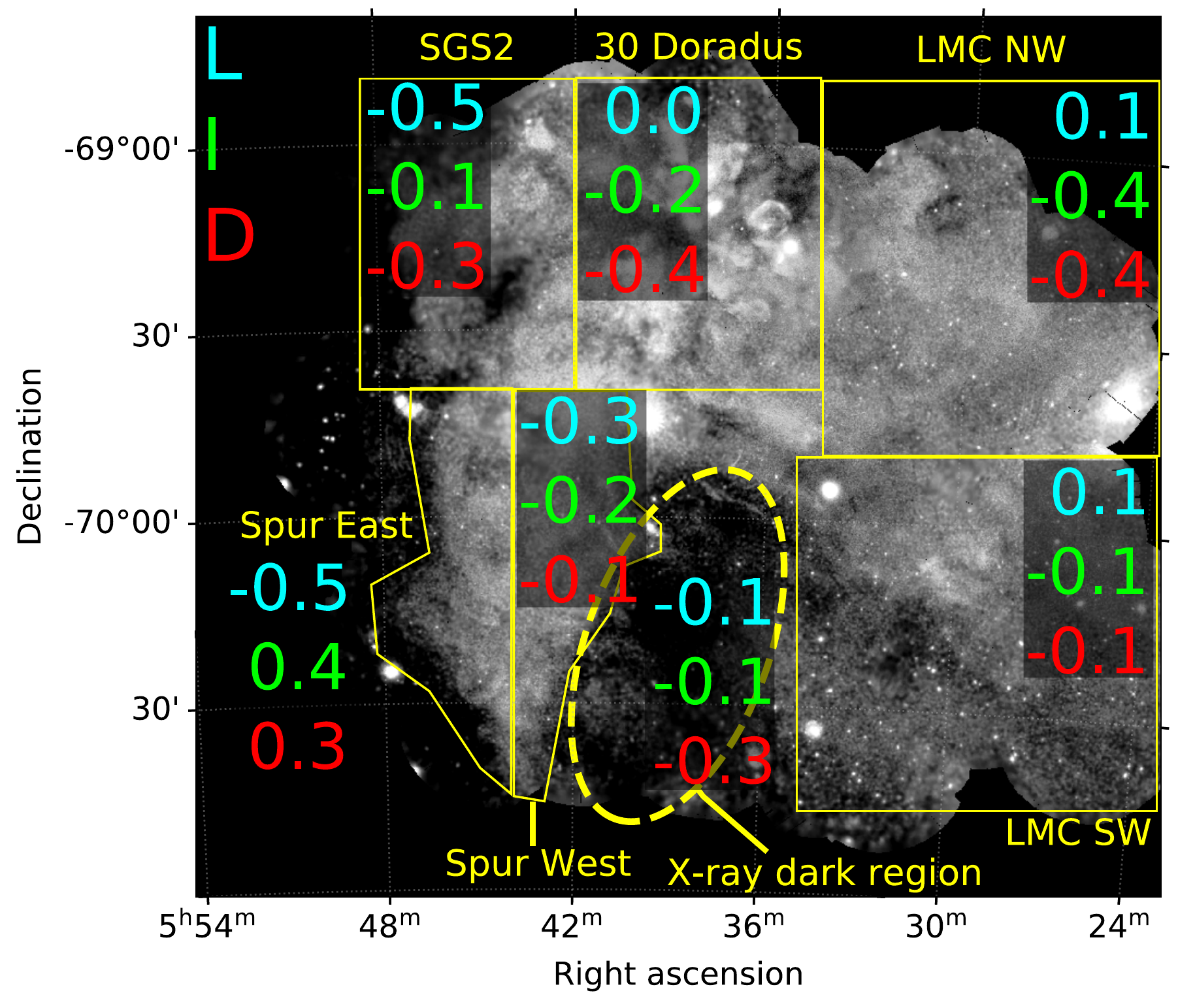}
		\caption{\label{fig:X-ray_HI_correlation_quadrants}Soft X-ray mosaic (0.4-1.25\,keV) with the individual regions (yellow) where we calculate the correlation coefficients of X-ray data with the \ion{H}{i} data. The correlation coefficients for each region are shown in green for the L-component, cyan for the I-component and red for the D-component. Positive coefficients indicate correlation while negative coefficients indicate anticorrelation (linear relationship).}
\end{figure}

In order to quantify the results of the previous section, we calculated correlation coefficients between the \ion{H}{i} and X-ray images. For each position of the sky we took the X-ray- and the \ion{H}{i} brightness values. The coefficients were calculated separately for each \ion{H}{i} component. We introduced thresholds in the calculations to reduce outliers due to image artifacts and noise. We calculated thresholds on the individual pixel brightness from the $0.1-95$ (X-rays) and $0.1-99.9$ (\ion{H}{i}) percentile range of the corresponding images. Additionally, the lower thresholds were set to at least above the respective noise level. If for any image the brightness was outside the thresholds, the sky pixel was discarded from the correlation coefficient calculations to reduce outliers. We divided the images into interesting regions as shown in Fig. \ref{fig:X-ray_HI_correlation_quadrants}. If a region had more then 100 valid sky pixels after threshold filtering, the correlation coefficient was calculated with the Pearson method\footnote{\url{https://docs.scipy.org/doc/scipy-0.14.0/reference/generated/scipy.stats.pearsonr.html}} (linear relationship). The range of correlation coefficients is $-1.0$ to $1.0$, where a negative coefficient indicates anticorrelation and a positive one correlation.

For the X-ray spur we obtain mixed coefficients depending on the \ion{H}{i} component. In the eastern part, we see the strongest anticorrelation between X-ray and the L-component in the whole area of the mosaic. On the other hand, the D- and especially the I-components seem to be correlated here. The anticorrelation of the L-component extends along the eastern end of the X-ray emission to the north (LMC-SGS~2 region) and confirms what we previously saw from the \ion{H}{i} contours. In the western part of the spur, all components seem to be weakly anticorrelated. In the X-ray dark region, all \ion{H}{i} components show anticorrelation with the X-ray emission, especially the D-component.

The 30 Dor region shows anticorrelation between X-rays and the I- and D-component. As shown previously the bright harder X-ray emission is surrounded by strong emission of the I- and D-component (Fig. \ref{fig:X_HI_cont}). 

In the northwestern region we see significant anticorrelation between the I- and D-component and X-ray brightness. We observe a similar trend for the southwestern region, albeit with weaker coefficients. We note that the L-component is very faint in the western part of the mosaic.
\label{sec:brightness_profiles}
\subsubsection{Brightness profiles}
\begin{figure*}
	\centering
		\begin{subfigure}[t]{0.49\textwidth}
			\includegraphics[width=\textwidth]{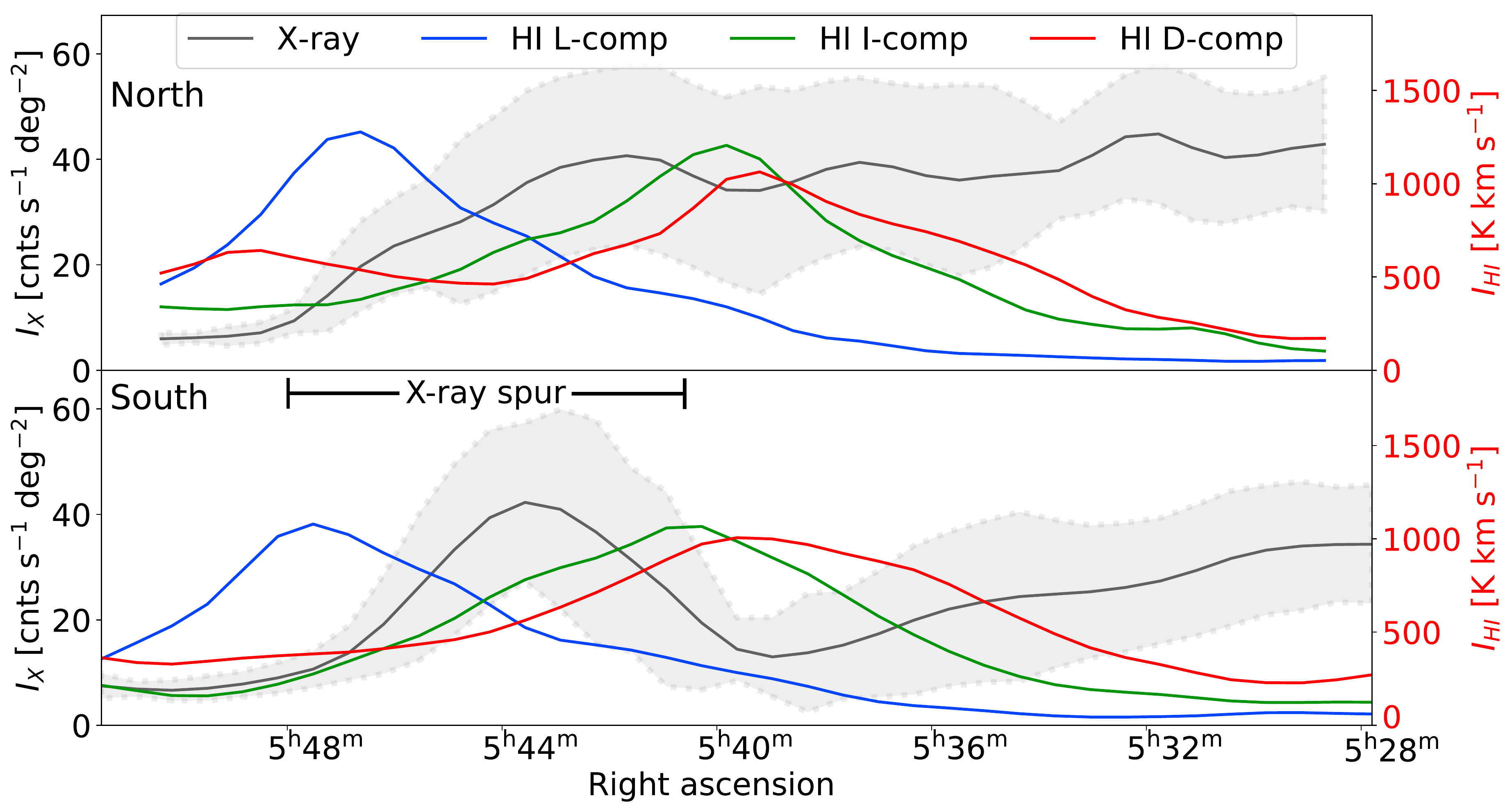}
			\caption{\label{fig:intensity_profile_ra}}
		\end{subfigure}
		\hfill
		\begin{subfigure}[t]{0.49\textwidth}
			\includegraphics[width=\textwidth]{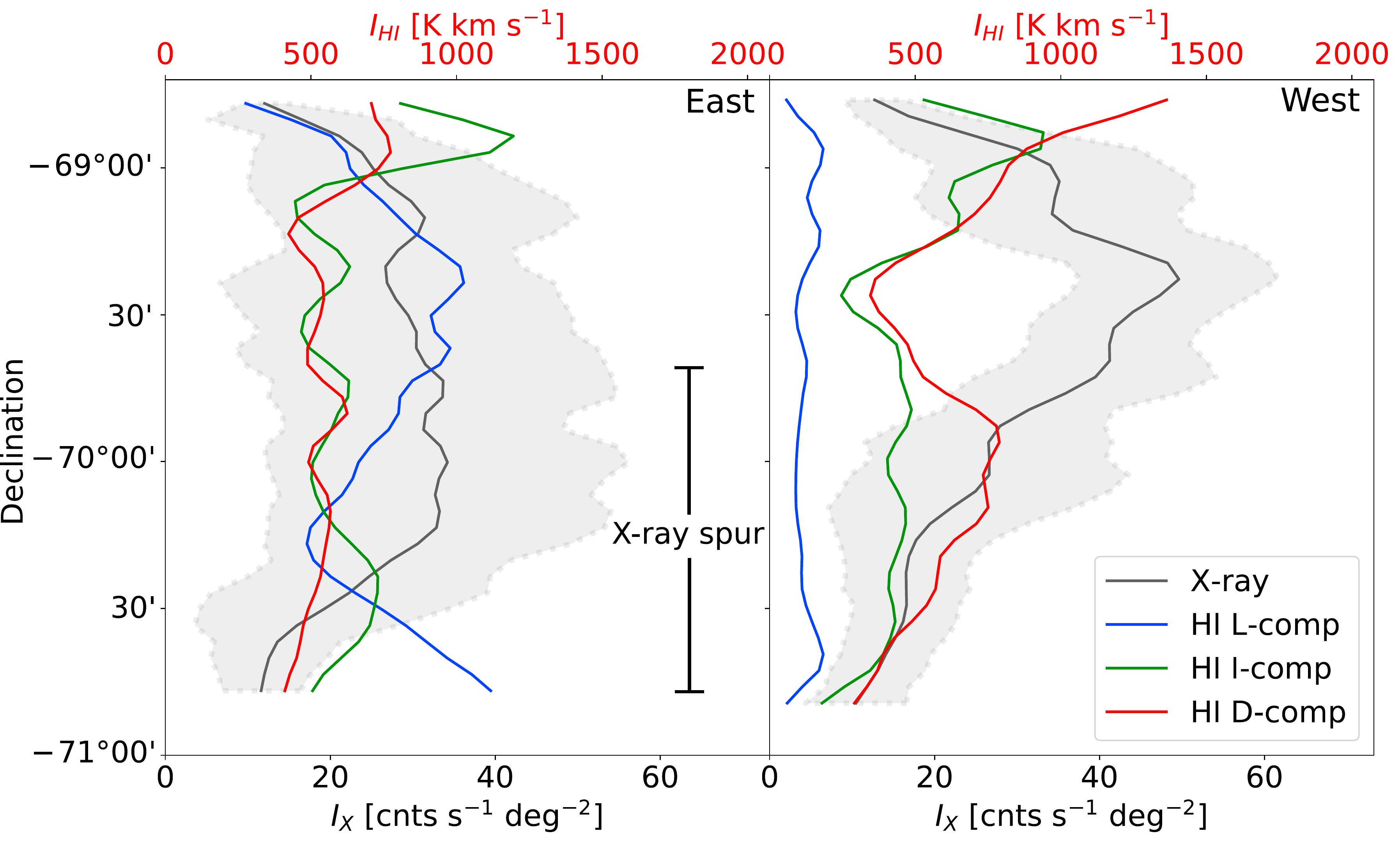}
			\caption{\label{fig:intensity_profile_dec}}
		\end{subfigure}
		\caption{\label{fig:intensity_profiles}(a) Right ascension brightness profiles for X-rays $I_X$ (0.4-1.25\,keV) and \ion{H}{i} $I_{\text{HI}}$. The integration range is Dec\,$=-68\degr 20'$ to $-69\degr 37'$ (north, upper panel) and Dec\,$=-69\degr 37'$ to $-70\degr 54'$ (south, lower panel). (b) Declination brightness profiles for X-rays (0.4-1.25\,keV) and \ion{H}{i}. The integration range is RA\,$=5^{\text{\text{h}}}49^{\text{\text{m}}}$ to $5^{\text{\text{h}}}40^{\text{\text{m}}}$ (east, left panel) and RA\,$=5^{\text{\text{h}}}40^{\text{\text{m}}}$ to $5^{\text{\text{h}}}28^{\text{\text{m}}}$ (west, right panel). The data were integrated in Dec (a) and RA (b) in slices of 				200\arcsec width. The curves were smoothed with a 1D-gaussian ($\sigma=1$) for visual purposes. The curve values were derived from the mean and the uncertainty band from the standard deviation of each slice. The extent of the 			X-ray spur is indicated with dashed lines.}
\end{figure*}

We also calculated brightness profiles of X-rays and the \ion{H}{i} components for a simpler visual representation of the correlation analysis. We divided each image into slices with a width of 200\arcsec. For each slice, we calculated the mean intensity and standard deviation. From these values, we calculated the brightness profiles in both RA and Dec (Fig. \ref{fig:intensity_profiles}). We set lower and upper thresholds on the \ion{H}{i} and X-ray brightness to the $1\mathrm{-} 95$ (X-ray) and $1\mathrm{-} 99$ (\ion{H}{i}) percentile range of the respective images. Values above or below these thresholds are rejected from the calculations. This reduced statistical scattering, especially for the X-ray brightness where very bright pixels are still present after data reduction. From the foregoing analysis, we saw large differences in the ISM conditions from east to west and north to south. Therefore, we divided the X-ray mosaic into north/south for the RA and east/west for the Dec profiles. Otherwise, averaging effects could smear out interesting features.
\paragraph{RA profiles:}
In the RA profiles, we note several interesting features. In the north the X-ray brightness stays very low for high L-component intensity in the east. Toward the west, the X-ray brightness rises and then  stays relatively constant, seemingly unaffected by the peak seen in the I- and D-component intensities at RA~$\approx 5^{\text{h}}40^{\text{m}}$. 

In the south, we see similar features to the east. The X-ray brightness is very low where the L-component is the strongest and peaks at RA~$\approx 5^{\text{h}}44^{\text{m}}$, where this component is much weaker. Again, at RA~$\approx 5^{\text{h}}40^{\text{m}}$ both the I- and D-component peak, but here the X-ray brightness drops significantly. The X-ray brightness only rises again to the west, where all \ion{H}{i} components become very weak.
\paragraph{Dec profiles:}
The Dec profiles reveal additional features about the different ISM components. In the east, there seems to be no clear trend between X-rays and \ion{H}{i}. Only for Dec~$< - 70\degr30'$ anticorrelation between the X-rays and L-component can be observed. 

In the west, the X-ray brightness peaks at Dec~$\approx -69\degr25'$. Around this position, \ion{H}{i} emission is faint, especially the D-component. At higher and lower declination, the D-component is significantly stronger, which seems to correlate with low X-ray brightness.
\subsection{CO emission}
\begin{figure}
	\centering
		\includegraphics[width=0.49\textwidth]{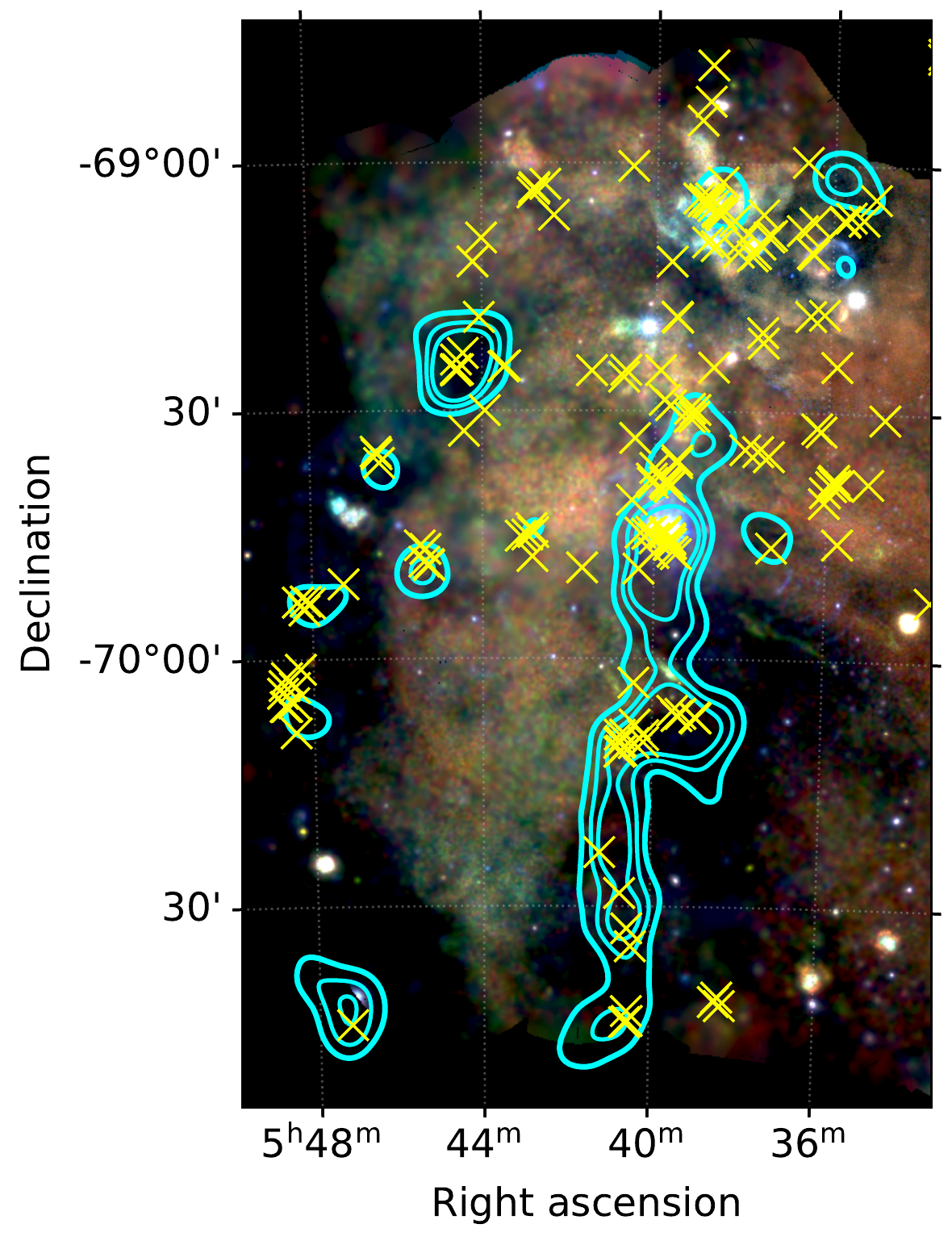}
		\caption{\label{fig:co_xray}  Soft XMM-Newton three-color mosaic with 0.4-0.7\,keV (red), 0.7-1.0\,keV (green) and 1.0-1.25\,keV (blue). Overlayed in cyan are the NANTEN CO data contours in the $95\mathrm{-} 99$ percentile range ($1\mathrm{-} 3\sigma$). The yellow crosses indicate positions of certain high- and intermediate-mass young stellar objects \citep[YSOs,][]{ysos_lmc}. }
\end{figure}

We show the NANTEN \element[][12]{CO} data ($J = 1 - 0$),  overlayed on the X-ray data, in Fig. \ref{fig:co_xray}. We also show the positions of high- and intermediate mass YSO candidates from the ``definite'' list of YSOs by \citet{ysos_lmc}. We see an extended strong CO ridge tracing the western edge of the X-ray spur at RA~$= 5^{\text{\text{h}}}41^{\text{\text{m}}}$ and Dec~$= -69\degr27\arcmin$ to Dec~$=-70\degr50\arcmin$. There are also smaller CO bright regions located in the spur (RA~$= 5^{\text{\text{h}}}45^{\text{\text{m}}}$, Dec~$= -69\degr49\arcmin$ and RA~$= 5^{\text{\text{h}}}43^{\text{\text{m}}}$, Dec~$= -69\degr44\arcmin$). East of the spur, we also see smaller regions of strong CO emission and an especially bright region near the center of LMC-SGS~2. To the north, we observe several CO clouds in the vicinity of 30 Dor. We observe many YSOs in the CO ridge at the western edge of the X-ray spur and also some YSOs in the smaller CO clouds in the northernmost part of the spur. However, for the most part the spur seems to be void of any YSOs or star-formation sites, in contrast to the X-ray bright region north of the X-ray spur.
\subsection{Dust}
\label{sec:Dust_analysis}
From the foregoing study, it is evident that the low X-ray brightness regions east and west of the X-ray spur are anticorrelated with the \ion{H}{i} emission. One possibility for the anticorrelation can be absorption of X-rays if we assume the absorbing material to be located in front of the plasma. As absorption by interstellar dust can be significant, we compare the infrared dust optical depth with the X-ray emission. X-rays can be directly absorbed by dust, mainly via photoelectric absorption \citep{rass_dust}. Also, the dust emission traces especially dense regions of the ISM where the absorbing column density caused by other material might be significantly enhanced.

\begin{figure*}
\begin{subfigure}[t]{0.49\textwidth}
	\centering
		\includegraphics[width=1.0\textwidth]{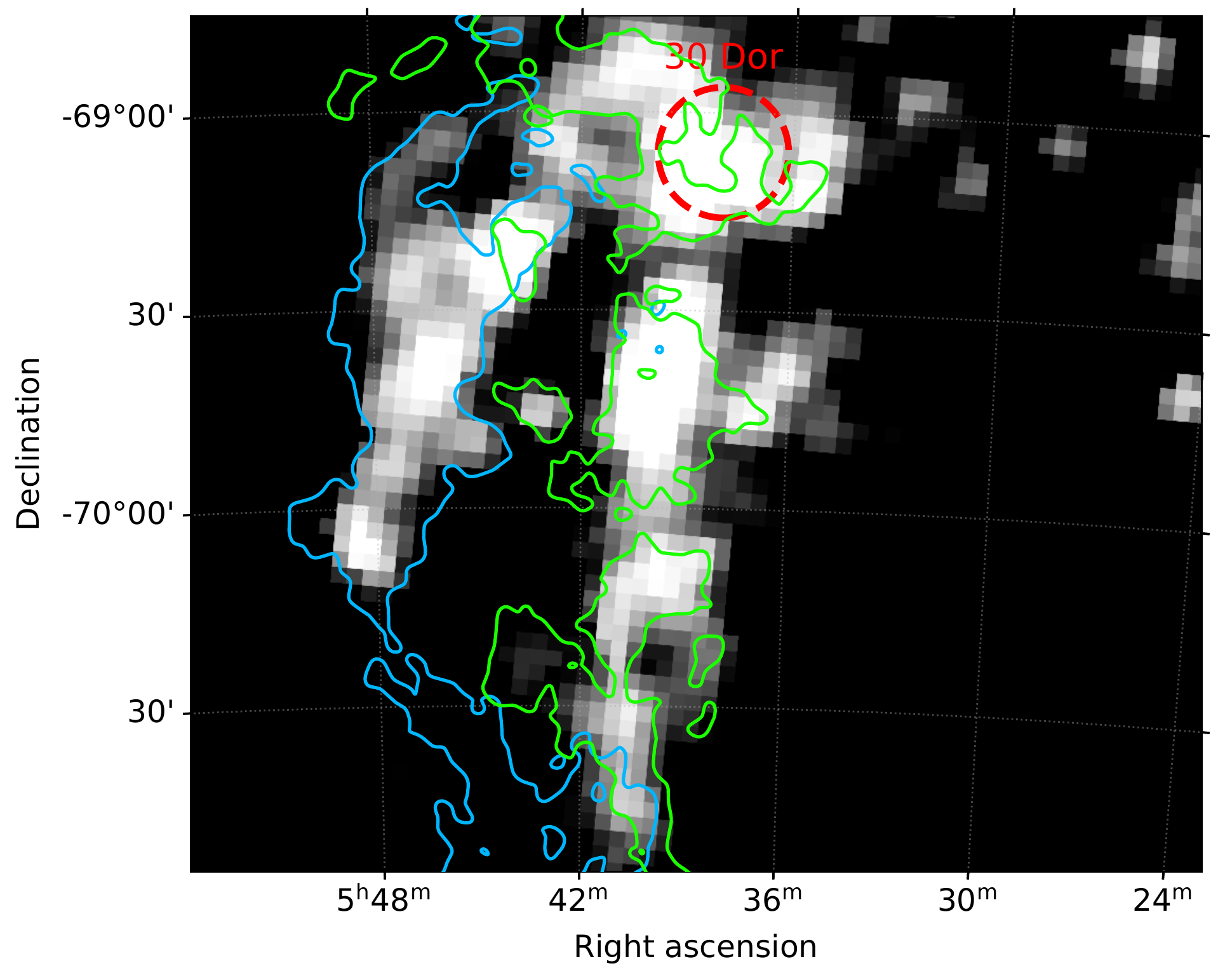}
		\caption{\label{fig:tau353_map}}
	\end{subfigure}
\hfill
\begin{subfigure}[t]{0.49\textwidth}
	\centering
		\includegraphics[width=1.0\textwidth]{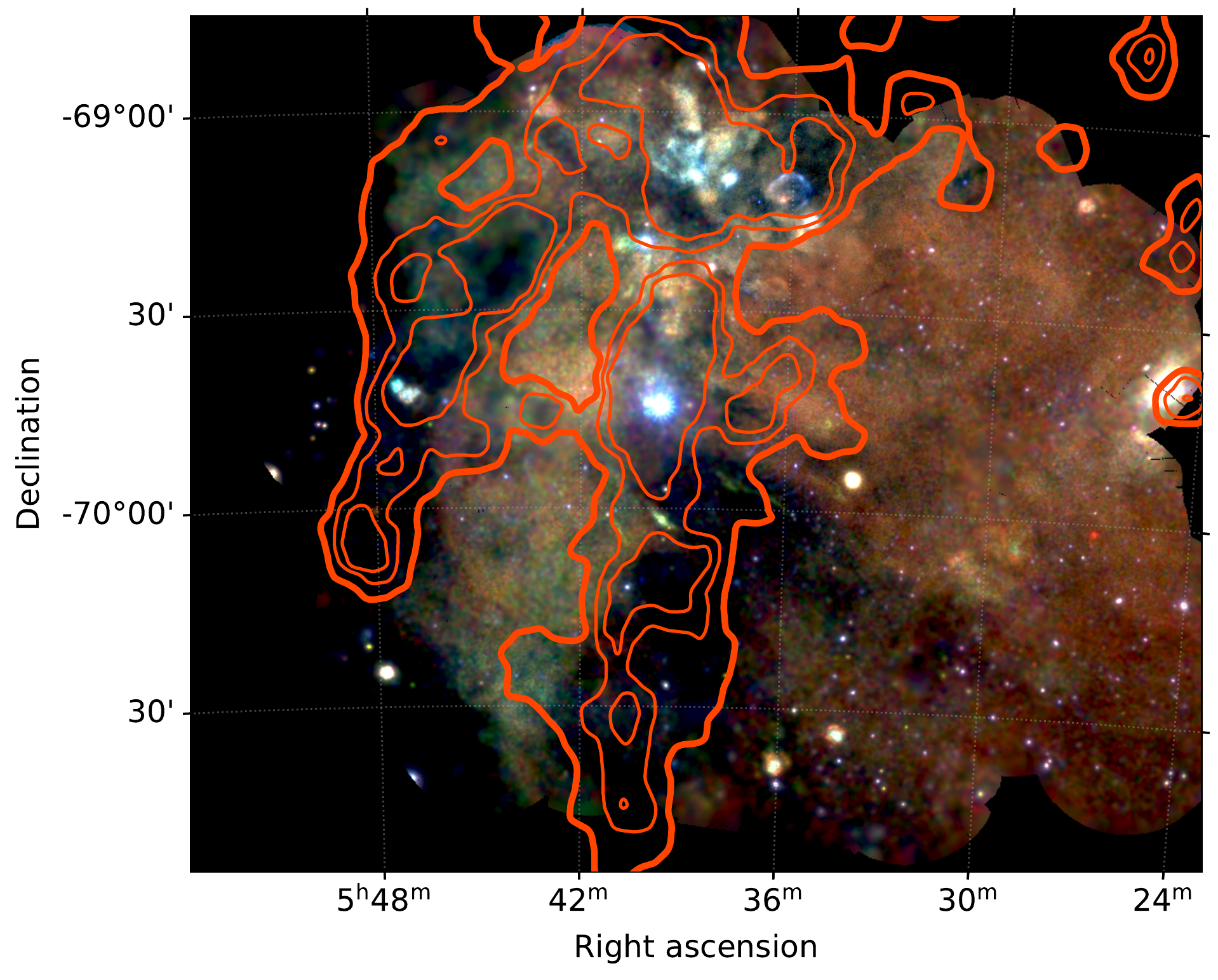}
		\caption{\label{fig:tau353_xrays_cont}}
		\end{subfigure}
		\caption{(a) Dust optical depth $\tau_{353}$ map from Planck data \citep{planck_dust} in the range of $\tau_{353} = 2.0\cdot 10^{-5}$ to $6.4\cdot 10^{-5}$ ($80\mathrm{-}98$ percentile range). Overlayed are contours of the \ion{H}{i} L-component (light blue) and I-component (green). The contour levels correspond to $870$~K km s$^{-1}$ and $930$~K km s$^{-1}$ for the L- and I-component, respectively. (b) Soft XMM-Newton three-color mosaic with 0.4-0.7\,keV (red), 0.7-1.0\,keV (green), 1.0-1.25\,keV (blue) and $\tau_{353}$ dust optical depth contours (orange) calculated with the same scale as in (a).}
\end{figure*}

We used the $\tau_{353}$ data described in Sect. \ref{sec:dust_data}.
The Galactic foreground absorption $\tau_{353}$ was estimated from the \ion{H}{i} data and subtracted from the maps. A detailed description of the procedure is given by \citet{n44_tsuge}.
We created contours from the dust optical depth map based on the image $80-98$ percentile range and show the comparison with \ion{H}{i} and soft X-rays in Fig. \ref{fig:tau353_map} and  Fig. \ref{fig:tau353_xrays_cont}, respectively. 
The dust optical depth distribution seems to be well correlated with the emission of the \ion{H}{i} L- and I-components. There also seems to be a correlation with X-rays near LMC-SGS~2 and the 30 Dor area. In contrast, at the X-ray dark region around RA\,$ = 5^{\text{h}}45^{\text{m}}\,$, Dec\,$= -69\degr 20'$, the optical depth appears to be complementary with the X-ray emission. Additionally, dust filaments are observed east and west of the spur. The distribution of dust also correlates well with the CO filaments and sites where YSOs are found (see Fig. \ref{fig:co_xray}). This indicates that in addition to \ion{H}{i} emitting gas, also denser ISM in the form of dust is present in those regions. In most parts of the spur, we observe no significant absorption by dust, confirming that the densities of the cold ISM in the spur must be smaller in comparison.
\subsection{Optical emission}
\label{sec:Optical_analysis}
\begin{figure*}
	\centering
	\begin{subfigure}[t]{0.49\textwidth}
		\includegraphics[width=1.0\textwidth]{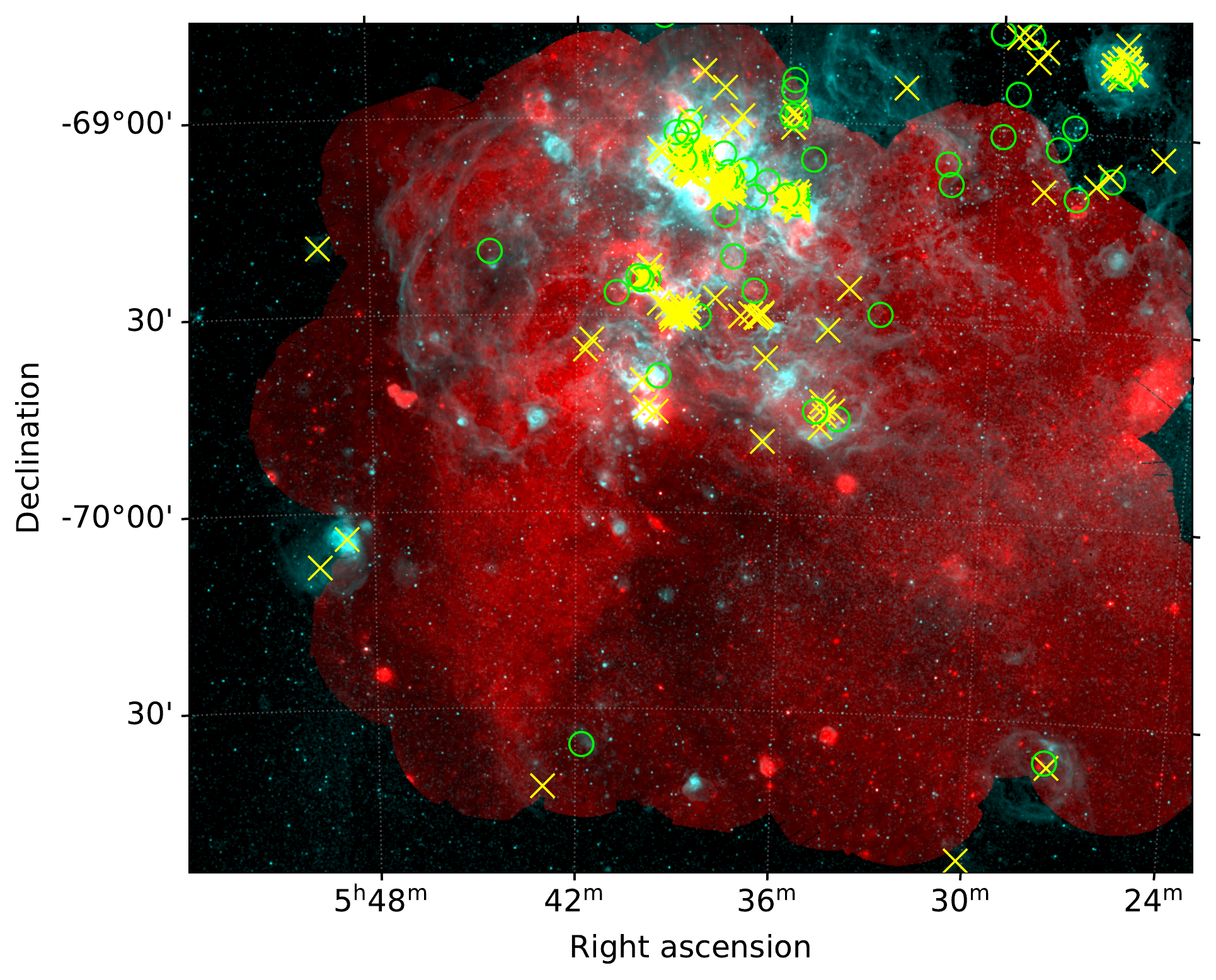}
		\caption{\label{fig:lmc_xray_ha_massive_stars}}
\end{subfigure}
\hfill
\begin{subfigure}[t]{0.49\textwidth}
	\centering
		\includegraphics[width=1.0\textwidth]{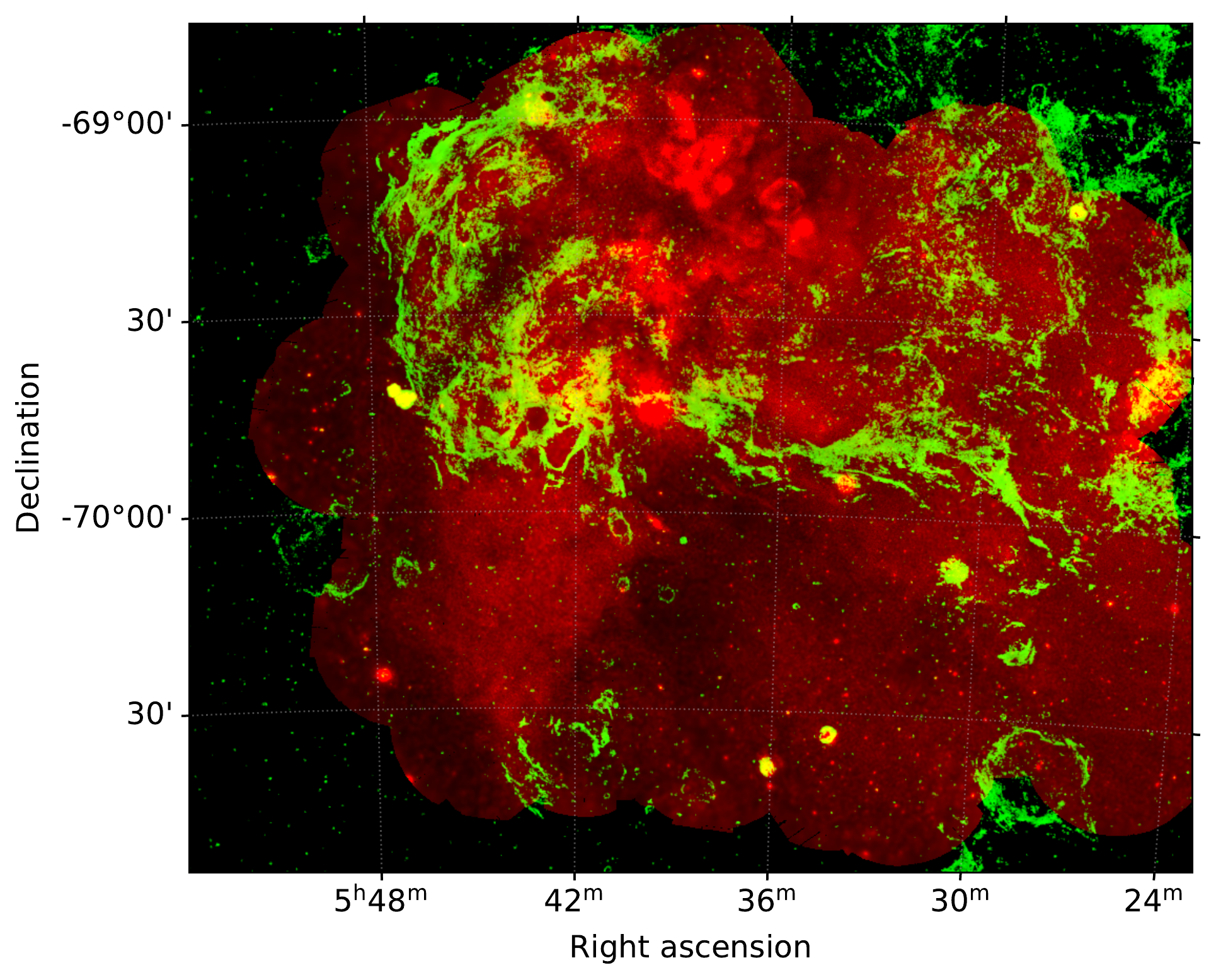}
		\caption{\label{fig:lmc_xray_sii2ha_ratio}}
		 \end{subfigure}
		 \caption{(a) Soft X-ray emission mosaic in the 0.4-1.25 keV range (red) and MCELS H$\alpha$ (cyan). Massive stars in the vicinity  are marked with yellow crosses (O stars) and green circles (WR stars) \citep{sage_massive_stars_lmc}. (b) Soft X-ray emission mosaic in the 0.4-1.25 keV range (red) and MCELS \mbox{[\ion{S}{ii}]/H$\alpha$ ratio} ratio in green. Only ratio values above the threshold of 0.4 are shown. The vicinity of 30 Dor appears blank since we have very strong H$\alpha$ emission here, thus yielding a very low ratio.}
\end{figure*}
We compared the optical data from the MCELS with the X-ray emission. We focused on the H$\alpha$ and \mbox{[\ion{S}{ii}]/H$\alpha$ ratio}, which gives us information about the dominant processes which produce line emission in the optical and is thus also crucial for the understanding of the X-ray emission. The comparison of X-rays and H$\alpha$ is shown in Fig. \ref{fig:lmc_xray_ha_massive_stars} and the comparison of X-rays and \mbox{[\ion{S}{ii}]/H$\alpha$ ratio} in Fig. \ref{fig:lmc_xray_sii2ha_ratio}.

In the H$\alpha$ comparison we note several interesting features. At the position of the X-ray spur we see very little H$\alpha$ emission. In contrast, the 30 Dor area shows very strong H$\alpha$ emission. We also see strong H$\alpha$ emission in the northeast which coincides very well with the soft X-ray structures in LMC-SGS~2. The northeast of 30 Dor and the northern X-ray spur also coincide with this structure. The morphology of the emission suggests, that the ISM condition in the southern X-ray spur is very different compared to the northern part of the mosaic. In Fig. \ref{fig:lmc_xray_ha_massive_stars} we also show the positions of the massive stars (O, WR) which are well correlated with the strong H$\alpha$ emission.

An enhanced \mbox{[\ion{S}{ii}]/H$\alpha$ ratio} indicates photoionization by ultra-violet (UV) photons and a higher temperature of the ISM compared to HII regions \citep{sii_raynolds}.
Even higher \mbox{[\ion{S}{ii}]/H$\alpha$ ratios} are most likely caused by shock-ionization \citep{sii_fesen}.
In the \mbox{[\ion{S}{ii}]/H$\alpha$ ratio} image we see high ratios in the north east related to the LMC-SGS~2 structure. To the northwest, the high ratios correspond to the structure LMC SGS-3. The very northern part of the X-ray spur also shows a high ratio. In contrast, the central- and southern part of the X-ray spur have an extremely low ratio. Here, we see neither H$\alpha$ emission nor high \mbox{[\ion{S}{ii}]/H$\alpha$ ratio}.

\begin{figure}
	\centering
		\includegraphics[width=0.4\textwidth]{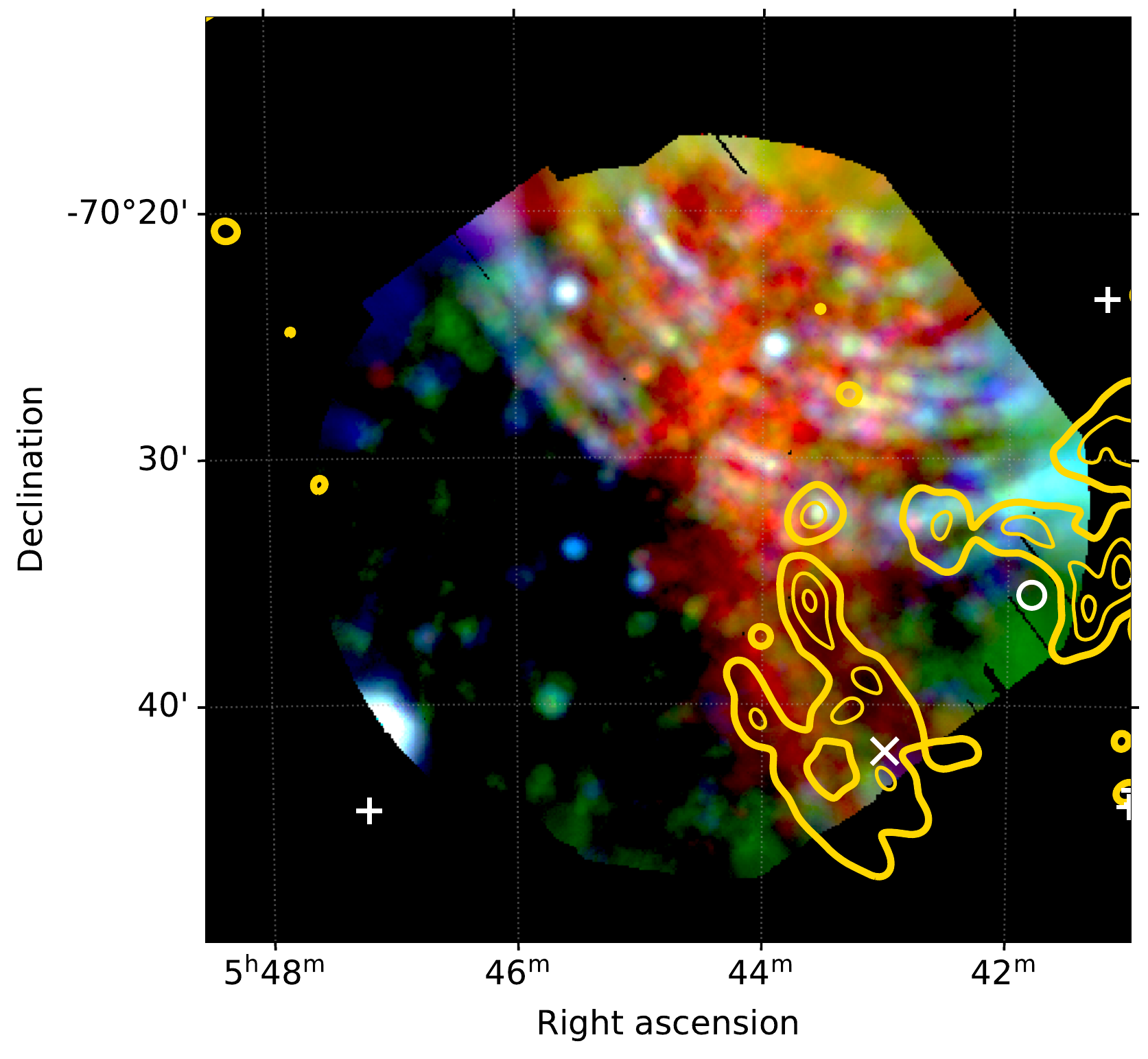}
		\caption{\label{fig:newobs_cont_ratio}  X-ray color composite with 0.4-1.25 keV (red), 1.25-2.00 keV (green) and 2.00-4.00 keV (blue) for the new observation of the spur (ObsID: 0820920101). The contours were calculated from the $85\mathrm{-}99$ percentile range ($0.68\mathrm{-}7.04$) of the \mbox{[\ion{S}{ii}]/H$\alpha$ ratio} MCELS map (see Fig. \ref{fig:lmc_xray_sii2ha_ratio}). The arc-shaped structures seen in the northern half in blue are caused by straylight from LMC X-1. Massive stars in the vicinity are marked by white Xs (O-stars), circles (WR-stars) and crosses (YSOs).}
\end{figure}

In the south, there are filaments which coincide with a WR (Brey 97) and an O (UCAC2 1449308) star at the souther tip of the X-ray spur.
  These filamentary structures, as seen in the \mbox{[\ion{S}{ii}]/H$\alpha$ ratio} in Fig. \ref{fig:lmc_xray_sii2ha_ratio}, are most likely caused by photoionization of the O- and WR-star. The arc-shaped southern half of the soft diffuse emission also seems to be correlated with a high \mbox{[\ion{S}{ii}]/H$\alpha$ ratio} as shown in Fig. \ref{fig:newobs_cont_ratio}.
%
%
%
\subsection{Discussion}
\begin{figure}
	\centering
		\includegraphics[width=0.49\textwidth]{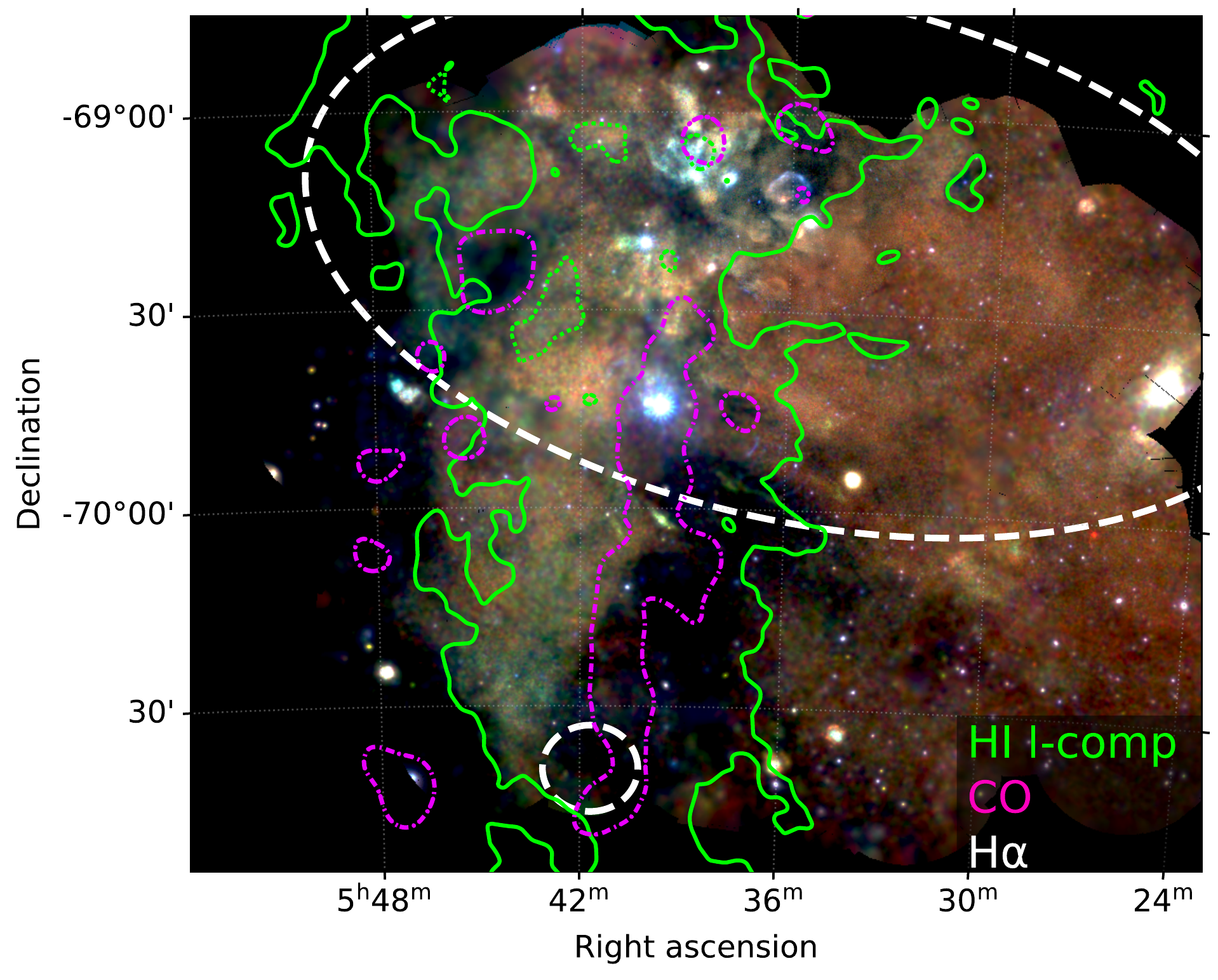}
		\caption{\label{fig:composite_discussion}Multiwavelength composite of the X-ray mosaic with \ion{H}{i}, CO and H$\alpha$ data. The \ion{H}{i} data are shown with contours in green (L-component) with the contour levels corresponding to the 90th percentile. Dashed contours indicate local minima. The magenta dash-dotted contours show the 95th percentile ($> 1\sigma$) of the CO data. The approximate area where significant H$\alpha$ emission is observed is marked with white dashed regions.}
\end{figure}
\subsubsection{Morphology}
In Fig. \ref{fig:composite_discussion} we show a multiwavelength view of the southeastern part of the LMC to highlight the morphology of the different ISM components. 
In this region of the LMC, ``bridge'' features - that are the I-component - between the \ion{H}{i} L- and D-components are measured in velocity space (\ref{fig:pv_diagram}). Additionally, the complementary spatial distribution suggests an ongoing collision between the \ion{H}{i} L- and D-component \citep{HI:fukui_R136,n44_tsuge}. The low metallicity of the L-component gas \citep{HI:fukui_R136,n44_tsuge,av_lmc} and hydrodynamical simulations \citep{tidal_simulation2, tidal_simulation1} suggest, that the gas was stripped from the SMC by a tidal interaction with the LMC in the past. The L-component collides with typical velocity differences of $\sim 50$~km\,s$^{-1}$ with the LMC disk. The high velocity and large scale ($\sim$~kpc) of the collision most likely triggered the formation of many massive stars by compressing the \ion{H}{i} gas. 
 Additional evidence of the collision are revealed by polarization measurements from radio continuum data by \citet{klein_1993}. The data show the existence of a giant magnetic loop south of 30 Dor, which appears to be stretched between the disk and the L-component. They discussed, that the movement of the L-component through the disk and the subsequent collision might have ``dragged the field along'' and thus caused the magnetic anomaly.

The direct comparison of the X-ray data and the \ion{H}{i} data has shown that on the eastern side, the X-ray spur follows the contours of the strongest L-component emission very closely in anticorrelation (green contours in Fig. \ref{fig:composite_discussion}, see also Fig. \ref{fig:X_HI_cont}a). 
 The V-shaped southernmost tip of the spur is also enclosed by strong L-component emission. On the western side, the X-ray spur's shape seems to be complementary with strong \ion{H}{i} emission of all components, especially that of the D-component (blue contours in Fig. \ref{fig:composite_discussion}, see also Fig. \ref{fig:X_HI_cont}c). These results indicate that the hot plasma in the X-ray spur could be surrounded by cold, dense ISM, resulting in the unusual triangular shape. While the large X-ray dark area west of the spur might be caused by very high column densities, the sharp dark edge east of the spur cannot be explained by absorption of \ion{H}{i} alone. Only the \ion{H}{i} L-component is strong here, while the \ion{H}{i} I-component seems to coincide with the hot plasma in the X-ray spur next to the \ion{H}{i} L-component. The position-velocity diagram of the \ion{H}{i} data shows that the \ion{H}{i} I-component is pronounced at the position of the X-ray spur (Fig. \ref{fig:pv_diagram}).

In addition to the \ion{H}{i} emission, we also observe relatively high dust optical depth $\tau_\text{353}$ northeast and west of the X-ray spur (Fig. \ref{fig:tau353_xrays_cont}). In the northeast, the optical depth appears complementary with the X-ray emission. The massive star cluster in 30 Dor seems to be deeply embedded in dust, while the massive heating processes inside the cluster cause the strong X-ray emission. We observe a similar correlation around LMC X-1, which also coincides with young stars (see Fig. \ref{fig:co_xray}). In contrast, the X-ray dark region just west of the spur is almost perfectly traced by high $\tau_\text{353}$ values. 
We also note a small overlap in the southwestern part of the X-ray spur, with high $\tau_\text{353}$ and no observable depression in X-ray brightness. A similar overlap is also observed for the \ion{H}{i} D-component. This suggests that, the southern part of the X-ray spur consists of less dense ISM and might be slightly located in front of the D-component and closer to the observer.

The radio continuum emission in the vicinity of the spur was shown to have a strong negative spectral index \citep{radio_conti}. This indicates stronger magnetic fields, hence nonthermal radio emission. The stronger magnetic fields could be a result of compression of the ISM caused by the massive collision of the \ion{H}{i} L- and D-components. In direct comparison, the radio continuum emission is stronger in the vicinity of the spur, but weaker in the spur itself. This indicates lower densities of the ISM in the spur compared to LMC-SGS~2 and 30~Dor.
\subsubsection{Spectral properties}
From the spectral analysis, we observe a higher temperature in the X-ray spur and 30 Dor area than for the rest of the diffuse emission. This is consistent with the harder color in the X-ray mosaic (Fig. \ref{fig:three-color_soft}). The spectra of the diffuse emission can also be fit with a two-component plasma model, in which the hotter component is only significantly necessary near 30 Dor and the X-ray spur. In the western part of the mosaic, the plasma temperature is much lower in comparison and there is no significant contribution of a second hot plasma component. When comparing the surface brightness $L_X^A$ of the hot component, we get on average values two times larger in the spur with $L_X^A \sim 5\times 10^{33}$ erg s$^{-1}$ arcmin$^{-2}$, compared to the northwest.
This indicates two different plasma components in the LMC: a low temperature component with $kT \sim 0.2$~keV which follows the disk and major stellar population of the galaxy, and a hot component with $kT > 0.6$~keV caused by energy input of massive stars and collisions. Similar plasma compositions have been also found in other galaxies, such as M31 \citep{m31_kavanagh_new}. 
The density $n_H$ of the hot plasma component also seems enhanced in the eastern part of the mosaic compared to the west. We obtain the lowest densities in both the X-ray dark region and the northwest. The density in the X-ray spur is significantly higher in comparison. This might be a further indication that part of the plasma was compressed by the tidal interactions caused by the colliding \ion{H}{i} components.

The absorption column in front of the plasma is higher in the eastern part of the X-ray mosaic. Toward 30 Dor, we obtain the highest absorption of about $N_{\mathrm{H}}^X\sim 1.0 \cdot 10^{22}$~cm$^{-2}$, while near the X-ray spur we obtain absorptions of $N_{\mathrm{H}}^X\sim 0.5-0.7 \cdot 10^{22}$~cm$^{-2}$ with lower values in the central and southern part of the X-ray spur. In the western part we obtain $N_{\mathrm{H}}^X\sim 0.4 \cdot 10^{22}$~cm$^{-2}$. This agrees well with a study of SNRs in the LMC by \cite{snrs_lmc_maggi}, which showed high $N_{\mathrm{H}}^X$ in front of the SNRs in the eastern part of the mosaic (Fig. \ref{fig:three-color_soft}) and lower $N_{\mathrm{H}}^X$ in the western part. The high absorption in the eastern part of the mosaic also correlates well with high dust optical depth $\tau_{353}$. The foreground $N_{\mathrm{H}}^X$ correlates weaker with the \ion{H}{i} gas distribution in comparison. This suggests, that the plasma east and west of the X-ray spur is observed through absorption by the cold ISM, with weaker absorption in the central X-ray spur. The background $N_{\mathrm{H,BG}}$  behind the soft diffuse plasma appears to be the highest in the X-ray spur and 30 Dor with $N_{\mathrm{H,BG}} \sim 1.3-1.4\cdot 10^{22}$~cm$^{-2}$. In the western part we obtain zero or lower absorption. 

We also calculated the ratio $N_{\mathrm{H}}^X/N_H$ to gain additional insight into the geometry. This ratio can indicate the location of the diffuse plasma in relation to the HI gas. The inferred X-ray absorption column is typically expressed as equivalent of a hydrogen column. However, lower column densities of $N_{\mathrm{H}}^X < 0.6\cdot 10^{22}$~cm$^{-2}$ and ratios $N_{\mathrm{H}}^X/N_H$ significantly higher than one are indicative of the presence of molecular gas \citep{nh_ratio}. 
 For the X-ray spur this ratio is the lowest with $N_{\mathrm{H}}^X/N_H\sim 1.1$.  For LMC-SGS~2 and 30 Dor we also obtain relatively low ratios, albeit larger compared to the spur. 
  In contrast, in the northwest the ratio is almost four times higher than in the X-ray spur, with $N_{\mathrm{H}}^X/N_H\sim 3.8$.
 Here, the plasma is most likely located behind most of the HI gas and also significant amounts of molecular gas. The anticorrelation with the HI I- and D-components and low $N_{\mathrm{H,BG}}$ values also support this.
 
 In the X-ray spur the ratio is very close to one, however, the distribution of CO and dust (Fig. \ref{fig:co_xray} \& Fig. \ref{fig:tau353_xrays_cont}) indicate a more complex geometry of the different ISM components. We only observe significant amounts of CO and dust in the western part of the X-ray spur, which explains elevated $N_{\mathrm{H}}^X/N_H$ values. Also, we observe anticorrelation of X-rays with all three HI components here, while in the eastern part we only obtain anticorrelation with the L-component (Fig. \ref{fig:X-ray_HI_correlation_quadrants}). High values of $N_{\mathrm{H,BG}}$, especially in the central and eastern part of the X-ray spur, indicate the presence of significant amounts of HI gas and other absorbing materials behind the plasma in the east. Therefore, it is likely that the eastern part is located slightly in front of the disk, closer to the plane of the L-component, while the western part of the X-ray spur seems to be located behind - or mixed in - with the HI I- and D-component.
\subsubsection{Constraints from the stellar energy input}
We compared the energy output of the stellar population into the ISM with the results derived from the spectral analysis to further constrain the origin of the hot plasma in the X-ray spur.
To obtain the stellar energy input, we used the Starburst99 synthesis code \citep{starburst99_01, starburst99_02, starburst99_03, starburst99_04}. We assumed a single starburst episode and used the star formation history  of the LMC derived by \citet{sfh_lmc} to calculate the total stellar mass of the initial population. The simulations were carried out for two regions. The first in the central spur with a box centered at RA~$= 5^{\text{h}}44^{\text{m}}02^{\text{s}}$, Dec~$= -70\degr 06\arcmin 04\arcsec$ with a diameter of $24\arcmin \times 24\arcmin$. The second in the part northwest of the spur - closer to the bar - which appears softer in X-rays. For the latter, we defined a box centered around RA~$= 5^{\text{h}}30^{\text{m}}02^{\text{s}}$, Dec~$= -69\degr 24\arcmin 07\arcsec$ with a diameter of $12\arcmin \times 24 \arcmin$ . We added up the best fit star formation rate of all quadrants calculated by \citet{sfh_lmc} for $\log( t\,[\mathrm{yr}] ) =  6.8$ that are located inside the respective defined region.  From this, the total stellar mass was calculated by assuming a constant star formation for the given time interval. For higher ages $\log( t\,[\mathrm{yr}] ) \geq  7.1$ there is a noticeable break in the star formation, therefore we limit the simulation to the most recent star formation episode. We then carried out the simulations assuming LMC metallicities and using an initial mass function (IMF) with two exponents after \citet{kroupa_imf}.
The energy input in the respective region was then determined by taking the accumulated total energy generated by the stellar winds and SNRs of the simulated population after $\log ( t\,[\mathrm{yr}] ) = 6.8 $. To compare these numbers with the results derived from the spectral analysis, we divided the total energy by the corresponding volume, assuming the same depths as for the calculations of the plasma properties. For a direct comparison, we also calculated the mean energy density values of the cold and hot plasma components inside the defined regions as described in Sect. \ref{sec:phys_prop}.

For the spur we obtain an energy density of $\varrho_{\mathrm{stellar}} = 1.97\cdot 10^{-12}$~erg\,cm$^{-3}$ injected by the stellar population into the ISM. Compared with the combined energy density of the cold and hot plasma component $\varrho_X = 1.94\cdot 10^{-11}$~erg\,cm$^{-3}$ ($f = 0.1$), the stellar input is lower by an order of magnitude. 
Even when we extend the simulations in the X-ray spur to a higher age $\log( t\,[\mathrm{yr}] ) = 7.1$, the difference is still a factor of $\sim 5$.
In contrast, for the region in the bar and northwest of the spur we obtain an energy density of $\varrho_{\mathrm{stellar}} = 2.42\cdot 10^{-11}$~erg\,cm$^{-3}$ from the simulations. Here, the energy density matches the value derived from the X-ray spectral analysis of $\varrho_X = 1.90\cdot 10^{-11}$~erg\,cm$^{-3}$ ($f = 0.1$) very well.
 The large discrepancy between the stellar energy input and the energy content of the hot plasma in the X-ray spur, in contrast to the consistent region in the bar, suggests that another mechanism for additional energy input into the system is needed. 
\subsubsection{Origin of the X-ray spur}
Near the massive stellar cluster R136, we obtain the highest plasma temperatures, which can be explained well with the standard scenario for stellar wind bubbles around massive stars. The strong H$\alpha$ emission also confirms strong photoionization in this area. The massive energy input into the ISM may also have caused shock ionization around 30~Dor and the stellar cluster R136 (Fig. \ref{fig:lmc_xray_sii2ha_ratio}).  We also obtain a relatively high foreground absorption of the X-ray emission there. This agrees well with high $\tau_{\text{353}}$ values and an overlap of all three \ion{H}{i} components. However, we do not observe similar indications for energy input by stellar winds in the X-ray spur.
Moreover, simulations of the stellar populations show that for the region west of 30~Dor and the X-ray spur the stellar energy output matches the energy contained in the plasma very well. For the X-ray spur however, the energy output of the stellar population is too small by an order of magnitude compared to the plasma energy content. This discrepancy suggests, that additional energy was introduced to the system by different means.

 We therefore propose that the ongoing collision of the L-component with the LMC disk caused compression of some ``relic'' plasma from previous heating processes, giving rise to enhanced plasma temperatures. The plasma condition before encountering the collision was most likely similar to the low temperature plasma observed in the western part of the mosaic, which appears unaffected by the collision. The collision then most likely compressed the ISM in the X-ray spur - thus increasing the plasma temperature - where densities of the L- and D-component were sufficient. This scenario would explain, why the entire eastern part of the X-ray mosaic appears greener (i.e., harder) in color and would also explain why the X-ray spur shows higher temperatures compared to the western part, without any presence of massive stars and their energy input. The \mbox{[\ion{S}{ii}]/H$\alpha$ ratio} comparison with X-rays (Fig. \ref{fig:lmc_xray_sii2ha_ratio}) shows that filaments are only found for the 30~Dor region, while in the X-ray spur we observe no filaments at all, despite similar high temperatures.  According to the numeric simulations by \citet{ccc_simulations_ana}, cloud-cloud collisions can lead to a shocked contact layer where the gas is compressed and heated. Although the simulations were performed on smaller scales, the simulation results indicate that the proposed scenario for the X-ray spur is possible.
\subsubsection{Collision geometry: The formation of the X-ray spur and the high-mass stars}
\begin{figure*}
	\centering
		\includegraphics[width=\textwidth]{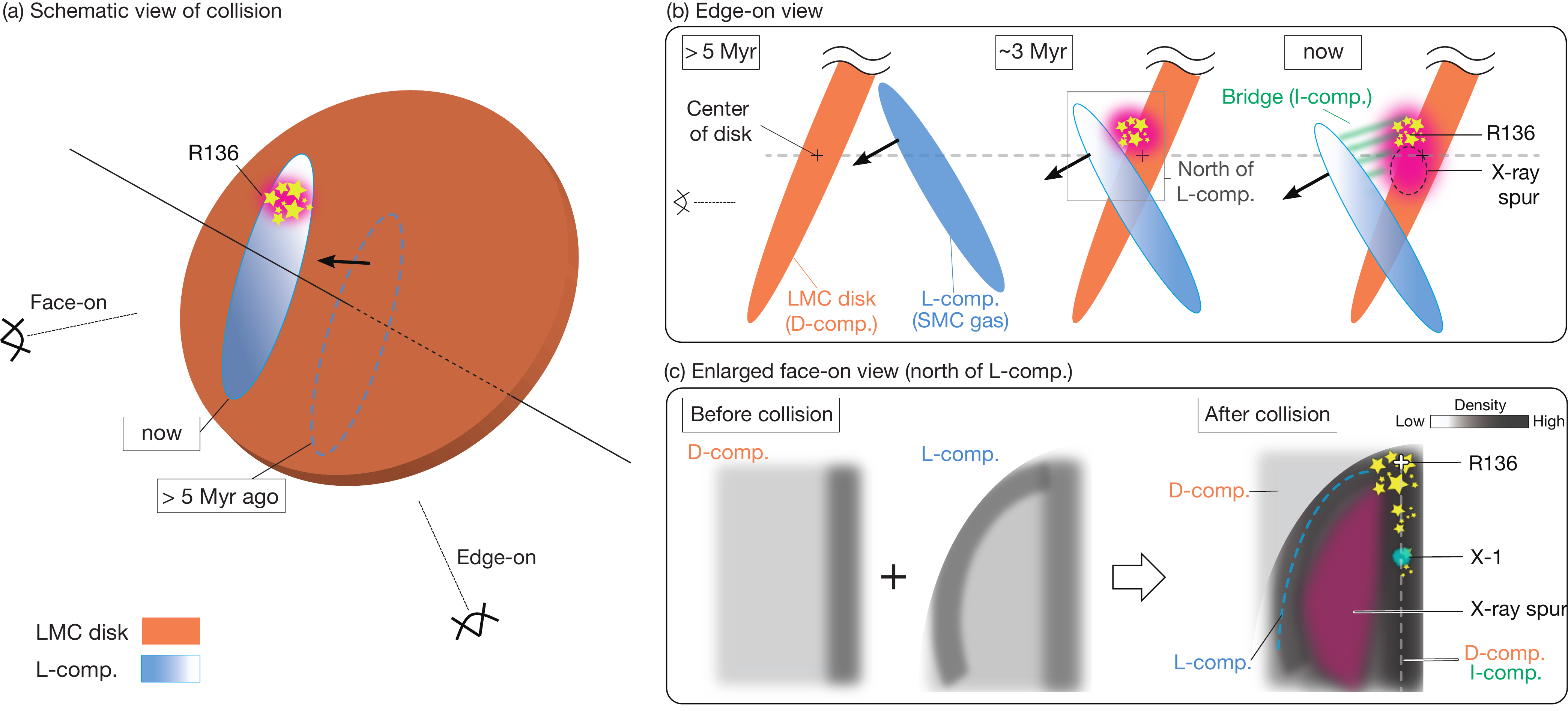}
		\caption{\label{fig:sketch} Schematic views of the collision scenario between the \ion{H}{i} L- and D-component. Orange represents the LMC disk, i.e., the \ion{H}{i} D-component, and blue the \ion{H}{i} L-component.
		(a) The collision as viewed by different observers with the L-component position 5\,Myr ago (dashed ellipse) and now (solid). The viewing directions for both the face-on and edge-on observer are indicated. (b) Edge-on view of the collision at $> 5$~Myr ago (left), $\sim 3$~Myr ago (middle) and now (right). The red colored area depicts ISM heated and compressed by massive stellar winds and/or the collision itself. (c) Detailed zoom-in of the face-on view near R136 and the X-ray spur. The panel depicts the densities of the different \ion{H}{i} components before (left) and after the collision (right). The gray scale correspond to the density of the \ion{H}{i} gas. The dark-red colored area indicates the compressed plasma in the X-ray spur as a result of the collision, giving rise to an enhanced plasma temperature.}
\end{figure*}

We discuss the interactions of the \ion{H}{i} L-, I-, and D-components in more detail. The present study indicates that the \ion{H}{i} components and the X-ray spur show complementary distribution on a kpc scale. This suggests that the X-ray spur is surrounded by the \ion{H}{i} components, and that the X-ray spur is shaped by absorption by the \ion{H}{i} components. 
We present a possible collision geometry as a schematic overview in Fig. \ref{fig:sketch}, where the X-ray spur and the high-mass stars were formed by the collision of the \ion{H}{i} L- and D-components. In a kpc scale, the L-component is bar-shaped and elongated in the north-south with a tilt to the LMC disk, where the northern end is closer to us as shown in Fig. \ref{fig:sketch}a (the stereogram) and Fig. \ref{fig:sketch}b (the edge-on view). At the present epoch the northern half of the L-component has penetrated through the disk with ~100 pc thickness, while the southern half is yet during or prior to collision. This configuration is suggested by the distribution of the I-component whose intensity decreases toward the south (Fig. \ref{fig:HI_images} or Fig. \ref{fig:X_HI_cont}). This tilt is also consistent with the $A_v$ distribution, which is large in the north and decreases to the south \citep{av_lmc}.
We postulate that the two clouds, the L- and D-components, prior to the collision have density distribution as shown in Figure \ref{fig:sketch}c (the enlarged face-on view of the north). The L-component is characterized by two filamentary features which are similar to the $\tau_{353}$ distribution, while the D-component has a gradual density decrease from the inner part to the outer part. The filamentary distribution of the L-component is possibly produced by the tidal interaction according to the simulation by \citet{tidal_simulation2}. The strong interaction will take place in a thin layer of the neutral gas in the disk, and the gas motion during the collision vertical to the line of sight may not be significant. The collision takes place in three ways in each part of the two-velocity components depending on density as follows:

Firstly, the eastern filament of the L-component collided with the edge of the LMC disk having low column density. The L-component is not decelerated by the collision and the gas compression is weak, causing no heating nor star formation.

Secondly, in the area surrounded by the two filaments, two velocity components have low column densities similar to each other, and the collision between the low density \ion{H}{i} gas compressed the plasma in the X-ray spur, increasing the plasma temperature, but not reaching the necessary densities to trigger star formation. Similar low column densities of the two components resulted in significant deceleration, converting the L-component into the I-component. The I-component is likely embedded in the X-ray spur, which can be thicker than the \ion{H}{i} components due to high sound speed more than 100 km/s, and explains the small X-ray absorption toward the I-component. It is possible, that the X-ray spur is shaped either by absorption and/or confinement by the \ion{H}{i} components, as well as the collision-induced compression and therefore heating of the plasma. The decrease of the X-ray spur to the south is explained due to the weak collision there, caused by the tilt of the L-component relative to the disk.

And thirdly, the western filament of the L-component collided with the dense part of the D-component, and the velocity was decelerated with strong compression. This leads to significant collisional deceleration and compression of the L-component which is converted to the I-component, and triggered the high-mass star formation including R136.

Theoretical studies of cloud-cloud collision were undertaken for the molecular clouds in the Milky Way with collision velocity of 5-20 km/s \citep{habe_ohta_ccc, ccc_simulations_ana, inoue_2013, takahira_2014, inoue_2018}, and they are not directly applicable to the present collision between \ion{H}{i} clouds at a velocity in the order of 100 km/s. MHD simulations of such a collision are under way by Maeda, Inoue, and Fukui, which will help to better understand the present collision.

%
\section{Summary \& conclusion}
We performed a large scale spectral analysis of new and archival XMM-Newton data in the southeastern part of the LMC, in particular around the X-ray spur. Utilizing the voronoi tessellation algorithm for spatial binning, we obtained detailed parameter maps for the X-ray emitting plasma properties. The spectra are well fit with a single APEC model for most regions. For the X-ray spur we obtain an average plasma temperature of $kT= 0.64^{+0.13}_{-0.05}$~keV. In comparison, the diffuse X-ray emission west of 30 Dor and the X-ray spur appears cooler and averages around $kT = 0.27^{+0.03}_{-0.01}$~keV. In the vicinity of 30 Dor and the massive stellar cluster R136, we obtain the highest temperatures with $kT= 0.66^{+0.05}_{-0.02}$~keV. 

In the east where 30 Dor, LMC-SGS~2 and the X-ray spur are located, a two component model for the diffuse emission fits the observed spectrum significantly better. One component tends to temperatures of $kT\sim 0.2$~keV and most likely corresponds to the diffuse emission consisting of the undisturbed hot ISM in equilibrium and unresolved stellar emission. For the hotter component, we obtain typical temperatures of $kT\sim 0.5\mathrm{-} 0.9$~keV. In the part west of the X-ray spur, we see no significant contribution of this component. The luminosity of the hot component is larger by at least a factor of two in LMC-SGS~2, 30 Dor region, and the X-ray spur compared to the west. Additionally, we get hydrogen densities $n_H$ which are higher in the spur with $n_H \sim 10^{-2} f^{-1/2}$ cm$^{-3}$ compared to the northwest $n_H \sim 0.7 \cdot 10^{-2} f^{-1/2}$ cm$^{-3}$.

 From the X-ray spectral analysis, we also obtained foreground absorbing column densities $N_{\mathrm{H}}^X$  for each tessellate region. Compared to the other regions, the column densities increase toward 30 Dor, where we get $N_{\mathrm{H}}^X\sim 1.0 \cdot 10^{22}$~cm$^{-2}$. In the vicinity of the X-ray spur, we obtain equivalent hydrogen column densities of $N_{\mathrm{H}}^X\sim 0.5 \cdot 10^{22}$~cm$^{-2}$. Directly east and west of the X-ray spur we obtain higher densities of $N_{\mathrm{H}}^X > 0.7 \cdot 10^{22}$~cm$^{-2}$. In the cooler part west of 30 Dor and the X-ray spur, the absorption column values are the lowest with $N_{\mathrm{H}}^X\sim 0.4 \cdot 10^{22}$~cm$^{-2}$. In contrast, the ratio $N_{\mathrm{H}}^X/N_H$ is significantly lower in the eastern part compared to the west, with the lowest ratio found in the X-ray spur. The ratios agree well with the absorption column $N_{\mathrm{H,BG}}$ behind the plasma, where the highest values are obtained in 30 Dor, LMC-SGS~2 and the X-ray spur. 
 
We also carried out a multiwavelength comparison with X-rays to study the ISM conditions in more detail. We focused on the comparison between \ion{H}{i} and X-rays, as indications for collisions between different \ion{H}{i} components were observed near the X-ray spur \citep{HI:fukui_R136}. We compared the intensity of X-rays with each \ion{H}{i} component and carried out a correlation analysis. The X-ray spur appears to be complementary with the L-component, indicating that the plasma emitting structure is surrounded by cold, dense ISM. For the I-component - a tracer for the interaction between L- and D-component - we actually obtain a correlation at the position of the spur. Together with the higher plasma temperatures this indicates that the fast collisions - up to 100\,km\,s$^{-1}$ - between the L- and D-component might have indirectly caused additional heating of an existing hot plasma ($kT \sim 0.2$~keV) and other gas via compression. We also compared X-rays with the dust optical depth at 353\,GHz ($\tau_{\text{353}}$), which also appears to be complementary for most parts of the X-ray spur. We obtain a good match of foreground absorbing column densities $N_{\mathrm{H}}^X$\  from the spectral fits and the optical depth $\tau_{\text{353}}$. 
We also compared the \element[][12]{CO} emission with the X-ray emission and see a strong CO ridge located on the eastern edge of the X-ray spur. The brightest CO regions correlate well with the known positions of YSOs, while they appear more numerous toward 30 Dor. In the spur YSOs are only observed in the northernmost part.
Additionally, we compared the optical line emission of H$\alpha$ and the [\ion{S}{ii}]/H$\alpha$ line ratio. We see strong H$\alpha$ emission and a high [\ion{S}{ii}]/H$\alpha$ line ratio in the vicinity of 30 Dor. The combined stellar winds and strong radiation field of the high number of massive stars in this region are most likely responsible for the emission. At the position of the X-ray spur we neither observe any significant H$\alpha$ emission nor [\ion{S}{ii}]/H$\alpha$ line ratio. We also find that no massive stars are located in the X-ray spur.

In addition, we placed constraints on the origin of the hot plasma in the X-ray spur. We simulated the stellar population in the central X-ray spur and its energy injected into the ISM via stellar winds and SNRs, and made comparisons with the derived energy content of the X-ray emitting plasma. We found that the stellar energy input is too low by an order of magnitude to have caused the heating of the X-ray plasma to higher temperatures. This suggests, that additional energy must have been introduced into the system via a different mechanism.

The collision between the \ion{H}{i} L- and D-components triggered the formation of the many high-mass stars including 30 Dor. While the velocities are not sufficient for directly heating the plasma, an already existing low temperature plasma might have been compressed significantly by the collision. It is likely, that this compression gave rise to an enhanced plasma temperature in the X-ray spur, while at the same time densities to trigger star formation were not reached. This scenario is consistent with the highly filamentary distribution of the dust optical depth $\tau_{353}$, which accompanies lower density inter-filament gas toward the X-ray spur. We therefore propose that the enhanced plasma temperature in the X-ray spur was caused by the collision between the \ion{H}{i} L- and D-components, as depicted in Fig. \ref{fig:sketch}. Future detailed multiwavelength observations of nearby galaxies might be able to tell us, if the physical conditions found in this study are typical for galaxies undergoing massive cloud-cloud collisions and tidal interactions.
\section*{Acknowledgments}
We thank the anonymous referee for the constructive and helpful feedback.
This study is based on observations obtained with XMM-Newton, an ESA science mission with instruments and contributions directly funded by ESA Member States and NASA.
This research has made use of data and/or software provided by the High Energy Astrophysics Science Archive Research Center (HEASARC), which is a service of the Astrophysics Science Division at NASA/GSFC and the High Energy Astrophysics Division of the Smithsonian Astrophysical Observatory. This research also made use of NASA's Astrophysics Data System; the VizieR catalogue access tool, CDS, Strasbourg, France; matplotlib, a Python library for publication quality graphics \citep{Hunter:2007}; XSPEC \citep{xspec_general}; Astropy, a community-developed core Python package for Astronomy \citep{2018AJ....156..123A, 2013A&A...558A..33A}; CIAO X-ray Analysis Software \citep{ciao}; ds9, a tool for data visualization supported by the Chandra X-ray Science Center (CXC) and the High Energy Astrophysics Science Archive Center (HEASARC) with support from the JWST Mission office at the Space Telescope Science Institute for 3D visualization. The acknowledgements were compiled using the Astronomy Acknowledgement Generator. M. S. acknowledges support by the DFG through the grants SA 2131/3-1, 5-1, 12-1. This work was partly supported by a grant-in-aid for Scientific Research
(\mbox{KAKENHI}~No.~JP15H05694) from MEXT (the Ministry of Education, Culture,
Sports, Science and Technology of Japan) and JSPS (Japan Society
for the Promotion of Science).
%

\bibliographystyle{aa}
\bibliography{literature}
\end{document}